**A 3D multimodal optical coherence tomography foundation model for retinal and systemic diseases with cross-cohort and cross-device validation.**

Zixuan Liu[1], Hanwen Xu[1], Addie Woicik[1], Linda G. Shapiro[1], Marian Blazes[2,3], Yue Wu[2,3], Verena Steffen[4], Catherine Cukras[4], Cecilia S. Lee[2,3], Miao Zhang[4]#, Aaron Y. Lee[2,3]#, Sheng Wang[1]#

[1]Paul G. Allen School of Computer Science and Engineering, University of Washington, Seattle, WA

[2]Department of Ophthalmology, University of Washington, Seattle, WA, USA

[3]Roger and Angie Karalis Johnson Retina Center, University of Washington, Seattle, WA, USA

[4]Genentech Inc, South San Francisco, CA, USA

[#]Email: zhang.miao@gene.com, leeay@uw.edu , swang@cs.washington.edu

## Abstract

Vision loss caused by retinal diseases remains a leading global cause of disability. Optical coherence tomography (OCT) is an essential imaging technique for diagnosing retinal diseases, resulting in computational models for various diagnostic and prognostic tasks using OCT images. However, most of the existing approaches overlook the rich 3D structure of OCT and are unable to jointly analyze other retinal imaging modalities with OCT images. Here, we present OCTCube-M, a 3D OCT-based multi-modal foundation model framework for jointly analyzing OCT and *en face* images. OCTCube-M first developed OCTCube, a 3D foundation model pre-trained on 26,605 3D OCT volumes encompassing 1.62 million 2D OCT images. It then exploits a novel multi-modal contrastive learning framework COEP to integrate other retinal imaging modalities, such as fundus autofluorescence Imaging (FAF) and infrared retinal imaging (IR), into OCTCube, efficiently extending it into multi-modal foundation models.

OCTCube achieves best performance on predicting 8 retinal diseases, and further demonstrates strong generalizability on cross-cohort, cross-device and cross-modality prediction. We further applied OCTCube to predicting nodule malignancy (CT) and low cardiac ejection fraction as well as systemic diseases, such as diabetes and hypertension, demonstrating its wide applicability beyond retinal diseases. Based on OCTCube and COEP, we develop OCTCube-IR, which continually trains OCTCube using 26,685 pairs of OCT and IR images. OCTCube-IR is able to accurately retrieve between OCT and IR images, allowing for joint analysis between two retinal imaging modalities.

Finally, we trained a tri-modal foundation model OCTCube-EF from 4 million 2D OCT images and 402,514 *en face* (EF) retinal images. OCTCube-EF attains the best performance on predicting the growth rate of geographic atrophy (GA) across datasets collected from 6 multi-center global trials conducted in 23 countries. This improvement is statistically equivalent to running a clinical trial with more than double the size of the original study. Our analysis based on another retrospective case study reveals OCTCube-EF's ability to avoid false positive Phase-III results according to its accurate treatment effect estimation on the Phase-II results. Collectively, OCTCube-M is a 3D multi-modal foundation model framework for integrating OCT and other retinal imaging modalities, achieving improvement on cross-cohort, cross-device, cross-modality, and systemic disease prediction, further and demonstrating benefits for applications in geographic atrophy clinical trials.

## Main

Vision loss caused by retinal diseases is a leading cause of disability worldwide, significantly impacting quality of life and posing a substantial economic burden.[1–4] A variety of retinal imaging techniques, such as optical coherence tomography (OCT), fluorescein angiography (FA), fundus autofluorescence Imaging (FAF), infrared retinal imaging (IR), OCT angiography (OCTA), and adaptive optic (AO) have been developed to offer different levels of details for assessing retinal conditions. Among these, optical coherence tomography is a 3D non-invasive imaging technique that enables volumetric imaging of the microstructure of the retina.[5] Because OCT images allow clinicians to visualize the distinctive layers of retina and quantify the thickness of these layers, it has become critical for the diagnostic assessments and treatments of many ophthalmic and retinal diseases,[6,7] including glaucoma,[8] diabetic macular edema,[9] age-related macular degeneration,[10] referable diabetic retinopathy,[11,12] retinal neovascularization,[13] as well as macular hole,[14] central retinal artery, and vein occlusion.[15,16] Manual interpretation and analysis of each OCT image in a volume is not only time-consuming but also prone to human error, necessitating the development of automated algorithms and machine learning models to enhance accuracy and efficiency in clinical practice.[17]

However, there are two challenges in developing machine learning methods for analyzing OCT images. First, it remains unclear how to jointly model OCT with other retinal imaging modalities, such as FAF and IR, which are often used together with OCT in clinics to provide complementary views for the retina. Second, existing approaches utilize 2D OCT images while overlook the 3D structure of an OCT volume.[18] In OCT imaging, the diseased macular areas might expand across the 3-dimensional fovea-centered spatial region because of the radial symmetry of the fovea anatomy.[19,20] Thus, modeling 3D OCT structure provides both front forward *en face* and depth information for a comprehensive characterization of the retina. Recently, foundation models have achieved state-of-the-art performance on various biomedical applications, especially on modeling biomedical images.[21–23] Foundation models exploit self-supervised learning to learn high-quality representations using large-scale unannotated images and then use these representations to train predictors for various downstream applications. Given the large numbers of OCT being collected in routine clinical practice, it is possible to develop large-scale foundation models to automate disease diagnosis from OCT images and other retinal images.

Here, we propose a multi-modal 3D foundation model framework OCTCube-M for jointly analyzing OCT and *en face* fundus images. Our framework has two steps: we first pre-trained a 3D OCT foundation model OCTCube based on the 3D Masked Autoencoder (**Fig. 1a**). We next propose COEP, a novel multi-modal contrastive learning technique (**Fig. 1b)** that can integrate other types of retinal images into OCTCube, effectively extending OCTCube to a multi-modal foundation model. By using this two-step framework, we are able to leverage large-scale OCT images that do not have matched *en face* (EF) images. It also eases the joint training of 3D OCT volumes and 2D *en face* images by developing a 3D model as the foundation for the joint training. By combining COEP and OCTCube, we develop two multi-modal foundation model: OCTCube-IR, which is a bi-modal model trained using 26,685 OCT volumes and 26,685 IR images, and OCTCube-EF, which is a tri-modal model trained using 71,462 OCT volumes, 170,832 IR and 231,682 FAF images. These resulting models present 3D solutions for mulit-modal retinal imaging that can be broadly applied to retinal diseases across devices, cohorts and modalities and systemic diseases.

We first evaluated OCTCube, the OCT foundation of OCTCube-M, on applications when only OCT is available, including retinal disease, cross-cohort, cross-device, cross-modality and systemic disease prediction. In prediction of retinal diseases, OCTCube improved existing methods from 0.77 to 0.81 AUPRC for eight retinal diseases on in-house benchmarks and from 0.76 to 0.83 AUPRC on four external benchmarks, indicating its broad applicability and strong generalizability. OCTCube also demonstrated stronger performance on cross-device prediction in analyzing OCT acquired with Topcon Maestro 2 and Zeiss Cirrus devices while the pre-training data were acquired using Heidelberg Spectralis devices. OCTCube achieved 6.4% AUPRC and 4.8% AUROC improvement compared to the best competing method, even when the training images used by the competing method were acquired from the more similar device as the test set. Finally, OCTCube achieved the best performance on predicting seven systemic diseases using OCT images, as well as nodule malignancy using CT image and low cardiac ejection fraction using ultrasound, underscoring its applicability beyond retinal diseases.

Next, we evaluated our multimodal framework OCTCube-M by developing two multimodal models OCTCube-IR and OCTCube-EF. OCTCube-IR was further developed based on OCTCube using 26,685 OCT volumes with paired IR images. It achieved accurate retrieval performance between OCT images and IR images, facilitating joint modeling even when one modality is missing. OCTCube-EF is a tri-modal model that integrates 71,462 OCT volumes, 170,832 IR and 231,682 FAF images. By modeling these three modalities together, OCTCube-EF achieved substantial improvement on predicting geographic atrophy (GA) progression across datasets collected from 6 multi-center global trials conducted in 23 countries. In particular, it improved the R-square from 0.485 to 0.531 on predicting future GA lesion growth rate from baseline images alone. This improvement is statistically equivalent to running a clinical trial with more than double the size of the original study[24], potentially saving millions of dollars from recruiting more patients as exemplified by a recent phase-II clinical trial (NCT03972709). Another retrospective case study demonstrates OCTCube-EF's ability avoid false positive Phase-III trial decisions by obtaining more accurate treatment effect estimation on the Phase-II results (NCT01229215), potentially saving up to hundreds of millions of USD and preventing patients from participating in similar trials with ineffective treatments in the future. Collectively, we present OCTCube-M, a 3D multi-modal foundation model for integrating optical coherence tomography and other fundus imaging modalities. Based on OCTCube-M, we developed three foundation models covering various retinal imaging modalities, achieving the best performance on cross-cohort, cross-device, cross-modality, and systemic disease prediction, further enhancing a clinical trial application for geographic atrophy through improved prognostic modeling and statistical integration.

## Results

### Overview of OCTCube-M

Our 3D multi-modal foundation model is developed based on two novel components, an OCT foundation model OCTCube and a novel multi-modal contrastive learning-based integrative loss COEP. OCTCube is a foundation model pre-trained using 26,605 3D OCT volumes. The key idea of OCTCube is to model the 3D anatomic structure during representation learning holistically instead of aggregating the learned features from each 2D slice (**Fig. 1a**). To achieve this, OCTCube exploits 3D masked autoencoders as the pre-training framework. In particular, OCTCube first splits an OCT volume into small 3D cubes. It then randomly masks 90% of cubes and uses an encoder-decoder architecture to reconstruct these masked cubes. By training this encoder-decoder architecture using 26,605 OCT volumes, OCTCube is able to obtain a high-quality encoder that can derive accurate representations for a new OCT volume. The decoder will be discarded for downstream applications while the parameters of the encoder will be updated for each downstream task according to the task-specific annotations, such as disease labels.

COEP is a contrastive learning-based framework that integrates other retinal imaging modality into OCTCube (**Fig. 1b**). The key idea is to first learn a separate encoder for each modality, which projects all modalities into the same space. It then exploits contrastive learning to align different modalities by encouraging images of the same eye to be similar while images of different eyes to be dissimilar in the embedding space. COEP can be applied to any number of imaging modality. By using OCTCube to initialize the encoder for OCT images, this alignment is grounded on OCT images, allowing us to incorporate other retinal imaging modalities while capturing the 3D structure of the retina.

We hypothesize that OCTCube-M is a broadly applicable and generalizable foundation model for various datasets, diseases and devices by effectively modeling 3D OCT volumes. We evaluate OCTCube-M on 32 tasks spanning retinal disease prediction, cross-cohort prediction, systemic disease prediction, cross-device prediction and cross-modality analysis (**Fig. 1c, Supplementary Fig. 1**). OCTCube-M demonstrates the best performance on 31 out of 32 tasks and comparable performance on the other two tasks.

### Proof-of-concept evaluation of 3D modeling over 2D

To investigate the advantage of 3D modeling over 2D modeling, we calculated the similarity between two slices in the same OCT volume. We found that nearby slices are more similar to slices far away to each other in terms of root mean square error (RMSE) (**Supplementary Fig. 2a**) and structural similarity index measure (SSIM) (**Supplementary Fig. 2b**), indicating the presence of repetitive patterns between nearby slices, suggesting the opportunities to exploit slices around the center slice to enhance the signal-to-noise ratio. To explore the potential of utilizing multiple slices for disease probability, we used RETFound to assign a disease probability of Age-related Macular Degeneration (AMD) for each slice. We then aggregated the predictions from $k$ slices around the center slice by averaging their predicted probabilities (**Supplementary Fig. 3c**). When $k$ is 0, this aggregation approach is the same as a 2D model that only considers the center slice. We found that the best AUROC and the best AUPRC were achieved when k was 14 and 10 respectively, indicating that considering more slices can boost the prediction performance (**Fig. 1d**). We observed similar patterns on seven other retinal diseases (**Supplementary Fig. 4**). Moreover, because RETFound is pre-trained and fine-tuned only using center OCT slices, when all slices are considered (*k*=60), the prediction performance could be worse than when only considering center slices. Furthermore, the efficacy of a 3D model that simply averages the predicted probability of all slices is compromised, as many diseases develop and progress at different locations and rates. This variability highlights the need for a more advanced 3D approach that can adapt to different disease areas.

To further support this conclusion, we present case studies of two AMD patients (**Fig. 1e,f, Supplementary Fig. 5a,b**) who were both predicted incorrectly as non-AMD patients by RETFound using the center slice (slice 30). In contrast, OCTCube used other nearby slices to predict AMD as drusens, small deposits in the retina, were observed in these slices and highlighted as the source of prediction by the corresponding saliency maps (**Fig. 1e,f, Supplementary Figs. 5c,d, 6-7**).

**OCTCubes offers accurate and interpretable retinal disease prediction**

We first evaluated OCTCube on the prediction of eight retinal diseases, including Primary Open-Angle Glaucoma (POAG), Diabetic Macular Edema (DME), Age-related Macular Degeneration (AMD), Epiretinal Membrane or Macular Hole (ERM/MH), Diabetic Retinography without Macular Edema (DR), Central Artery / Vein Retinal Occlusion (CRAO/CRVO), Posterior Vitreous Detachment (PVD) and Retinal Neovascularization (RNV). We exploited different sizes of fine-tuning sets (20% data in **Fig. 2a-b**, 80% data in **Supplementary Figs. 8-9**) to fine-tune our pretrained model for each disease and evaluate it on the held-out test set. We set the pre-training patient collection, the fine-tuning patient collection, and the test patient collection to be independent from each other to ensure a practical inductive learning setting. Competing methods are fine-tuned and evaluated using the same patient collection.

We compared OCTCube to supervised models that did not leverage unannotated images for pre-training and found that OCTCube substantially outperformed them with an average 53.8% AUPRC and 45.4% AUROC relative improvement, indicating the importance of pre-training (**Fig. 2a-b**). RETFound, which is pre-trained on 2D images, also outperformed supervised models, reemphasizing the effectiveness of pre-training. Next, we assessed the advantage of considering the entire 3D volumes by comparing OCTCube to RetFound (center), which uses the embedding of the center slice, and RETFound (all), which averages the embeddings of slices in the same volume (**Supplementary Fig. 2a,b**). We found that OCTCube significantly outperformed both variants of RETFound on 7 out of the 8 diseases (paired t-test p-value < 1e-3), demonstrating the effectiveness of modeling the 3D volume. The improvement of OCTCube is more prominent on POAG among the eight diseases. Because POAG diagnosis relies on measuring the change of layer thickness throughout the entire macula, requiring the global context of the full volume at the same time to determine the diagnosis,[25] this larger improvement further reveals OCTCube's ability to leverage retinal patterns beyond the fovea. Our visualization further showed that OCTCube exploited different retinal regions to make predictions for different diseases (**Fig. 2c)**, such as the peripapillary region for glaucoma, geographic atrophy for AMD. Interestingly, the supervised (all) model did not outperform the supervised (center) model, suggesting that the benefit of 3D modeling mostly arose from the pre-training stage. In conclusion, our experiments on eight retinal disease predictions in the inductive learning setting demonstrate the effectiveness of OCTCube by leveraging the 3D structure of OCT volumes.

To further understand the improvement of OCTCube, we visualized the saliency maps of an AMD patient on multiple slices around the center slice, reflecting the image regions that the model used to make the prediction (**Fig. 2d**). We first found that both OCTCube and RETFound capture relevant clinicopathologic features such as the drusen to predict AMD. Nevertheless, OCTCube utilized consistent regions across slices, while RETFound used relatively different regions across slices, suggesting that OCTCube considers the 3D spatial retinal structure. We further visualized the saliency map of the OCT volume across the slow-scan dimension (**Supplementary Fig. 10**), and again observed clinically relevant saliency maps generated by OCTCube at the level of the retinal pigment epithelium in the perifoveal region. In contrast, RETFound demonstrated less meaningful saliency maps in this dimension partially due to its pre-training only using the slow-scan and depth dimension, indicating the broad applicability of OCTCube to analyze OCT images from different dimensions.

**OCTCube has strong generalizability across cohorts, organs, modalities and devices**

The superior performance of OCTCube on in-house benchmarks further motivates us to evaluate its generalizability on new cohorts and on images acquired using different devices. We collected four independent benchmarks that neither RETFound nor OCTCube has access to in the pre-training stage, enabling us to more rigorously examine the performance under potential cohort variance (**Fig. 3a**). Similar to our observation on the UW Ophthalmology dataset, OCTCube outperformed 2D models with a large margin (21.3% AUPRC improvement). RETFound (all) also outperformed Retfound (center), highlighting the effectiveness of considering the entire 3D volume. Importantly, OCTCube still outperformed RETFound (all) by 8.3% AUPRC and 8.7% AUROC improvement, demonstrating the effectiveness of 3D pre-training. OCTCube also significantly outperformed a recent OCT foundation model SLIViT[26] on all tasks. Collectively, OCTCube significantly outperformed both variants of RETFound and SLIViT on all 4 independent datasets, demonstrating that OCTCube could be a more generalizability tool for retinal disease prediction across cohorts.

Since independent datasets might have varying numbers of annotated data, we next investigated the performance of OCTCube under different ratios of training data (**Fig. 3b**). We found that while OCTCube achieved the best performance on different training data ratios, its improvement was larger when there are fewer annotations, indicating its generalizability in the low-data setting or with uncommon diseases. For example, RETFound (center) achieves 0.86 AUROC that is comparable to OCTCube (0.87 AUROC) using 70% of training data, but its performance drops substantially to 0.61 AUROC when only using 20% of training data, whereas OCTCube still maintains an AUROC of 0.77.

Next, we examined a more challenging task of cross-device prediction (**Fig. 3c**), where the test data at the fine-tuning stage are collected from Zeiss Cirrus in the GLAUCOMA dataset and Topcon Maestro 2 devices[29] in the AI-READI dataset.[30] RETFound was trained on Topcon 3D OCT-2000 SA devices[31] and Topcon Triton devices,[29] while OCTCube was trained solely on the more different Heidelberg Spectralis devices.[32] A few examples of images acquired by these different devices are illustrated in **Supplementary Fig. 11,** showing clear visual distinctions between the same OCT volume from different devices. Even though RETFound was trained on Topcon images and OCTCube was not, OCTCube still significantly outperformed RETFound (center) by 29.4% AUPRC and RETFound (all) by 4.3% AUPRC on the Topcon test images, demonstrating its strong generalizability across devices. OCTCube also significantly outperformed RETFound (center) by 15.8% AUPRC and RETFound (all) by 8% AUPRC on the Zeiss Glaucoma test images, underscoring its robust and promising performance on images acquired by different devices.

Finally, we examined a cross-organ cross-modality setting following a recent work[26] that transfers the OCT model to chest CT and ultrasound (**Fig. 3d**). We studied three tasks: classifying nodule malignancy on lung CTs[27], classifying low ejection fraction ($EF_b$) using Ultrasound videos, and predicting the ejection fraction value (EF) using Ultrasound videos.[28] Overall, OCTCube obtained consistent improvement compared to SLIViT and RETFound. In particular, on nodule malignancy classification, OCTCube achieved 4.5% AUPRC improvement on the CT malignancy classification compared to the best competing method. On ejection fraction prediction, OCTCube achieved the best performance with a 10.2% AUPRC improvement, reassuring the strong transferability by OCTCube to other 3D medical imaging modalities.

**OCTCube enables the prediction of systemic diseases**

After confirming the performance of OCTCube on predicting retinal diseases, we next investigated whether OCTCube can be used to predict systemic diseases related to retinal structure by exploiting OCT volumes. We selected 7 diseases based on International Statistical Classification of Diseases (ICD-9[33] and ICD-10[34,35]) codes that are frequently found among people with retinal conditions. Among them, hypertension was diagnosed in 3,440 patients, occurring in 35% of the patients with disease records included in our study, diabetes was diagnosed in 2,420 patients, occurring in 24.7% of the included patients. We found that OCTCube significantly outperformed RETFound on 5 out of 7 disease labels in terms of AUPRC (**Fig. 4a**) and 7 out of 7 in terms of AUROC (**Fig. 4b**) on predicting concurrent systemic diseases, demonstrating the advantage of modeling 3D OCT volumes for predicting systemic diseases.

To further understand how OCTCube successfully predicts systemic diseases, we examined the performance of the aggregation approach (**Supplementary Fig. 3c**) that averages the prediction scores over multiple slices (**Fig. 4c, Supplementary Fig. 12**). We found that aggregating more slices can improve the prediction performance on all seven diseases, necessitating the modeling of 3D structures. Moreover, the improvement of OCTCube over RETFound (center) and the improvement of the aggregation approach over RETFound (center) are highly correlated (Pearson correlation 0.86 for AUROC and 0.78 for AUPRC), indicating that both methods leverage similar 3D patterns to enhance the prediction performance (**Fig. 4d,e**). Nevertheless, OCTCube consistently outperformed the aggregation approach under various numbers of slices, demonstrating its effectiveness in modeling 3D structures.

Since OCTCube achieved the largest improvement on predicting diabetes diagnosis, we have included a case study of a patient with diabetes to understand how OCTCube leverages OCT volumes for diabetes prediction (**Fig. 4f-h**). We observed many hard exudates, indicators of early diabetic retinopathy, present in various OCT scans of the right eye (**Fig. 4f**). However, hard exudates were not present in the center slice (slice 30). As a result, OCTCube was able to predict diabetic retinopathy but RETFound could not based on the OCT scans of the right eye. In contrast, a larger burden of hard exudates were found in the OCT scans of the left eye (**Fig. 4g, Supplementary Fig. 13**) as well as in both eyes during the 1 year follow up visit (**Fig. 4h, Supplementary Fig. 14**). The hard exudates are more severe in the left eye (OS) on the same acquired date and in the right eye acquired during the one-year follow-up visit, therefore both OCTCube and RETFound correctly predicted the diagnosis of diabetes in these instances. In conclusion, RETFound can only make correct predictions when patterns are visible and present on the center slice, whereas OCTCube is able to correctly predict diabetes when patterns are less visible on a single slice, leading to potential early disease detection.

**OCTCube-IR shows transferability between OCT and IR images**

After confirming the superior performance of OCTCube on analyzing OCT images, we next examined its applicability on multi-modal analysis by integrating OCT with infrared retinal (IR) images. IR is usually taken together with the OCT imaging as the navigation map and provides a much larger 2D *en face* field of view (FOV) of multiple fundus components,[32] such as optic disc, vessel direction, and the projected structure of macula. In practice, it is challenging and tedious to jointly analyze OCT volumes and IR pairs even for human experts.[36]

To this end, we trained OCTCube-IR by using COEP to continually train OCTCube using 26,605 pairs of OCT and IR images (**Fig. 5a**). We first studied two retrieval tasks: find the most relevant OCT volume given an IR image and vice versa (**Fig. 5b**). OCTCube is able to retrieve relevant images across modalities more accurately with a 0.64 Recall@1, substantially higher than the 0.46 Recall@1 attained by RETFound, indicating the effectiveness of OCTCube on aligning OCT and IR images. RETFound (all) also outperformed RETFound (center), again demonstrating the benefits of modeling OCT data in the 3D space for multi-modal analysis (**Fig. 5c, Supplementary Fig. 15a-c**). Next, we studied the retrieval performance on AI-READI where the alignment of both RETFound and OCTCube were fine-tuned using the UW Ophthalmology dataset. OCTCube again outperformed RETFound (all) and RETFound (center) on both retrieving OCT to IR and IR to OCT, and the improvement was much larger on this cross-cohort setting, reflecting its generalizability in cross-modality analysis. Furthermore, we used laterality to evaluate this cross-modality alignment by examining whether OCTCube can retrieve images with the same laterality (right (OD) or left (OS) eye). OCTCube achieved the best performance on both IR to OCT and OCT to IR laterality prediction with an average 0.97 accuracy@1 (**Fig. 5d, Supplementary Fig. 15d-f**).

Finally, we presented four case studies from the UW Ophthalmology (**Fig. 5e,f**) and AI-READI[30] (**Fig. 5g,h**) datasets and found that the improvement of OCTCube came from successfully matching fundus structure between modalities. For example, OCTCube retrieved similar IR images for the OCT volume that is optic-disc centered (**Fig. 5e**), while both RETFound (all) and RETFound (center) failed to retrieve optic-disc centered IR images. Additional examples showed that OCTCube better understands macular changes and laterality (**Fig. 5f,h**), illumination difference (**Fig. 5e,f,g**) and blood vessel structure (**Fig. 5f,g,h**), demonstrating its ability to capture spatial fundus structures through mutli-modal modeling.

**OCTCube-EF utilizes three modalities for Geographic Atrophy (GA) prognostic and structural functionality analysis**

Geographic atrophy is an advanced form of age-related macular degeneration that affects 5 million people around the world with irreversible vision loss and poses tens of billions of USD of economic burden.[2,37,38] The long progression period (up to 7 years) and limited availability of approved effective treatments result in a pressing need for GA structural functionality and prognostic analysis.[39–42] Two metrics essential to the research of treating GA are the growth rate of GA, which can be defined as the enlargement of the hypo-autofluorescence lesion area on FAF, and the loss of best corrected vision acuity (BCVA). The GA lesion growth rate is a primary clinical endpoint approved by the Federal Drug Agency (FDA) to measure the efficacy of new treatments under investigation.[24] Because GA area can be characterized and segmented using multiple 3D and *en face* (EF) retinal imaging modalities, including OCT, IR and FAF modalities,[39] there is a pressing need to develop a multi-modal model that can perform comprehensive analysis using multiple modalities jointly.

To overcome this difficulty, we proposed OCTCube-EF (**Fig. 6a**), a tri-modal ophthalmic clinical model based on OCTCube, to predict GA growth rate. Based on OCTCube, OCTCube-EF exploits COEP to train on 71,462 OCT volumes, 170,832 IR and 231,682 FAF images collected from 6 multi-center global clinical trials conducted in 23 countries (See **Data availability** for more details). These images were mainly from patients diagnosed with AMD, leading to stronger visual reasoning ability for AMD patients, despite that the disease label of these patients were not used in the continual-training stage.

We first evaluated OCTCube-EF on predicting GA growth rate and regressing BCVA on four benchmarks consisting of data from 6 clinical trials. These benchmarks are designed to evaluate different settings: Lampa as an in-house benchmark, ProximaB as a cross-cohort benchmark, GAllego as a prospective analysis benchmark, and Mahalo as a cross-device benchmark. OCTCube-EF achieved consistent improvement on both GA lesion growth rate prediction (**Fig. 6b**) and BCVA (**Fig. 6c**) regression across all benchmarks. Specifically, OCTCube-EF achieved an average of 0.534 $R^2$ on lesion growth rate prediction for four test sets, significantly outperforming the previous state-of-the-art CNN-based DenseNet[44,45] model that only considers FAF images (0.485 in average), demonstrating the effectiveness of COEP to integrate different modalities. OCTCube-M also reveals significant superiority of using OCT for this task, in contrast to traditional methods that only use FAF images, validating its ability to better capture the photoreceptor status around the fovea. While we found the original OCTCube also achieves competitive performance, aligning with other modalities further improves the adaptability and generalizability of OCTCube-EF.

This prognostic OCTCube-M model results in an effective sample size increase (ESSI) of 124% over no covariate adjustment in the clinical usage (**Fig. 6d**), which is equivalent of running a clinical trial with more than double the size of the original study.[46] Taking the prospective GAllego Phase-II trial as an example, OCTCube-EF achieved an ESSI of 61.2% (0.4 $r^2$ on the treatment arm and 0.36 $r^2$ on the control arm) significantly outperformed an ESSI of 44.9% (0.33 $r^2$ on the treatment arm and 0.29 $r^2$ on the control arm) from the DenseNet FAF model (**Supplementary Fig. 16**). Specifically, adjustment with OCTCube-EF prediction carried the same effectiveness as recruiting 269 more patients compared to the non-adjusted trial, 72 more patient recruitment with adjustment with the FAF model, saving a potential extra cost from tens of million USD compared to recruit actual patients to reach same statistical power (actual saving varies from specific clinical design and outcome).

To further validate the applicability of OCTCube-EF to clinical trial, we utilized the prediction from OCTCube-EF as a baseline covariate to apply covariate adjustment on the retrospective Phase-II Mahalo trial[44], which revealed false positive effectiveness signals for Lampalizumab (**Fig. 6e,f**). The adjusted results using OCTCube-EF prediction leads to better precision and hence tighter confidence intervals for the treatment effect estimate. Specifically, the estimated mean difference in GA lesion area decreases from 20% to 4.5% and becomes non-significant (i.e., larger than the usual Phase-II trial p-value threshold of 0.2). The adjusted subgroup analysis also revealed less favoritism to advance to the confirmatory Phase-III stage for the studied molecule, providing more accurate evidence to avoid false positive decisions costing up to hundreds of millions of USD and save patients the burden from participating in similar trials with ineffective treatments in the future.

Finally, by visualizing the saliency map, we validated that OCTCube-EF successfully captures the sub-healthy tissue near the GA lesion for GA lesion area growth prediction and fovea region for BCVA prediction (**Fig. 6g,h, Supplementary Figs. 17,18**), consistent with the recent research findings[47]. We further utilized the saliency map to examine how multimodal information is effectively integrated and contributed into the prediction (**Fig. 6i**). The saliency map of IR and FAF images of the first visit matches well to the lesion growth region after 18 months on each modality. Since the images after 18 months have never been seen by OCTCube-EF, this visualization results again demonstrating the superiority of OCTCube-EF on integrating and analyzing multiple fundus imaging modalities.

## Discussion

OCTCube-M is related to previous efforts in computational ophthalmology, especially previous works that develop deep learning-based approaches for diagnosing and predicting retinal diseases from OCT images. In particular, image processing techniques, such as contrast-limited adaptive histogram equalization (CLAHE),[51] speckle noise reduction,[52] layer delineation,[53] and layer segmentation,[54–56] and deep learning approaches, such as convolutional neural networks,[10,17,48,57] generative adversarial networks[50] and vision transformers,[58] have been exploited to develop supervised learning models for OCT-based diagnosis. In contrast to these approaches, OCTCube-M has been developed as a foundation model using self-supervised learning, which consists of a pre-training stage and a fine-tuning stage. As a result, unlabeled images can be utilized by OCTCube to provide accurate initialization for the supervised fine-tuning stage. Recently, RETFound[59] has been proposed as a retinal foundation model, which is trained on 1.6 million 2D retinal images including 736,442 2D OCT images (foveal and OCT volume center slice B-scans) and achieves state-of-the-art performance in the prediction of retinal diseases.[60] In addition, this foundation model was able to predict several systemic diseases such as myocardial infarction and ischemic strokes.[60] In contrast to RetFound, OCTCube directly modeled the 3D structure of OCT volumes. Moreover, we proposed COEP to integrate other imaging modalities into OCTCube, making it a multi-modal 3D foundation model. Our extensive experiments show that OCTCube outperforms RetFound in various settings, demonstrating the importance of 3D volumes.

The superiority of OCTCube paves the way for building OCTCube-M, a multimodal ophthalmic clinical AI model for geographic atrophy clinical development continually trained on more than 70,000 OCT volumes and 402,514 *en face* retinal images. Affecting 5 million people around the world, effectively treating geographic atrophy is a demanding need with limited solutions. OCTCube-EF reveals the potential of AI-assisted multi-modal analysis for better GA prognosis with superior disease progression analysis performance and significant budget saving benefits (up to the effectiveness of unadjusted clinical trials with 124% more patient recruitment with no extra cost) for clinical trial designs, accelerating the treatment development for GA. In clinical development, if applying OCTCube-EF for covariate adjustment in a trial like the Lampalizumab Phase 2 study Mahalo, the increased precision in the treatment effect estimate equivalent to more than doubling the number of patients (ESSI=212.5%) leads to better decision making whether a molecule should advance to the confirmatory Phase 3 stage, without the need to spend tens of millions of USD to recruit actual patients. Furthermore, in this disease area, this stage involves trials that cost up to hundreds of millions of USD and avoiding false positive decisions can save patients the burden of participating in trials with ineffective treatments and pharmaceutical companies a significant amount of money.

There are a few limitations we would like to address for OCTCube-M in future work. First, despite being equipped with the interpretable method, OCTCube-M could be more clinically useful if we are able to identify the most important cubes that contribute to the final disease predictions and filter out less important cubes to improve efficiency. We plan to explore advanced interpretable methods, such as SHAP[67] and RELPROP,[68] to identify such cubes and enable better interpretation and efficacy in the future. Second, a retinal patient could have multiple visits, presenting a longitude data containing multiple 3D volumes. We plan to incorporate such temporal information into OCTCube-M by extending it to 4D space where time is the fourth dimension.[69,70] While incorporating temporal information into medical imaging foundation models has not been well studied before, we plan to use more computationally efficient neural network architectures to allow the model to consider a time-series of volumes at the same time. Third, OCTCube-M could be further enhanced by extending its capabilities to dense prediction tasks such as segmentation and integrating additional 3D modalities, such as OCTA, to broaden its clinical utility. We plan to develop methods to handle multimodal and temporal data efficiently, enabling the model to process longitudinal information and support broader applications in ophthalmology. Finally, OCTCube-M could be expanded to support image-text multimodal learning by integrating textual clinical notes or reports with imaging data. We plan to explore this direction to enhance the model’s ability to generate richer insights and improve its clinical decision-making capabilities.

We have developed OCTCube-M, a 3D foundation model framework for optical coherence tomography and *en face* images. We demonstrate the advantage of modeling 3D volumes holistically instead of pre-training on 2D image sets by comparing OCTCube-M to 2D OCT foundation models on multiple independent datasets. We found that OCTCube-M outperformed comparison approaches consistently in all 32 tasks with significant improvement on 31 tasks (**Fig. 1b, Supplementary Fig. 1**), including retinal disease prediction, cross-cohort prediction, cross-device prediction, systemic disease prediction, cross-modal transfer learning and cross-modality retrieval, indicating its accurate performance and strong generalizability. The strong predictive performance of OCTCube-M on both retinal diseases and systemic diseases indicates its potentially broad applicability. OCTCube-M led to the successful development of a new state-of-the-art GA prognostic model OCTCube-EF that achieves superior prognosis, visual acuity prediction performance and large benefits reducing patient recruitment challenges. OCTCube-M may be used as a general tool for analyzing OCT data, paving the path for AI-based retinal diagnostic and prognostic applications.[71,72]

**Figure legends**

**Fig. 1 OCTCube model overview. a,** OCTCube exploits 3D Masked Autoencoders (MAE) for the pre-training. MAE uses an encoder-decoder architecture where it randomly masks 90% of the cubes in each volume at the encoding stage and is optimized to reconstruct them at the decoding stage. The encoder and the decoder are implemented using a 3D Vision Transformer (ViT). FlashAttention is further used to reduce the GPU memory usage when modeling 3D cubes. **b,** Illustration of the cross-modality alignment using COEP, a novel contrastive self-supervised learning-based framework for integrating OCT and *en face* retinal images such as IR and FAF images, after the 3D MAE pre-training. COEP utilized contrastive loss to align 3D OCT volume and *en face* images. **c,** Radar plot comparing the performance of OCTCube and competing methods on 32 tasks, including eight retinal disease prediction tasks in the inductive learning setting, four retinal disease prediction tasks in the cross-cohort learning setting, seven systemic disease prediction tasks, three cross-organ transferring prediction tasks, eight cross-modality retrieval tasks and two cross-device prediction task. Recall@1 is used as the metric for the cross-modal retrieval tasks, Acc@1 is used as the metric for the cross-modal laterality prediction tasks, coefficient of determinant (R2) is used as the metric for EF(R) ejection fraction prediction task, and AUROC is used as the metric for the other tasks. UW-Oph is the abbreviation for UW Ophthalmology dataset. EF (C) is the low ejection fraction classification task, and EF (R) is the ejection fraction regression task. **d,** Plot showing the AUROC and AUPRC of an aggregation approach (**Supplementary Fig. 3c**) averages the prediction probabilities of $k$ slices around the center slices, where $k$ is shown in the x-axis and the prediction probability is derived using RETFound. The metric AUROC and AUPRC are the abbreviation of Area under the Receiver Operating Characteristic Curve and the Area under the Precision-Recall Curve. RETFound, as a 2D approach, corresponds to $k = 0$. **e,f**, Visualization of the center OCT slice (slice 30) and the OCT slice near the center slice, including the corresponding prediction probabilities of two AMD patients (**e** for patient 1 and **f** for patient 2). Red boxes highlight the small drusen that occurs at the slice 31 of patient 1 and the slice 28 of patient 2, indicating signals for AMD.

**Fig. 2 Evaluation on retinal disease prediction in the inductive learning setting. a,b,** Barplots comparing OCTCube and competing methods on disease classification of 8 retinal diseases on UW Ophthalmology dataset in terms of AUPRC (**a**) and AUROC (**b**). Inductive learning setting is used to ensure that test OCT volumes are not seen by OCTCube in the pre-training stage. The train:validation:test split is set to be 20%:60%:20%. POAG, DME, AMD, ERM/MH, DR, CRAO/CRVO, PVD, RNV denote primary open-angle glaucoma, diabetic macular edema, age-related macular degeneration, epiretinal membrane or macular hole, diabetic retinography without macular edema, central retinal vein / artery occlusion, posterior vitreous detachment, and retinal neovascularization, respectively. Supervised approaches do not have a pre-training stage. RETFound (all) and Supervised (all) average the embeddings of all slices within a 3D volume. * indicates the significance level at which OCTCube outperforms the best-competing method, with paired t-test p-value $< 5\times10^{-2}$ for *, p-value $< 1 \times 10^{-2}$ for **, p-value $< 1 \times 10^{-3}$ for ***. **c,** Saliency maps for eight different ophthalmic diseases captured by OCTCube. OCTCube successfully revealed the diseased area such as the peripapillary region for glaucoma, hyperreflective area around retinal pigment epithelial layer for AMD. Red pixels indicate higher importance determined by the model, while blue pixels indicate lower importance. Images are resized to (256, 256) for visualization. **d**, Visualization of multiple slices across the slow-scan dimension from a single OCT volume with the sampling location (**1st row**) in the corresponding IR en face image, OCT slices (**2nd row**), saliency maps based on the prediction of RETFound (center) (**3rd row**), and saliency maps based on the prediction of OCTCube (**4th row**). OCTCube provides a more coherent saliency map across slices in the diseased area, indicating the effectiveness of the 3D modeling. Green lines in the first row are drawn with the consideration of pixel spacing of sampled OCT slices. Red pixels in the third and the fourth rows indicate higher importance determined by the model, while blue pixels indicate lower importance. Images are resized to (256, 256) for visualization.

**Fig. 3 Evaluation on cross-cohort, low-resource, cross-device and cross-organ prediction. a,** Evaluation on cross-cohort analysis. Bar plots comparing OCTCube and competing methods on predicting retinal disease in terms of AUPRC and AUROC in the cross-cohort setting. The cross-cohort OCT volumes are collected from four independent datasets. **b,** Plots comparing OCTCube and competing methods across different sizes of fine-tuning data in terms of AUROC and AUPRC. **c,** Evaluation on cross-device generalization. Bar plots comparing OCTCube and competing methods in the cross-device setting in terms of AUROC and AUPRC on Zeiss Cirrus devices from the Glaucoma dataset and Topcon Maestro 2 devices from the AI-READI dataset. OCTCube is pre-trained on Heidelberg Spectralis OCT data. * indicates the significance level at which OCTCube outperforms the best-competing method, with paired t-test p-value $< 5\times10^{-2}$ for *, p-value $< 1 \times 10^{-2}$ for **, p-value $< 1 \times 10^{-3}$ for ***. **d**, Evaluation on cross-organ prediction. Bar plots comparing OCTCube and competing methods on cross-organ transfer learning on 3D lung CT and Echo Cardiac Ultrasound video in terms of AUPRC, AUROC and coefficient of determinant score ($R^2$) on classification and regression tasks. * indicates the significance level at which OCTCube outperforms the best-competing method, with paired t-test p-value $< 5\times10^{-2}$ for *, p-value $< 1 \times 10^{-2}$ for **, p-value $< 1 \times 10^{-3}$ for ***.

**Fig. 4 Evaluation on systemic disease prediction. a,b,** Barplots comparing OCTCube and competing methods on predicting seven systemic diseases in the UW Ophthalmology dataset in terms of AUPRC (**a**) and AUROC (**b**). Inductive learning setting is used to ensure that test OCT volumes are not seen by OCTCube in the pre-training stage. * indicates the significance level at which OCTCube outperforms the best-competing method, with paired t-test p-value $< 5\times10^{-2}$ for *, p-value $< 1 \times 10^{-2}$ for **, p-value $< 1 \times 10^{-3}$ for ***. **c,** Plots showing the AUROC performance of an aggregation approach that averages the predicted probabilities of k slices around the center slices on seven systemic diseases, where k is shown in the x-axis. The prediction probability is derived using RETFound. The RETFound (center) model, as a 2D approach, corresponds to k = 0 (**Supplementary Fig. 3a**). Different from directly averaging predictions, the RETFound (all) model uses a neural network to aggregate features (**Supplementary Fig. 3b**). The improved performance by considering more slices necessitates the development of 3D models. **d,e,** Scatter plots comparing the relative improvement over RETFound (center) by OCTCube and by the aggregation approach in terms of AUROC (**d**) and AUPRC (**e**) across seven diseases. The high Pearson correlation indicates that the improvement of OCTCube is from considering the 3D structure. **f,** Visualization of OCT slices from the right eye (OD) of the patient with diabetes at the first visit. Hard exudates are observed in several slices (marked by red boxes) but are not clearly seen in the center slice. RETFound thus failed to predict diabetes by only using the center slice. In contrast, OCTCube successfully predicted diabetes with probability of 0.72 by considering the entire volume. **g,h**, Visualization of representative OCT slices from the left eye (OS) acquired at the same day (**g**) and the right eye acquired after 1 year (**h**) for the same patient. Macular edema and hard exudates can be observed more clearly compared to the slices in OD (**f**), indicating diabetic retinopathy. RETFound successfully predicts diabetes using the center slice of OS (**g**) and OD 1 year (**h**), but fails using OD (**f**). In contrast, OCTCube successfully predicts diabetes using either OS, OD, or OD 1 year.

**Fig. 5 OCTCube-IR and evaluation on cross-modality analysis. a,** Illustration of the development of OCTCube-IR using COEP to integrate OCT and IR retinal images. COEP utilized contrastive loss to align 3D OCT volume and IR *en face* image pairs. OCTCube is used to initialize the OCT encoder and RETFound is used to initialize the *en face* encoder. **b,** Illustration of cross-modality retrieval evaluation. After the alignment, the OCT to IR retrieval task is evaluated by retrieving the most similar IR image based on the queried OCT volume. The IR to OCT retrieval task is evaluated similarly. **c-d** Barplots comparing OCTCube and competing methods on cross-modality retrieval and laterality prediction on UW Ophthalmology and AI-READI dataset in terms of recall@1 (**c**) and accuracy@1 (**d**) on OCT to IR retrieval and IR to OCT retrieval. * indicates the significance level at which OCTCube outperforms the best-competing method, with paired t-test p-value < $5{\times}10^{-2}$ for *, p-value < $1 \times 10^{-2}$ for **, p-value < $1 \times 10^{-3}$ for ***. **e-h**, Visualization of the top retrieved IR images given four different queried OCT volumes. Ground truth IR images are excluded. OCTCube retrieves more similar images that have similar structures in the retina, such as the optic disc (**f,h**), macula changes, lateralities (**e,f,g**) and illumination difference (**f,g,h**).

**Fig. 6 OCTCube-EF and evaluation on GA prognostic / visual functionality analysis. a**, Illustration of the pipeline that develops OCTCube-EF for GA prognosis and functional prediction. OCTCube-EF first exploited COEP to jointly embed multiple 3D and 2D retinal modalities and utilized the aggregated embeddings to predict targeted outcomes, including GA lesion growth rate and best corrected visual acuity (BCVA). **b-c,** Barplots comparing OCTCube-EF and competing methods on GA prognosis and visual functionality analysis in terms of the square of Pearson coefficient ($R^2$) on GA lesion growth rate (**b**) and BCVA (**c**) prediction. * indicates the significance level at which OCTCube outperforms the best-competing method, with paired t-test p-value < $5{\times}10^{-2}$ for *, p-value < $1 \times 10^{-2}$ for **, p-value < $1 \times 10^{-3}$ for ***. **d,** Illustration of applying covariate adjustment with OCTCube-EF GA prognostic model to increase the statistical power of the treatment effect measurement in randomized clinical trials (RCTs). By subtracting the predicted GA lesion growth rate as a baseline covariate, the treatment measurement can obtain better precision, leading to an effective sample size increase (ESSI) without recruiting more patients. A higher Pearson correlation between predicted and observed growth rate will lead to a higher ESSI (see **Methods**). **e,** Retrospective treatment effect analysis on Phase-II Mahalo trial with covariate adjustment using OCTCube-EF GA prognostic model. The adjusted results are more precise and reveal a non-positive treatment effect, in favor of not going into Phase-III (Chroma & Spectri), which showed negative treatment effects of Lampalizumab. **f,** Analysis of the treatment effect by dose arm with covariate adjustment using the OCTCube-EF GA prognosis model. **g-h,** Saliency map of the center OCT B

scan extracted from the OCTCube-EF GA prognosis (**g**) and BCVA (**h**) model. Both models capture anatomical regions that have correlation with the GA lesion growth rate, such as the boundaries of the GA lesion area and the ellipsoid zone (EZ) / retinal pigment epithelium (RPE) losses around fovea, verified by the recent research results. Red pixels in (**g,h**) indicate higher importance determined by the model, while blue pixels indicate lower importance. **i**, The OCT (center B scan), en face IR and FAF images from a GA patient pictured at screening visit and the follow-up visit after 18 months. The saliency maps are generated by OCTCube-EF using the images from screening visits. Red pixels in (**i**) indicate higher importance determined by the model, while blue pixels indicate lower importance. The gray dashed line is drawn based on the edge of the GA lesion in the screening visit. The red marks indicate the GA lesion growth after 18 months.

## Methods

### Details of UW Ophthalmology dataset

**Dataset overview**. The UW Ophthalmology dataset contains 3D macula OCT volumes, paired Infrared Retinal images from the medical screening process of 17,214 patients, along with the diagnosis codes (ICD-9 and ICD-10 code) across the UW-medicine system. In this study, we utilize this dataset for pre-training, within-dataset ophthalmic disease prediction, systemic cross-disease prediction, and cross-modality prediction. We only include the first screening of each patient, to avoid distribution shifts brought by the longitudinal follow-up screening results. We include data samples from both eyes of a patient when available. This results in 33,262 3D OCT volume and IR image pairs. Each macula OCT volume and its paired IR images were extracted by the Heidelberg Spectralis (Heidelberg Engineering) imaging device[32] from 2006 to 2023. The IR images are not used in the pre-training stage and only used for the cross-modality analysis. Each OCT volume is a composition of either 60 or 61 slices, and the default digital resolution of each slice is 496 by 768. The absolute pixel spacing varies for each dimension and each instance, and the average pixel spacing area is 7.51 by 1.88 by 8.66 mm. The paired IR image is taken together with the OCT volume acquisition, and the typical digital resolution is 768 by 768. This study was approved by the Institutional Review Board of the University of Washington (UW) and was in adherence with the tenets of the Declaration of Helsinki and the Health Insurance Portability and Accountability Act.

**Data split**. We split the dataset at patient level, with a ratio of 80% training and 20% test for the within-dataset evaluation. This results in 26,605 training and 6657 test sample volumes from 13,771 and 3,443 patients. For all the conducted experiments, the held out test set is only used for evaluation. In the pre-training stage, only the 3D macula OCT volumes are included and associated disease labels for these volumes are not seen by the model. In the within-dataset retinal and systemic cross-disease prediction, the OCT volumes are used to predict diagnosis information provided as label supervision. In the cross-modality prediction, both the OCT volumes and IR images are provided to perform alignment training and evaluation. Among all 17,214 patients, 12,830 patients had clinical records available. We thus use this subset to construct the data for the within-dataset retinal and systemic cross-disease prediction task.

**Labels for retinal disease prediction.** We picked 8 retinal diseases based on ICD code (**Supplementary Fig. 19, Supplementary Table 1**): Primary Open-Angle Glaucoma (POAG, 1,248 patients), Diabetic Macular Edema (DME, 881 patients), Age-related Macular Degeneration (AMD, 1,888 patients), , Diabetic Retinopathy without Macular Edema (DR, 1,694 patients), Epiretinal Membrane or Macular Hole (ERM/MH, 1,482 patients), Central Retinal Artery or Vein Occlusion (CRAO/CRVO, 204 patients), Posterior Vitreous Detachment (PVD, 1,725 patients), Retinal Neovascularization (RNV, 299 patients). We then set up the rest of the recorded patients as the non-diseased cohort (3,936 patients). We also included patients with multiple retinal diseases, resulting in a multi-label and imbalanced dataset. We followed the train-test set split in the pre-training stage to hold out the test samples that are not seen in the pre-training stage for evaluation. For all the tasks being evaluated, we further split the training subset into train and validation subsets as 75% to 25%, resulting in a data split of 60% training, 20% validation, 20% test split.

**Labels for systemic cross-disease prediction.** We aimed to examine if OCTCube is able to predict other systemic diseases using OCT volumes. We extracted the ICD code of systemic diseases for the recorded patients and counted their frequencies. We merged ICD-9 code to ICD-10 code and used the first level code of the ICD-10 code system. We then extracted all the level 1 disease codes with frequency larger than 100, resulting in 455 different diseases over 9,801 patients. The disease distribution can be seen at (**Supplementary Fig. 20**). We used these samples to construct the multi-label systemic cross-disease dataset. We followed the same train and validation split as we did in preparing data for retinal disease prediction, ensuring samples in the held out test set are never seen in the pre-training stage.

**Details of OCTCube**

**Detailed overview of OCTCube**. In this section we present the detailed architecture, design, and training of the OCTCube. OCTCube is based on attention-based Vision Transformer, which treats OCT volumes as a long sequence of continuous cube feature vectors and learns to generate summarized representations from the sequence based on stacked multi-head attention and non-linear transformations. It is composed of a heavy transformer encoder and a lightweight transformer decoder, and is trained with 3D masked autoencoder objectives.[73] The encoder first indexes the spatial position of each cube, and then randomly selects most of the cube tokens and masks them out from the long sequences, and processes the rest feature sequence together with their indexed position information. After getting the output from the encoder, the lightweight decoder will insert a learnable embedding vector named <mask> as the placeholder token for each masked-out cube at its original indexed position and try to reconstruct the whole sequence. The reconstruction is guided by minimizing the mean squared error (MSE), thus not requiring label supervision at all. Prior works[73–75] have shown that this process will lead to a good encoder for downstream tasks. After the training, the decoder will be discarded and the encoder is taken as the basic encoder for the downstream tasks.

**Decomposing the volume into the sequence of 3D cubes.** Unlike a plain 2D Vision Transformer,[76] the OCTCube is a designated 3D-aware Vision Transformer model that can take 3D OCT volumes or 2D OCT slices with arbitrary sizes. To achieve this, OCTCube utilizes CubeEmbed, a non-overlapped 3D convolution layer that can split the OCT volumes into small cubes and project each cube to an embedding token vector. Let (z, h, w) be the cube size pre-chosen in the CubeEmbed layer, applying this operation to the whole 3D OCT volume with Z slices and (H, W) resolution will result in a long sequence of cube embedding with length of $(Z * H * W) / (z * h * w)$. It is thus crucial to pick an appropriate cube size in order to both maintain flexibility and avoid heavy computation. For the OCT modality, we propose to use a smaller size on the 3rd z-axis and a larger window size on the 2D slices for three reasons. First, depending on the scanning mode, the z-axis might have a much sparser sampling region; a smaller z will lead to similar pixel spacing to the other two dimensions. Second, a smaller z will provide flexibility to handle volumes with less number of slices. To the extreme, 2D slices can be treated as the volume with z=1, and since the slices will commonly be converted to a duplicated 3-channel rgb image, we set the cube size to be 3 in OCTCube. This makes OCTCube capable of taking both 2D slices and 3D volumes as inputs, improving its flexibility on downstream applications. We also make a 2D positional embedding for the h and w axis and a separate positional embedding for the 3rd z-axis, to help the model adapt to variant volume size. What's more, setting z=3 has another important benefit to accelerate the learning, which is being easy to utilize pre-trained 2D checkpoints to warm up the 3D-aware model, with simple conversion from PatchEmbed to CubeEmbed. This allows OCTCube to quickly adapt from 2D to 3D with much less computation effort.

**Incorporating FlashAttention into OCTCube.** One big drawback of 3D modeling in medical imaging is the drastically increased computational cost. Compared to the pure convolutional neural network, the plain vision transformer suffers more on the increased sequence length, as the multi-head attention induces $O(L^2)$ space and time complexity, where L is the sequence length. As an example, an OCT volume with resize resolution 48 x 256 x 256 with cube size 3 x 16 x16 leads to a sequence with length of 4,096, and fine-tuning a ViT-large encoder with a linear projection head with this single sequence will use up more than 50 GB GPU memory, which exceeds the memory of most of the GPUs. To relieve this issue and make OCTCube more efficient and affordable to end users, we incorporate FlashAttention-2[77,78] into the Vision Transformer structure. FlashAttention-2 is an advanced technology that helps to reduce the GPU memory costs by 5~20 times and enable 2~4x training / inference speed for transformer structure without accuracy lost, by optimizing the computation of attention. Because FlashAttention-2 reduces the space complexity of attention from $O(L^2)$ to O(L), including FlashAttention-2 reduces the GPU memory usage of the above example to 10.52 GB, which is more affordable on modern GPUs. Moreover, the computing speed also achieves at least 2 times improvement. This greatly improves the efficiency of OCTCube.

**Prediction head design.** In the fine-tuning stage for downstream tasks, a light weight multi-layer perceptron (MLP) head is used to map the representation of the OCT volumes to the outcome.
After getting the output representation sequence from the pre-trained encoder, we perform average pooling for representation at all spatial positions and acquire the overall embedding. We use slightly different heads for disease prediction tasks and the cross-modality prediction task.
For the disease prediction tasks, we use a MLP layer with layer normalization to map the representation to the number of classes, depending on the tasks. Dropout[76,79] is set for all disease prediction experiments with a rate of 0.5 to avoid over-fitting. For the cross-modality prediction tasks, the dropout operation is removed for full feature utilization. We add one more MLP layer with the same dimension of the embedding dimension, and a GELU[80] non-linear activation to map the representation to the aligned space.

**Extending RETFound to handle 3D OCT volumes.** The original RETFound encoder only takes 2D OCT slices as input (referred to as RETFound (center)). We extend RETFound to handle 3D OCT volumes in a multi-instance learning manner: Given an OCT volume with size (Z, H, W), we input all Z slices into the RETFound and acquire the embedding for each slice. We then aggregate the embedding together with average pooling for the final representation of the volumes (referred to as RETFound (all)). We maintain the prediction head design to be the same as OCTCube for these two major baselines for a fair comparison.

Another critical change we made for RETFound is to equip the model also with FlashAttention-2. Without such modifications, RETFound will fail in most of the experiments. Interestingly, we observe that, when equipped with FlashAttention-2, RETFound (all) will consume more GPU memory when taking the same size of OCT volumes. This is because OCTCube can better benefit from the space complexity improvement in terms of sequence length, as it takes the whole volume as a single sequence. On the other hand, the extended RETFound is equivalent to multi-slice learning with a large batch size of short slice sequence. This makes the OCTCube encoder not only capable of leveraging information across the volume, but also able to enjoy higher computational efficiency. As an example, fine tuning a 48 by 256 by 256 OCT volume with the OCTCube encoder will take less than 25 GB GPU memory, while the RETFound model with FlashAttention-2 will take more than 60.52 GB GPU memory.

**Pre-training data processing and other implementation details.** For the 3D OCT volume data for pre-training OCTCube, we resize the images to 60 by 256 by 256. We then normalize the voxel to the region of 0 to 1. We perform very lightweight augmentation, such as random flip. We discard the ImageNet mean and standard deviation normalization as the 3D OCT volumes do not have 3 channels and only do [0, 1]-normalization. We set the cube size to be 3 by 16 by 16, this gives sequences with length of in total 20 by 16 by 16 = 5,120. We therefore set up the 2D position embedding to be the shape of 16 by 16, and the z-axis position embedding length to be 20.
We set up the encoder of OCTCube to be ViT-large, with 24 layers, 16-heads and 1024-dim embedding for each layer, and the decoder to be ViT-small, with 8 layers, 16-heads and 512-dim embedding for each layer. Different from RETFound, we used a mask ratio of 90%, as it has been proved to be an optimal setting in 3D MAE training.[73] As discussed before, setting z=3 enables us to borrow from other pretrained 2D checkpoints. We therefore set up the RETFound OCT ViT-large model as the initialized checkpoint, with moderate modification. We trained OCTCube using a batch size of 4 (effective 2D batch size = 240) on 4 NVIDIA A100 80 GB devices with 50 epochs. We used the AdamW optimizer with a base learning rate of $1.6 \times 10^{-3}$ with cosine annealing. The first 5 epochs are set to be the warm up epochs with a linear increase of lr from 0 to the peak learning rate. The whole training usually takes 5 days.

**Details of within-dataset retinal disease prediction**

We formulate the disease prediction of the eight diseases as eight binary classification tasks, and a multi-task setting by jointly predicting all eight diseases. For the binary disease prediction, the non-diseased cohort is treated as the negative controls. For the multi-task setting, the non-diseased cohort is treated as the negative controls for all diseases. For each disease, only patients with the record of the matched disease code were included in the training and the evaluation. For each experiment, we trained the model for 10 epochs with a learning rate of $5 \times 10^{-3}$ and a batch size of 1, and applied the learning rate warming up (from 0 to $5 \times 10^{-3}$) strategy to the first epoch and a cosine annealing schedule for the rest of the epochs. Similar to the setting used by RETFound, we performed label smoothing with smoothing factor 0.1. The validation set is used to perform model selection, where the model with highest AUPRC will be used for evaluation on the held out test set. For the RETFound (center) model, only the center slice was extracted from the OCT volume and served as the input. We set up the batch size to be 16 and the number of epochs to be 50, following the original RETFound setting for a fair comparison. For the transformer encoder, we set up layer-wise learning rate decay with a factor of 0.65, to achieve better trade-off between utilizing learnings from the pre-training in the early layers and flexible weight adjustment for downstream tasks in the layers close to the prediction head.

For the interpretable visualization of OCTCube and RETFound (center), we leveraged gradient-based visualization to better understand the pixel-level importance to disease prediction, instead of visualizing the attention map that only assigns importance at patch or cube level. Specifically, we chose Grad-CAM++[81] as the base method, which utilizes second-order gradients to help generate saliency maps. We adopted and modified Pytorch-grad-cam software[82] to instantiate Grad-CAM++ for both OCTCube and RETFound (center). For RETFound (center), since the prediction was made based on each slice, a saliency map would thus be generated when hooking the model with the visualization method. For OCTCube, directly hooking the software did not work, because the prediction was made based on the whole volume. We therefore adapted the software to allow gradient flow pass through the 3D-aware ViT and reorganized the shape to generate saliency map for each slice, or for the other perspective that visualizes depth and slice sampling direction. For the slice perspective, we visualized the saliency map on the slice with a resized resolution of 256 by 256. For the other perspective, we adjusted the image from 61 by 496 to 61 by 256 and then visualized the saliency map with adjustment of aspect ratio to view it as a 256 by 256 image. We used the grayscale saliency and converted the intensity distribution to color distribution using the software for better visualization. For the sampling location and pixel spacing of each OCT slice on IR image (marked as red line), we extracted such information from the metadata of the imaging process and removed the out-of-FOV sampling region. We used a red line with different linewidth to denote the real pixel spacing in different directions, as the slice perspective usually has larger pixel spacing compared to the other perspective.

**Details of cross-cohort, low-data and cross-device prediction**

We leverage 6 public cross-sectional 3D OCT volumetric disease datasets for cross-cohort and cross-device evaluation. These datasets target on multiple ophthalmological and systemic, such as Age-related Macular Degeneration (AMD), Glaucoma, Diabetic Macular Edema (DME), multi-stage Macular Hole (MH), Multiple Sclerosis (MS), and Type-2 Diabetes Mellitus (T2DM).

We formatted the disease prediction task on these datasets as binary or categorical classification depending on the number of classes. We summarized the datasets below:

**UMN dataset**[83]: This dataset contains 29 patients with DME and 24 patients with AMD. OCTCube is used to predict binary classification. The number of slices for each volume is 25, and the digital resolution is 496 x 1024. Each slice is resized to 256 x 256 for the downstream evaluation.

**DUKE 14 dataset**[83,84]: This dataset contains in total 45 patients either being healthy, with AMD, or with DME. Each class has 15 patients. OCTCube is used to predict 3-class classification. The number of slices for each volume is 50, and the digital resolution is 496 x 512. Each slice is resized to 256 x 256 for the downstream evaluation.

**HCMS dataset**[85]: This dataset contains 14 healthy patients and 21 patients with multiple sclerosis. OCTCube is used to predict binary classification. The number of slices for each volume is 49, and the digital resolution is 496 x 1024. Each slice is resized to 256 x 256 for the downstream evaluation.
**OIMHS dataset**[14]: This dataset contains in total 3,859 OCT slices extracted from 125 eyes of 118 patients that have either 4 stages of macular holes. We exclude the stage 1 class because the number of patients is 1. Among the rest, 16 volumes are with stage 2, 34 volumes are with stage 3 and 74 volumes are with stage 4. OCTCube is used to predict 3-class stage classification. The screening of each eye results in a 3D OCT volume with a variant number of slices from 17 to 97. For the evaluation on this dataset, we picked up the number of slices to be included for each volume as 17. For the digital resolution, 220 images has a resolution of $384 \times 496$ pixels, 3002 images has a resolution of $512 \times 496$ pixels, and 637 images has a resolution of $768 \times 496$ pixels. Each slice is resized to 256 x 256 for the downstream evaluation.
**Glaucoma dataset**[25]: This dataset contains in total 1,110 Patients, where 263 of them are healthy, and the rest 847 of them are with glaucoma. The OCT volumes were Zeiss Cirrus devices. The original volume size is 200 x 1024 x 200, captured from the optic disc area. OCTCube is used to predict binary classification. The number of slices for each volume is 64, and the digital resolution is 128 x 64. Each slice is resized to 128 x 128 for the downstream evaluation.
**AI-READI dataset**[30]: This dataset is a recently-released flagship dataset focusing on Type 2 Diabetes Mellitus. The current released version (v1.0.0) is cross-sectional and contains 204 patients, with 4 diagnosis outcomes: healthy (74 patients), pre-diabetes (48 patients), Type-2 diabetes with oral medication and/or non insulin injectable medication controlled (57 patients), Insulin-dependent T2DM (25 patients). We used the OCT volumes dataset acquired by TopCon Maestro 2 imaging devices to test the cross-device disease prediction performance. For the cross-device disease prediction task, we clustered healthy and pre-diabetes patients into one group, and the other two diabetes-diagnosed into another, and then performed binary classification between these two groups. The resulting control and diseased group size are 122 and 78, respectively. The number of slices for each volume is 128, and the digital resolution is 885 x 512. We filter out patients with problematic z-axis coherence and missing data, resulting in 167 patients included, resulting in 330 OCT volumes.

For all the public datasets except DUKE14, we set up multi-fold splits to evaluate the effectiveness of OCTCube. For UMN, Glaucoma and OIMHS, we set up 10-fold splits and test the label-restricted setting. For this type of split, we perform training using only 1 fold (10% data as training), and evaluate the performance on the rest of the 90% subset. For DUKE 14, we do a 5 fold split and use 20% data as a training set, as 10% training data can not always include all three classes into every split. For HCMS, we understand the model performance under label-rich settings, and partly also because the dataset size is limited to be 35. We set up a standard 5-fold cross validation for this setting, where for each iteration, one of the folds will be treated as the validation set, and the other four folds will be treated as the training set. We also evaluated the cross-device performance on the Glaucoma and AI-READI using this setting.

For the low-resource prediction study on OIMHS dataset, we use k% of the data to be set as the training set. We set up k% to be 10%, 20%, 30%, 40%, 50%, 60%, 70%, and 80%. For each k, we randomly select 5 splits based on the ratio and report the validation performance. For the cross-device study on AI-READI dataset, we utilize the same split for dataset from all three devices. Because the pixel spacing, intensity distribution and sampling frequency on three axes are different across device type and settings, we perform a pixel spacing adjustment for data taken by Topcon Maestro 2 and Topcon Triton devices as OCTCube is pre-trained on data taken from Heidelberg Spectralis device. Specifically, we perform foreground extraction on the y-axis (the depth axis) using the OTSU filter[86] across the volume to adjust the pixel spacing. We then perform intensity thresholding to ground the extreme dark and light pixels. For each volume, We set up the pixel value at $1 \times 10^{-4}$ and 0.9999 percentile to be the grounding pixel value, the pixel smaller or larger than the threshold will be converted to the threshold, respectively. This helps to adjust the intensity distribution to be more aligned with the intensity distribution of the pre-trained dataset. We then interpolate the z-axis to 60 to align with the pre-training data distribution, as well as reducing the computational cost brought by the large z-axis sampling frequency. For the RETFound (center) and 3D baselines on the cross-device disease prediction experiments on AI-READI, we found the models achieve inferior performance (slightly better than random) with standard [0,1]-normalization for the data collected by Topcon Maestro 2 devices. We therefore specifically adjust the normalization to be both [0,1]-normalization and ImageNet normalization to help not cover their potential. It is worth noting that we didn't encounter this problem when experimenting with other datasets.

For the model training, we set up a longer training recipe for the public dataset, with in total 100 epochs and 10 epochs for warming up the learning rate from 0 to the peak, as the dataset size is relatively smaller. We follow the similar setting as we did in the within-dataset retinal disease prediction.

**Details of systemic cross-disease prediction**

We set up the multi-label task using the constructed ICD10-based disease labels. We did not utilize any other prior knowledge about diseases that are correlated with retina, but directly leveraged OCTCube to extract possible predictable signals. Specifically, we used the same training recipe as we did in the within-dataset retinal disease prediction. We calculated the macro AUROC and AUPRC over all classes and used this macro AUPRC for model selection. After picking up the best checkpoint, we calculated the class-wise metrics over all 455 diseases, and found out the diseases that have predictable signals. We reported the 7 diseases that have balanced accuracy significantly higher than 0.5 (measured by two-sided t-test) using the best checkpoint. They are Diabetes (E11), Hypertension (I10), Joint pain (M25), Hyperlipidemia (E78), Soft tissue disorders (M79), Back pain (M54), and general Pain (G69). Diseases such as diabetes and hypertension have been shown to have correlation with various retinopathy, validating our prediction signal extraction process.

**Details of disease prediction by extending and aggregating slices**

For the feature aggregation study presented in **Fig. 4c-e and Supplementary Figs. 3,4,12**, we extracted the logits of the selected diseases from the RETFound (center) model for all the slices of all volumes in the held out test set. We then performed the prediction aggregation, by averaging the prediction of k slices besides the center slice (slice 30) in both directions. A choice of k indicates an aggregation of predictions from 2k + 1 slices. We reported the standard deviation based on 5 runs of RETFound (center) model, and reported the mean value of OCTCube and RETFound (all) model (the same as in **Fig. 2a-b** and **Fig. 4a-b**).

We calculated the correlation between the relative improvements brought by aggregation and OCTCube and reported the Pearson correlation coefficient. Both correlation achieved a p-value < 0.05.

**Details of the cross-organ CT and Ultrasound transferability analysis using OCTCube**

The task of the cross-organ transferability analysis of OCTCube on CT and Ultrasound modalities are tested on the same benchmark used by SLIViT.[26] Specifically, for CT modalities we tested on the NoduleMNIST3D dataset[27] and performed the binary malignancy classification. Among 1,633 samples, 401 are labeled as malignant. For Ultrasound modalities, we performed two tasks on the Echonet-dynamic dataset,[28] a regression task predicting actual ejection fraction (EF), and a binary classification task predicting the low ejection fraction (<50%, EF_b), containing 2,246 samples from in total 10,030 samples. For the classification task on both modalities, we observe significant class imbalance (24.5% on CT and 22.3% on Ultrasound).

For the CT classification task, we picked the newer version of CT volumes with the shape of (64, 64, 64), and adapted it to (60, 256, 256) to apply OCTCube and the other baselines. For the Ultrasound tasks, each volume is a video with T frames of (112, 112, 3) B Scans with variant T. We resized the height and width to (256, 256) for all baselines. For OCTCube, we picked the first channel and resized the temporal dimension to 60 uniformly. For other baselines except for SLIViT, we picked all of the 3 channels as they can inherently take 3-channel input. For SLIViT, we followed the default setting to equally sampled 32 frames across the video. Results from SLIViT were reproduced and reported with the average of running with 5 different random seeds.

For the classification task, we maintained the same structure as we did in other disease prediction tasks for both OCTCube and RETFound (all). For the regression task, we replaced the two-layer MLP head with the SLIViT head to have a more flexible structure for the cross-modality adaptation with both OCTCube and RETFound. We used a default setting for the SLIViT head with a depth of 5 and number of heads to be 20.

We used the AdamW optimizer for all tasks. For the CT classification tasks, we fine-tuned OCTCube for 50 epochs with a learning rate of $5 \times 10^{-3}$. For the Ultrasound classification and regression tasks we finetuned OCTCube for 20 epochs with a learning rate of $1 \times 10^{-2}$. For the regression task, the loss is set to be L1 loss. We set up the prediction of left ventricular end systolic volume (ESV) and left ventricular end diastolic volume (EDV) as auxiliary regression tasks to help both OCTCube and RETFound to better adapt to the ultrasound modality when predicting EF. The loss weight is set to be 0.2. Empirically we found it helpful to the ViT-based encoder for both OCTCube and RETFound, but observed slightly downgraded performance for SLIViT. We thus reported the SLIViT performance with the default setting. The model with the best validation AUPRC / coefficient of determination ($R^2$) performance for the classification / regression task was selected and tested on the test set. The average results of running with 5 different random seeds were reported.

**Details of OCTCube-M**

**COEP: Contrastive OCT volume-Enface image pre-training.** Given an OCT volume and its paired *en face* fundus image (IR, FAF, etc), it is challenging to measure the similarity between them from pure vision. Directly compressing OCT to an `averaged' en face map is inferior, as (i) the FOV of OCT volumes is much smaller than the fundus image, (ii) discrepancies between OCT slices might exist, and (iii) some detailed structures such as vessel might be missing after the compression. While modern imaging devices such as the Heidelberg Spectralis imaging device may provide scanning coordinates in IR image for each OCT slice, it's challenging to incorporate such anchoring information into multimodal representation learning, and even harder to align with other *en face* imaging modalities. One key challenge is that the unsupervised self mask-and-reconstruct 2D / 3D MAE objectives used to pre-train OCTCube are not directly applicable for the cross-modal analysis. We therefore propose COEP, a large-scale contrastive learning strategy at the volume-image level between OCT volume and *en face* images, on the UW Ophthalmology dataset. COEP is different from 2D or 3D MAE, serving as a secondary representation learning procedure after the first-stage pre-training. The idea of COEP is inspired by Contrastive language-image pre-training (CLIP),[87] which tries to learn a multimodal embedding space by pushing positive image-text pairs together, and pulling the negative image-text apart. The key advantage of this method is the contrastive loss being used only requires positive-negative pair information, which is naturally lied in the OCT-Enface pair datasets. We thus replace the image-text design with 3D OCT volume-2D *en face* fundus image and propose COEP. Next, we provide two implementations of COEP, which focus on two or more than two modalities.

**Bi-COEP: Bi-modalities Contrastive OCT volume-Enface image Pre-training.** Given a batch of N (OCT, En face) image pairs, COEP aims to train two separate encoders $Enc_O$ and $Enc_I$ for each modality. Let $O_j$ and $I_j$ be the embedding of the j-th OCT volume and *en face* image pair output by the encoder, COEP optimize the symmetric cross-entropy style INFONCE[88] loss:

$$\mathcal{L}_{O,I}^{COEP} = -\frac{1}{2N}\left(\sum_{i=1}^{N} \log \frac{e^{cos(\mathbf{O}_j,\mathbf{I}_j)/\tau}}{\sum_{k=1}^{N} e^{cos(\mathbf{O}_j,\mathbf{I}_k)/\tau}} + \sum_{i=1}^{N} \log \frac{e^{cos(\mathbf{O}_j,\mathbf{I}_j)/\tau}}{\sum_{k=1}^{N} e^{cos(\mathbf{O}_k,\mathbf{I}_j)/\tau}}\right)$$

Here, $\cos(O_j , I_k)$ is the cosine similarity of the j-th OCT embedding $O_j$ and the k-th *en face* image embedding $I_k$, $\tau$ is a learnable scaling temperature factor.

**Tri-COEP: Three-modalities Contrastive OCT volume-Enface image Pre-training**. Given a batch of N (OCT, En face 1, En face 2) image triplets, Tri-COEP is a natural-yet-novel extension of COEP that can learn to jointly embed all three modalities into one representation space. Let $O_j$, $I_j$, and $E_j$, be the embedding of the j-th OCT volume, the first and the second *en face* image triplet output by the encoder; Tri-COEP optimize the symmetric cross-entropy style INFONCE[88] loss between each of the two modalities:

$$\mathcal{L}_{O,I,E}^{COEP-3} = \frac{1}{3}\left(\mathcal{L}_{O,I}^{COEP} + \mathcal{L}_{O,E}^{COEP} + \mathcal{L}_{I,E}^{COEP}\right)$$

**Adapting OCTCube to boost COEP.** We aim to evaluate how OCTCube will help COEP training. Specifically, we set the initial OCT volume encoder to be the pre-trained encoder of OCTCube. Directly setting Vision Transformer as the OCT volume encoder has several computational challenges. First, the sequence length of the 3D volume will be much longer. Second, the InfoNCE loss used in COEP suggests a large batch size, this prerequisites brings extra constraint for the GPU space consumption. As an example, the original OCTCube with fully fine-tunable weights can only use a batch size of 8 on a NVIDIA A100 GPU. To relieve this challenge, we make a critical modification to adapt OCTCube for affordable and efficient COEP training. Specifically, we set up a more aggressive layer-wise learning rate decay strategy by freezing the first two third layers (in total 16) of the transformer encoder and layer decay of 0.65 for the last 8 layers. This makes the actual gpu memory consumption to be only at most one third of the original cost, providing a more efficient training recipe. It is worth noting that such strategy, being compromised to sup-optimal performance compared to using a fully fine-tunable model, is much less harmful to OCTCube, thanks to the 3D MAE pre-training. For the RETFound baselines, we make the strategy the same for a fair comparison. We maintain the volume size to be 60 x 256 x 256, the same as what we have done in the 3D MAE pre-training.

**Other implementation details**. COEP was developed based on an open sourced distributed contrastive language-image pre-training (OpenCLIP)[89] based on PyTorch.[90] We performed moderate adaptation to set up two image encoders for COEP.

We first introduced the details in the cross-modal alignment task between OCT and IR images from the UW Ophthalmology dataset, resulting in the OCTCube-IR model. Since our goal is to understand how well OCTCube is as the base OCT volume encoder, for the *en face* image encoder, we chose not to pre-train on IR images but directly use the ViT-large model checkpoint trained on Color Fundus Photography by RETFound, to serve as a relatively acceptable initialization for IR images. We set up the *en face* encoder to be fully fine-tunable, in order not to restrict the COEP model with sub-optimal IR embedding. We set up the digital resolution of IR images to be 224 x 224, following the original design of RETFound.

For the training recipe, we wanted to validate the hypothesis that OCTCube has captured meaningful embeddings for OCT volumes. We therefore set up a relatively short 50 epoch training procedure. We used the AdamW optimizer, set the learning rate to be $1 \times 10^{-4}$, and linear warming up steps to be 200.

We further optimized the GPU memory cost with several techniques, namely gradient checkpointing, automatic mixed precision (AMP) with float16 (whenever possible) and sharded contrastive loss, which provides identical gradient, but removes redundant intermediate cross-GPU similarities computation and only computes similarities between local relevant features. All of these improvements led to an achievable single-GPU batch size of 32 on 4 NVIDIA A100 80GB GPUs. To further enlarge the effective batch size, we set up an accumulated gradient with a step of 4, resulting in a batch size of 512 over all GPUs. We maintained the setting of RETFound baselines to be the same. The whole training usually takes less than 2 days.

**Evaluation of COEP within UW Ophthalmology dataset.** We evaluated the retrieval performance of bi-COEP on the held out test set as discussed before. Following the protocol of previous work,[91] we performed cosine similarity based nearest neighbor search at the whole test set level, i.e., retrieval from in total 6,647 images / volumes. Given a test set with size of T, we first computed the embedding of all OCT volumes and IR images in the test set, respectively. We then calculated the pairwise cosine similarity of all pairs, resulting in a T x T similarity matrix. Given an OCT volume embedding, the top-K most similar IR embedding will be retrieved, and vice versa. We then reported top-K recall (referred to as R@K), which counts if the correct paired images / volumes are within the top-K retrieved list. Besides, we report mean rank, i.e., the average ranking of the ground-truth images / volumes. We performed sub-training and sub-validation splits similar to what we did previously and used the validation subset for model selection. The best model checkpoint was picked based on the average of the R@1 for both modalities on the validation set. For the zero-shot laterality retrieval task, we removed the ground truth from the candidate list to better understand if the embedding space is able to cluster images / volumes by laterality. For each query IR image / OCT volume, we retrieve the volumes / images with top-K similarity to calculate the accuarcy@K (Acc@K), by computing how many of them have the same laterality compared to the query.

**Evaluation of COEP cross AI-READI dataset.** To further evaluate OCTCube-IR, we leveraged the Heidelberg Spectralis OCT-IR paired subset from the AI-READ dataset to test the cross-cohort multimodal alignment performance. The Heidelberg Spectralis OCT-IR subset covered almost the same patient cohort compared to the Topcon Maestro 2 subset used in cross-device disease prediction. Among all 204 patients, 172 of them took the data collection process using Heidelberg Spectralis imaging devices. We used all the available 344 macula OCT-IR pairs to construct the cross-cohort test set. For each OCT / IR query sample, the candidate retrieval set is the whole set of the other modality. We set up the preprocessing for OCT volumes and IR images to be the same as the training pipeline for a fair evaluation. We followed the same evaluation protocol of the within-dataset setting. For the zero-shot laterality retrieval task, we follow the setting used in the evaluation of UW Ophthalmology dataset.

**Setting of OCT-IR case study demonstration**. For the qualitative case studies reported, we set up the retrieval task of retrieving the most possible IR image given a query OCT volume. Because of the challenges to demonstrate visual similarity of a 3D OCT volume and its paired IR image, we chose to visualize the ground truth IR image for the query. To better understand how different models perform retrieval, we modified the quantitative evaluation protocol discussed above slightly by removing the ground truth paired IR image from the candidate set. While it is impossible to retrieve the correct IR image, the compromised retrieval results reveal the importance of different anatomic structures in the retina for the retrieval, such as the optic disc, blood vessels and their locations / directions, through the lens of qualitative similarity of visual patterns.

**Details of the OCTCube-EF design and the Geographic Atrophy growth rate and structural functional BCVA analysis**

**Task motivation and formulation**. Geographic atrophy (GA), as an advanced stage of Age-related macular degeneration, is a long progressive disease with increasing GA lesion area causing vision loss. The GA lesion can be revealed and detected from multiple retinal imaging modalities, including OCT, IR and fundus autofluorescence (FAF) modalities. FAF is another non-invasive imaging modality, but different from IR, it gets contrast from natural and pathological occurring fluorophores in the retina,[92,93] especially the retinal pigment epithelium (RPE) layer. It is thus helpful to observe RPE layer loss across the macular and quantify the area of the GA lesion.

Combining the OCT, FAF, and IR modalities is a natural need for the GA diagnosis and prognosis. The growth rate of the hypo-autoflourscent lesion area (indicating retinal pigment epithelium loss) on FAF is the FDA approved anatomical clinical endpoint for GA. It is noteworthy that when humans grade the GA area on FAF images, experts also utilize imaging modalities that can compliment the FAF information especially in certain areas such as the fovea. IR and OCT images help to determine foveal involvement. On OCT images, the RPE layer loss is highly correlated with the FAF GA area. In addition, the 3D information inherent in OCT imaging allows inspection of multiple layers of the retina, beyond RPE loss, that is relevant to GA disease progression.
We introduced three clinical needs performed in this study: predicting visual acuity from retinal structure, predicting GA growth rate from baseline imaging, and quantifying the area of RPE loss from an image. The BCVA is an integer that can range from [0, 90]. The growth rate of GA lesion measurement for a certain eye requires reading out the quantified GA lesion area measurement (usually based on a segmentation technique) from at least three longitudinal FAF / OCT screenings to regress out the slope of the changing lesion area. The GA growth rate is thus a real number ranging from [0, 8]. The prediction tasks are thus designed as regression tasks and are evaluated based on the $R^2$ metric.

**Data collection**. We collected a large scale multi-modal ophthalmic imaging data (OCT, FAF, IR) from multiple global, multi-center, retrospective and prospective phase-II and phase-III clinical trials, including Mahalo (phase 2 of Lampa, GA, NCT02247479)[44], Lampa (Chroma & Spectri, phase 3, GA, NCT02247479 and NCT02247531)[43], Proxima A/B (observational, GA, NCT02399072)[41], GAllego (phase 2, GA, NCT03972709)[41]. The trial design and patient inclusion / exclusion criteria are significantly diverse and wide to capture the AMD population. Among all clinical trials, OCT volumes from Heidelberg Spectralis were used except for Mahalo, as the OCT volumes were collected from Zeiss Cirrus devices.
The data is processed to have two usages, the continual unsupervised pre-training and the GA growth rate and BCVA prediction. For the continual pre-training data, no label information is required or included, and all the data was used for the continual pre-training, except Proxima B and Mahalo, to serve as the inductive held out test sets. The total images (counting in 2D) used for training OCTCube-EF is larger than 4 million, including more than 70,000 OCT volumes and more than 402,514 en face images. The final dataset used was screened and volumes / images with lower imaging quality were excluded.
For the data cohort to fine-tune OCTCube-EF for GA lesion growth rate / BCVA prediction, we leveraged the first visit imaging data paired with the quantified GA lesion growth rate and BCVA measurements from a subset of Chroma, Spectri and Proxima A. 20% of the data in Chroma and Spectri were held out to construct the held out `Lampa' test set. We also constructed a test set from GAllego studies, which can be treated as a prospective clinical trial as it is still processing.
We thus constructed four test sets, namely Lampa (in-distribution), GAllego (prospective), Proxima B (held out in pre-training as a full inductive setting), Mahalo (held out in pre-training, cross-device). Collectively, these four evaluation sets cover a wide range of potential distribution shifts settings.

**OCTCube-EF development and implementation details.** *Architecture*. The OCTCube-EF shares almost the same model architecture as used in the cross-modal alignment study. The OCT encoder is set up the same as the OCTCube. The *en face* encoder has the same Vision Transformer structure (ViT) with a modality-specific projection head design. In this task the number of heads is 2 for FAF and IR.

*Pre-training*. We first continually pre-trained the OCTCube on the collected data using 3D MAE. The newer training lasts for 50 epochs with a batch size of 1 on an 8 40GB NVIDIA A100 computing platform, with the effective batch size to be 8. The newer volume size in the pre-training stage shifts to (60, 256, 384), to better handle pixel spacing for volumes with different fields of view.

We then continually trained the *en face* encoder with another 200 epochs using 2D MAE objectives on the collected *en face* data from the cross-modal alignment checkpoint. The image size is enlarged to (384, 384). The batch size is set to be 8 on the same platform. Both en face modalities shared the ViT weight and used different projection heads before coming to the decoder.

*Multi-modal alignment using Tri-COEP*. Compared to IR, integrating OCTCube with the FAF modality is more challenging because of spatial misalignment given that it is not acquired in the same OCT scan session and therefore is not aligned, and because integrating 3 modalities adds additional complexity beyond pairwise cross-modal alignment. We therefore applied Tri-COEP to jointly align all three modalities after the continual MAE pre-training for both encoders. The training recipe is almost the same as with the cross-modal alignment tasks; the only difference is a different batch size and accumulation step adjusted for the new tasks.

*Downstream finetuning*. For the downstream GA lesion growth rate and BCVA regression tasks, we separately trained a model for each task and forced the model to predict the current GA lesion area as an auxiliary task. We found adding this task effectively improved the performance for both metrics. The loss objective for each metric is set to be a linear combination of L2 and L1 loss with both weights equal to 1. The auxiliary area prediction loss has an extra weight of 0.1. The total loss is thus:

$$\mathcal{L}_{comb}(GT, Pred) = \mathcal{L}_1(GT, Pred) + \mathcal{L}_2(GT, Pred)$$

$$\mathcal{L} = \mathcal{L}_{comb}(GT_{bcva/GAGrowth}, Pred_{bcva/GAGrowth}) + 0.1 \times \mathcal{L}_{comb}(GT_{GAarea}, Pred_{GAarea})$$

We set up a 5-fold cross-validation training strategy for the model selection. Specifically, the training set was split into 5 folds, and for each validation fold, the model with best validation performance will be saved, leading to 5 models in total. We calculated the final performance based on the average prediction of the 5 chosen models and performed the same strategy for the other baseline methods. For each fold, the model was trained for 50 epochs with the AdamW optimizer with cosine annealing scheduling strategy. The learning rate was set to be $2 \times 10^{-5}$, and the weight decay was set to be 0.2 for all the parameters except for the gain and bias parameters.

*Baseline implementation*. For the basic OCTCube and the RETFound baseline, the pretrained checkpoint is directly loaded to instantiate the model. RETFound here refers to the RETFound all model, using all OCT slices in the fine-tuning. The training recipe, optimizer and scheduling strategy stayed the same as the OCTCube-EF. For the DenseNet[45] baseline, the FAF used a 2D DenseNet while the OCT volume used a 3D DenseNet. The learning rate was set to be $1 \times 10^{-3}$ and the rest of the training recipe stayed the same as OCTCube-EF.

*Interpretability*. For the interpretable visualization of OCTCube-EF, we followed a similar procedure to the one used for visualizing the saliency map of OCTCube retinal disease prediction model. The grayscale saliency map was adjusted to 256 x 384, and 61 x 384 for the slice and the slow scanning perspective accordingly. For the saliency map of the *en face* image, we instantiated a replica of the *en face* encoder to separately count the saliency of the FAF and IR images to avoid mixed saliency issues because of the weight sharing nature. For the

**Estimation of ESSI and patient recruitment number reduction using prognostic model for covariate adjustment.** To estimate the potential effective sample size increase (ESSI) for a clinical trial, we assumed the randomization ratio to be 1:1 with equal marginal variances across the treatment arms.[46] Let X be the baseline covariate, $Y_{treatment}$ and $Y_{control}$ be the outcome of the treatment arm and control arm, let $r_{treatment}$ and $r_{control}$ be the correlation between the outcome and covariate in the treatment and control arm; the general formula to calculate ESSI will be:

$$ESSI = (\frac{1}{1-(\frac{r_{control}+r_{treatment}}{2})^2} - 1) \times 100\%$$

Here we introduce the calculation of the potential control cohort sample size increase as an example. Let $N$ be the number of patients in both arms, let M be a prognostic model that can achieves $r_{control}$ on the control arm and $r_{treatment}$ the on treatment arm and let $r_{avg} = (r_{control} + r_{treatment})/2$, the adjusted results effectively indicates the same statistical power as an unadjusted results with $N \times (r^2_{avg}/(1 - r^2_{avg})$ more patients. Therefore, the total potential number of patient recruitment needed for an unadjusted analysis will be $N(1/(1 - r^2_{avg}))$, and the difference of the number of patient recruitment reduction between two prognostic model $M_1$ and $M_2$ that can achieve $r_{avg_1}$ and $r_{avg_2}$ ($r_{avg_2} < r_{avg_1}$) on the control arm will be:

$$Diff(M_1, M_2) = N\frac{1}{\frac{1}{1-r^2_{avg_1}} - \frac{1}{1-r^2_{avg_2})}} = N\frac{(1-r^2_{avg1})(1-r^2_{avg2})}{r^2_{avg1} - r^2_{avg2}}$$

For the results reported in the main paper, we chose the ongoing Phase-II clinical trial for Galegenimab (RG6147, see https://clinicaltrials.gov/study/NCT03972709)[94] as an example. We included $N = 440$ to calculate the $r^2$. OCTCube-EF achieved $r^2_{control} = 0.36$ and $r^2_{treatment} = 0.4$, while the DenseNet FAF model achieved $r^2_{control} = 0.29$ and $r^2_{treatment} = 0.33$ (**Supplementary Fig. 16**). Plugging these into the formula above we got the effective number of patient gains to be 709 and 637 for the OCTCube-EF and FAF model. This indicates a 72 more patient gain achieved by OCTCube-EF compared to the FAF model and a 269 more patient gain compared to unadjusted results. For the potential cost benefit, a rough estimate is provided to reveal the potential benefit. The actual cost varies depending on the specific clinical design.

**Evaluation standards and statistical analysis**

For the inhouse retinal and systemic cross-disease prediction tasks, we set up 5 random sub-training and sub-validation splits and report the average performance and standard deviation. For the cross-cohort retinal disease prediction task, we run the 10-fold 10% training data and 5-fold cross validation setting with 3 random splits. We use two-sided t-tests to test OCTCube and the most comparable baseline to show the significance.

**Data Availability**

The ophthalmology imaging data used for the pretraining and internal evaluation were collected from UW-Medicine Ophthalmology. The data used for the continual training of OCTCube-EF and downstream GA prognosis were collected from multiple clinical trials owned by companies. Specifically, data from the following clinical trials are included: NCT02247479; NCT02247531; NCT02479386; NCT02699450; NCT03972709; and an exploratory study.[95] All participants provided written informed consent for future medical research and analyses and the data is anonymized for secondary use. For up-to-date details on Roche's Global Policy on the Sharing of Clinical Information and how to request access to related clinical study documents, (see here: https://go.roche.com/data_sharing). For eligible studies, qualified researchers may request access to the clinical data through a data request platform. At the time of writing, this request platform is Vivli (https://vivli.org/ourmember/roche/). For the imaging data underlying this publication, requests can be made by qualified researchers, subject to a detailed, hypothesis-driven proposal and necessary agreements. Results from SLIViT were reproduced using the provided publicly available codebase (https://github.com/cozygene/SLIViT). The AI-READI dataset can be accessed through https://aireadi.org/. The Glaucoma dataset can be accessed through: https://zenodo.org/records/1481223. The DUKE14 dataset can be accessed through: https://people.duke.edu/~sf59/Srinivasan_BOE_2014_dataset.htm. The UMN dataset can be accessed through: https://people.ece.umn.edu/users/parhi/.DATA/. The HCMS dataset can be accessed through: https://iacl.ece.jhu.edu/index.php?title=Resources. The OIMHS dataset can be accessed through: https://springernature.figshare.com/collections/OIMHS_An_Optical_Coherence_Tomography_Image_Dataset_Based_on_Macular_Hole_Manual_Segmentation/6662624/1.

**Acknowledgements**

We thank Genentech Ophthalmology FAIR data team for data anonymization and preliminary data curation; thank Aicha BenTaieb, Christina Rabe, Daniela Ferrara, Hao Chen, Julia Cluceru, Kenta Yoshida, Ling Ma, Neha Anegondi for their previous relevant work and insightful discussions. We are grateful for all participants of the trials for their contribution to medical research and global health.

We also thank Tong Chen, Xiao Wang, Yu Zhang, Tangqi Fang, Yongkang Li, Zhiping Xiao, Yuguang Li from University of Washington, Can Chen from University of Mcgill, Dr. Yong Chen from University of Pennsylvania, Dr. Qingyu Chen from Yale University for their insightful discussions.

**Reference**

1. Assi, L. *et al.* A global assessment of eye health and quality of life: A systematic review of

Systematic Reviews: A systematic review of systematic reviews. *JAMA Ophthalmol.* **139**, 526–541 (2021).

2. Paudel, N. *et al.* Economic burden of late-stage age-related macular degeneration in Bulgaria, Germany, and the US. *JAMA Ophthalmol.* (2024) doi:10.1001/jamaophthalmol.2024.4401.

3. Flaxman, S. R. *et al.* Global causes of blindness and distance vision impairment 1990-2020: a systematic review and meta-analysis. *Lancet Glob Health* **5**, e1221–e1234 (2017).

4. Forrest, S. L. *et al.* Does the Current Global Health Agenda Lack Vision? *Glob Health Sci Pract* **11**, (2023).

5. Bernardes, R. & Cunha-Vaz, J. *Optical Coherence Tomography: A Clinical and Technical Update*. (Springer Science & Business Media, 2012).

6. Burlina, P., Paul, W., Liu, T. Y. A. & Bressler, N. M. Detecting Anomalies in Retinal Diseases Using Generative, Discriminative, and Self-supervised Deep Learning. *JAMA Ophthalmol.* **140**, 185–189 (2022).

7. Campbell, J. P. *et al.* Detailed Vascular Anatomy of the Human Retina by Projection-Resolved Optical Coherence Tomography Angiography. *Sci Rep* **7**, 42201 (2017).

8. Park, K. H. & Kim, T.-W. *OCT Imaging in Glaucoma: A Guide for Practitioners*. (2021).

9. Parravano, M. *et al.* Multimodal imaging in diabetic retinopathy and macular edema: An update about biomarkers. *Surv. Ophthalmol.* (2024) doi:10.1016/j.survophthal.2024.06.006.

10. Li, X., Jia, M., Islam, M. T., Yu, L. & Xing, L. Self-Supervised Feature Learning via Exploiting Multi-Modal Data for Retinal Disease Diagnosis. *IEEE Trans. Med. Imaging* **39**, 4023–4033 (2020).

11. Tao, Y. *et al.* Exploration on OCT biomarker candidate related to macular edema caused by diabetic retinopathy and retinal vein occlusion in SD-OCT images. *Sci. Rep.* **14**, 14317 (2024).

12. Shanthini, A., Manogaran, G. & Vadivu, G. *Deep Convolutional Neural Network for The Prognosis of Diabetic Retinopathy*. (Springer Nature, 2022).

13. Prenner, V. *et al.* Advancing the visibility of outer retinal integrity in neovascular age-related

macular degeneration with high-resolution OCT. *Can. J. Ophthalmol.* (2024) doi:10.1016/j.jcjo.2024.05.014.

14. Ye, X. *et al.* OIMHS: An Optical Coherence Tomography Image Dataset Based on Macular Hole Manual Segmentation. *Sci Data* **10**, 769 (2023).

15. Razavi, P. *et al.* Changes in wider field swept-source OCT angiography vascular metrics with anti-vascular endothelial growth factor therapy in central retinal vein occlusion. *Graefes Arch. Clin. Exp. Ophthalmol.* **262**, 2111–2120 (2024).

16. Hong, S. H. & Kim, H. D. Central retinal artery occlusion after intravitreal brolucizumab injection for treatment-naïve neovascular age-related macular degeneration; a case report. *BMC Ophthalmol.* **24**, 200 (2024).

17. De Fauw, J. *et al.* Clinically applicable deep learning for diagnosis and referral in retinal disease. *Nat. Med.* **24**, 1342–1350 (2018).

18. Curcio, C. A., Kar, D., Owsley, C., Sloan, K. R. & Ach, T. Age-Related Macular Degeneration, a Mathematically Tractable Disease. *Invest. Ophthalmol. Vis. Sci.* **65**, 4 (2024).

19. Lin, A. C., Lee, C. S., Blazes, M., Lee, A. Y. & Gorin, M. B. Assessing the Clinical Utility of Expanded Macular OCTs Using Machine Learning. *Transl. Vis. Sci. Technol.* **10**, 32 (2021).

20. Park, S.-J., Ko, T., Park, C.-K., Kim, Y.-C. & Choi, I.-Y. Deep Learning Model Based on 3D Optical Coherence Tomography Images for the Automated Detection of Pathologic Myopia. *Diagnostics (Basel)* **12**, (2022).

21. Ikezogwo, W. O. *et al.* Quilt-1M: One Million Image-Text Pairs for Histopathology. *Adv. Neural Inf. Process. Syst.* **36**, 37995–38017 (2023).

22. Xu, H. *et al.* A whole-slide foundation model for digital pathology from real-world data. *Nature* **630**, 181–188 (2024).

23. Huang, Z., Bianchi, F., Yuksekgonul, M., Montine, T. J. & Zou, J. A visual-language foundation model for pathology image analysis using medical Twitter. *Nat. Med.* **29**, 2307–2316 (2023).

24. Center for Drug Evaluation & Research. Adjusting for Covariates in Randomized Clinical Trials for Drugs and biological products. *U.S. Food and Drug Administration* https://www.fda.gov/regulatory-information/search-fda-guidance-documents/adjusting-covariates-randomized-clinical-trials-drugs-and-biological-products (2024).

25. Maetschke, S. *et al.* A feature agnostic approach for glaucoma detection in OCT volumes. *PLoS One* **14**, e0219126 (2019).

26. Avram, O. *et al.* Accurate prediction of disease-risk factors from volumetric medical scans by a deep vision model pre-trained with 2D scans. *Nat Biomed Eng* (2024) doi:10.1038/s41551-024-01257-9.

27. Yang, J. *et al.* MedMNIST v2 - A large-scale lightweight benchmark for 2D and 3D biomedical image classification. *Sci Data* **10**, 41 (2023).

28. Ouyang, D. *et al.* Video-based AI for beat-to-beat assessment of cardiac function. *Nature* **580**, 252–256 (2020).

29. Ko, T. H., Chisholm, C. M. & Chen, M. H. 11. Optical coherence tomography angiography imaging on the Topcon Triton and Maestro2 systems. *Optical Coherence Tomography Angiography of the Eye* (2024).

30. AI-READI Consortium. Flagship dataset of type 2 diabetes from the AI-READI project. FAIRhub https://doi.org/10.34534/1 (2024).

31. Chen, B., Chen, H., Zheng, C. & Zhang, M. Performance of Topcon 3D optical coherence tomography-2000 in re-analyzing OCT-1000 raw data. *Exp. Ther. Med.* **17**, 4395–4402 (2019).

32. Barteselli, G. *et al.* Accuracy of the Heidelberg Spectralis in the alignment between near-infrared image and tomographic scan in a model eye: a multicenter study. *Am. J. Ophthalmol.* **156**, 588–592 (2013).

33. Buck, C. J. *2015 ICD-9-CM for Hospitals, Volumes 1, 2 and 3 Standard Edition - E-Book*. (Elsevier Health Sciences, 2015).

34. Cartwright, D. J. ICD-9-CM to ICD-10-CM Codes: What? Why? How? *Adv. Wound Care* **2**, 588–592

(2013).

35. American Medical Association. *ICD-10-CM 2020 the Complete Official Codebook*. (American Medical Association Press, 2019).

36. Chen, M. *et al.* Automated diagnosis of age-related macular degeneration using multi-modal vertical plane feature fusion via deep learning. *Med. Phys.* **49**, 2324–2333 (2022).

37. Bourne, R. R. A. *et al.* Causes of vision loss worldwide, 1990-2010: a systematic analysis. *Lancet Glob Health* **1**, e339–49 (2013).

38. Boyer, D. S., Schmidt-Erfurth, U., van Lookeren Campagne, M., Henry, E. C. & Brittain, C. THE PATHOPHYSIOLOGY OF GEOGRAPHIC ATROPHY SECONDARY TO AGE-RELATED MACULAR DEGENERATION AND THE COMPLEMENT PATHWAY AS A THERAPEUTIC TARGET. *Retina* **37**, 819–835 (2017).

39. Anegondi, N. *et al.* Deep Learning to Predict Geographic Atrophy Area and Growth Rate from Multimodal Imaging. *Ophthalmol Retina* **7**, 243–252 (2023).

40. Fleckenstein, M. *et al.* The progression of geographic atrophy secondary to age-related macular degeneration. *Ophthalmology* **125**, 369–390 (2018).

41. Holekamp, N. *et al.* Natural history of geographic atrophy secondary to age-related macular degeneration: results from the prospective Proxima A and B clinical trials. *Ophthalmology* **127**, 769–783 (2020).

42. Shen, L. L., Sun, M., Grossetta Nardini, H. K. & Del Priore, L. V. Progression of Unifocal versus Multifocal Geographic Atrophy in Age-Related Macular Degeneration: A Systematic Review and Meta-analysis. *Ophthalmol Retina* **4**, 899–910 (2020).

43. Holz, F. G. *et al.* Efficacy and Safety of Lampalizumab for Geographic Atrophy Due to Age-Related Macular Degeneration: Chroma and Spectri Phase 3 Randomized Clinical Trials. *JAMA Ophthalmol* **136**, 666–677 (2018).

44. Sivaprasad, S. *et al.* Reliability and Construct Validity of the NEI VFQ-25 in a Subset of Patients With Geographic Atrophy From the Phase 2 Mahalo Study. *Am J Ophthalmol* **190**, 1–8 (2018).

45. Huang, G., Liu, Z., van der Maaten, L. & Weinberger, K. Q. Densely connected convolutional networks. in *Proceedings of the IEEE Conference on Computer Vision and Pattern Recognition* (2017).

46. Schiffman, C., Rabe, C. & Friesenhahn, M. stats4datascience - How to get the most out of prognostic baseline variables in clinical trials. https://stats4datascience.com/posts/covariate_adjustment/index.html (2022).

47. Coulibaly, L. M. *et al.* Progression dynamics of early versus later stage atrophic lesions in nonneovascular age-related macular degeneration using quantitative OCT biomarker segmentation. *Ophthalmol. Retina* **7**, 762–770 (2023).

48. Lee, C. S., Baughman, D. M. & Lee, A. Y. Deep learning is effective for the classification of OCT images of normal versus Age-related Macular Degeneration. *Ophthalmol Retina* **1**, 322–327 (2017).

49. Rasel, R. K. *et al.* Assessing the efficacy of 2D and 3D CNN algorithms in OCT-based glaucoma detection. *Sci. Rep.* **14**, 11758 (2024).

50. Sun, L.-C. *et al.* Generative adversarial network-based deep learning approach in classification of retinal conditions with optical coherence tomography images. *Graefes Arch. Clin. Exp. Ophthalmol.* **261**, 1399–1412 (2023).

51. Opoku, M., Weyori, B. A., Adekoya, A. F. & Adu, K. CLAHE-CapsNet: Efficient retina optical coherence tomography classification using capsule networks with contrast limited adaptive histogram equalization. *PLoS One* **18**, e0288663 (2023).

52. Wong, A., Mishra, A., Bizheva, K. & Clausi, D. A. General Bayesian estimation for speckle noise reduction in optical coherence tomography retinal imagery. *Opt. Express* **18**, 8338–8352 (2010).

53. Sun, J. Q. *et al.* Comparison of the Iowa Reference Algorithm to the Heidelberg Spectralis optical coherence tomography segmentation algorithm. *J. Biophotonics* **13**, e201960187 (2020).

54. Garvin, M. K. *et al.* Automated 3-D intraretinal layer segmentation of macular spectral-domain optical coherence tomography images. *IEEE Trans. Med. Imaging* **28**, 1436–1447 (2009).

55. Quellec, G. *et al.* Three-dimensional analysis of retinal layer texture: identification of fluid-filled regions in SD-OCT of the macula. *IEEE Trans. Med. Imaging* **29**, 1321–1330 (2010).

56. Zhang, M. *et al.* Advanced image processing for optical coherence tomographic angiography of macular diseases. *Biomed Opt Express* **6**, 4661–4675 (2015).

57. Li, Q. *et al.* DeepRetina: Layer Segmentation of Retina in OCT Images Using Deep Learning. *Transl. Vis. Sci. Technol.* **9**, 61 (2020).

58. Cai, L. *et al.* Classification of diabetic maculopathy based on optical coherence tomography images using a Vision Transformer model. *BMJ Open Ophthalmol* **8**, (2023).

59. Zhou, Y. *et al.* A foundation model for generalizable disease detection from retinal images. *Nature* **622**, 156–163 (2023).

60. Zhang, J. *et al.* RETFound-enhanced community-based fundus disease screening: real-world evidence and decision curve analysis. *NPJ Digit Med* **7**, 108 (2024).

61. Zhao, H. *et al.* MMICL: Empowering Vision-language Model with Multi-Modal In-Context Learning. *arXiv [cs.CL]* (2023).

62. Li, C. *et al.* Llava-med: Training a large language-and-vision assistant for biomedicine in one day. *Adv. Neural Inf. Process. Syst.* **36**, (2024).

63. Wu, R. *et al.* MM-Retinal: Knowledge-enhanced foundational pretraining with Fundus Image-text expertise. *ArXiv* **abs/2405.11793**, (2024).

64. Spaide, T. *et al.* Geographic Atrophy Segmentation Using Multimodal Deep Learning. *Transl. Vis. Sci. Technol.* **12**, 10 (2023).

65. Niemeijer, M., van Ginneken, B., Staal, J., Suttorp-Schulten, M. S. A. & Abràmoff, M. D. Automatic detection of red lesions in digital color fundus photographs. *IEEE Trans. Med. Imaging* **24**, 584–592 (2005).

66. Fernando Arévalo, J. *Retinal Angiography and Optical Coherence Tomography*. (Springer Science & Business Media, 2008).

67. Lundberg, S. M. & Lee, S.-I. A unified approach to interpreting model predictions. *Adv. Neural Inf. Process. Syst.* 4765–4774 (2017).

68. Chefer, H., Gur, S. & Wolf, L. Transformer Interpretability Beyond Attention Visualization. in *2021 IEEE/CVF Conference on Computer Vision and Pattern Recognition (CVPR)* (IEEE, 2021). doi:10.1109/cvpr46437.2021.00084.

69. Zhao, Q., Liu, Z., Adeli, E. & Pohl, K. M. Longitudinal self-supervised learning. *Med. Image Anal.* **71**, 102051 (2021).

70. Liu, Z. *et al. ForeSeer: Product Aspect Forecasting Using Temporal Graph Embedding*. (2023).

71. Grzybowski, A. *Artificial Intelligence in Ophthalmology*. (Springer Nature, 2021).

72. Ting, D. *et al.* Artificial intelligence and deep learning in ophthalmology. *Br. J. Ophthalmol.* **103**, 167–175 (2018).

73. Feichtenhofer, C., Fan, H., Li, Y. & He, K. Masked Autoencoders As Spatiotemporal Learners. *Adv. Neural Inf. Process. Syst.* **abs/2205.09113**, (2022).

74. Tong, Z., Song, Y., Wang, J. & Wang, L. VideoMAE: Masked autoencoders are data-efficient learners for self-supervised video pre-training. *Adv. Neural Inf. Process. Syst.* **abs/2203.12602**, (2022).

75. He, K. *et al.* Masked autoencoders are scalable vision learners. in *Proceedings of the IEEE/CVF conference on computer vision and pattern recognition* 16000–16009 (2022).

76. Dosovitskiy, A. *et al.* An image is worth 16x16 words: Transformers for image recognition at scale. *Int Conf Learn Represent* **abs/2010.11929**, (2020).

77. Dao, T., Fu, D. Y., Ermon, S., Rudra, A. & R'e, C. FlashAttention: Fast and memory-efficient exact attention with IO-awareness. *Adv. Neural Inf. Process. Syst.* **abs/2205.14135**, (2022).

78. Dao, T. FlashAttention-2: Faster attention with better parallelism and work partitioning. *ArXiv*

**abs/2307.08691**, (2023).

79. Srivastava, N., Hinton, G. E., Krizhevsky, A., Sutskever, I. & Salakhutdinov, R. Dropout: a simple way to prevent neural networks from overfitting. *J. Mach. Learn. Res.* **15**, 1929–1958 (2014).

80. Hendrycks, D. & Gimpel, K. Gaussian Error Linear Units (GELUs). *arXiv [cs.LG]* (2016).

81. Chattopadhay, A., Sarkar, A., Howlader, P. & Balasubramanian, V. N. Grad-CAM++: Generalized gradient-based visual explanations for deep convolutional networks. in *2018 IEEE Winter Conference on Applications of Computer Vision (WACV)* (IEEE, 2018). doi:10.1109/wacv.2018.00097.

82. Gildenblat, J. & Contributors, D. PyTorch library for CAM methods. Preprint at https://github.com/jacobgil/pytorch-grad-cam (2021).

83. Rashno, A. *et al.* Fully-automated segmentation of fluid regions in exudative age-related macular degeneration subjects: Kernel graph cut in neutrosophic domain. *PLoS One* **12**, e0186949 (2017).

84. Srinivasan, P. P. *et al.* Fully automated detection of diabetic macular edema and dry age-related macular degeneration from optical coherence tomography images. *Biomed. Opt. Express* **5**, 3568–3577 (2014).

85. He, Y. *et al.* Retinal layer parcellation of optical coherence tomography images: Data resource for multiple sclerosis and healthy controls. *Data Brief* **22**, 601–604 (2019).

86. Otsu, N. A threshold selection method from gray-level histograms. *IEEE Trans. Syst. Man Cybern.* **9**, 62–66 (1979).

87. Radford, A. *et al.* Learning Transferable Visual Models From Natural Language Supervision. in *Proceedings of the 38th International Conference on Machine Learning* (eds. Meila, M. & Zhang, T.) vol. 139 8748–8763 (PMLR, 18--24 Jul 2021).

88. van den Oord, A., Li, Y. & Vinyals, O. Representation learning with Contrastive Predictive Coding. *ArXiv* **abs/1807.03748**, (2018).

89. Creators Ilharco, Gabriel Wortsman, Mitchell Carlini, Nicholas Taori, Rohan Dave, Achal

Shankar, Vaishaal Namkoong, Hongseok Miller, John Hajishirzi, Hannaneh Farhadi, Ali Schmidt, Ludwig. *OpenCLIP*. doi:10.5281/zenodo.5143773.

90. Paszke, A. *et al.* PyTorch: An imperative style, high-performance deep learning library. *Adv. Neural Inf. Process. Syst.* **abs/1912.01703**, (2019).

91. Zhang, S. *et al.* BiomedCLIP: a multimodal biomedical foundation model pretrained from fifteen million scientific image-text pairs. *arXiv preprint arXiv:2303. 00915* (2023).

92. Lois, N. & Forrester, J. V. *Fundus Autofluorescence*. (Lippincott Williams & Wilkins, 2012).

93. Holz, F. G. *et al.* Imaging Protocols in Clinical Studies in Advanced Age-Related Macular Degeneration: Recommendations from Classification of Atrophy Consensus Meetings. *Ophthalmology* **124**, 464–478 (2017).

94. A Study Assessing the Safety, Tolerability, and Efficacy of Galegenimab (FHTR2163) in Participants With Geographic Atrophy Secondary to Age-Related Macular Degeneration (AMD) (GALLEGO). *ClinicalTrial.gov*.

95. Willis, J. R. *et al.* Feasibility study of a multimodal, cloud-based, diabetic retinal screening program in a workplace environment. *Transl. Vis. Sci. Technol.* **10**, 20 (2021).

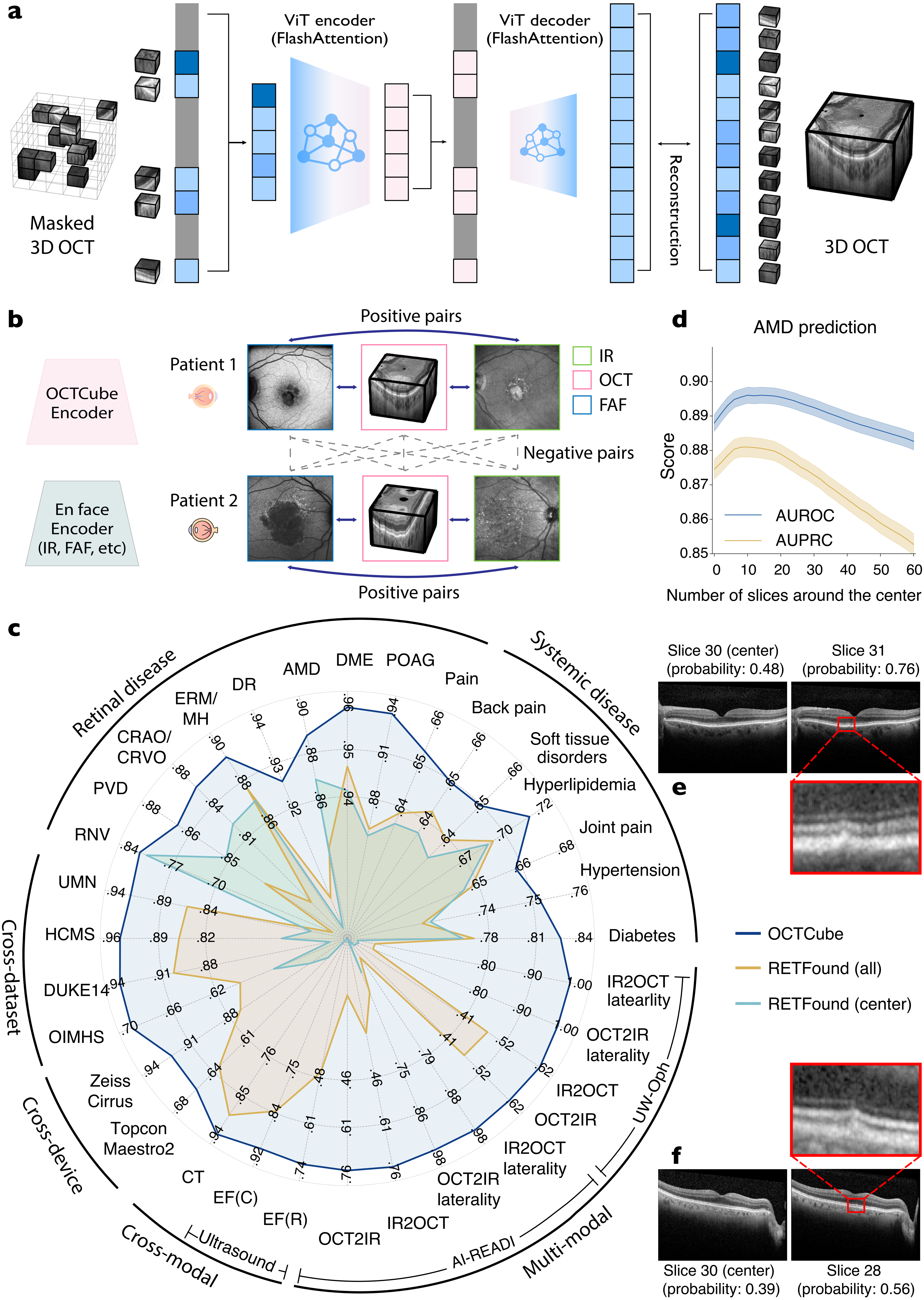

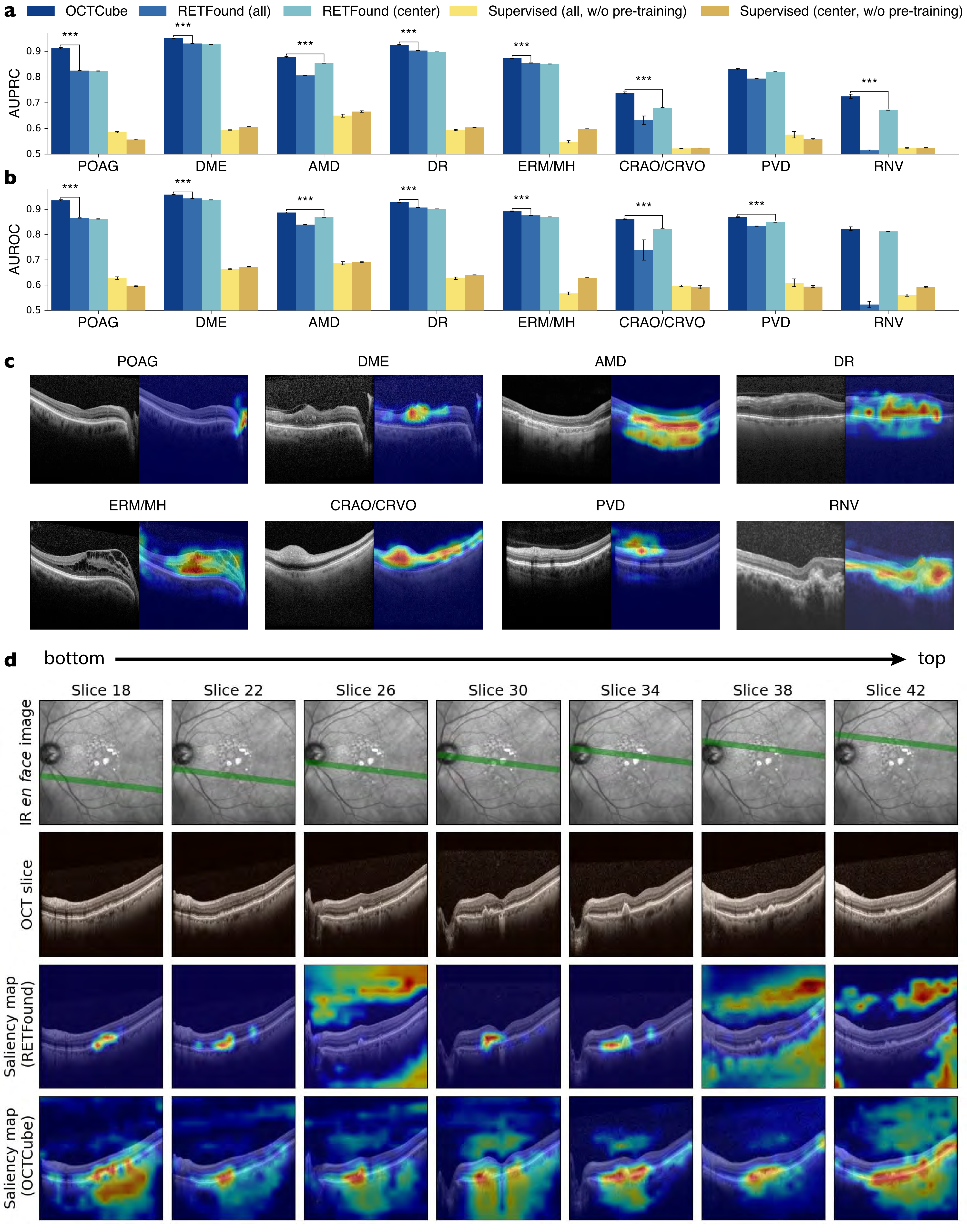

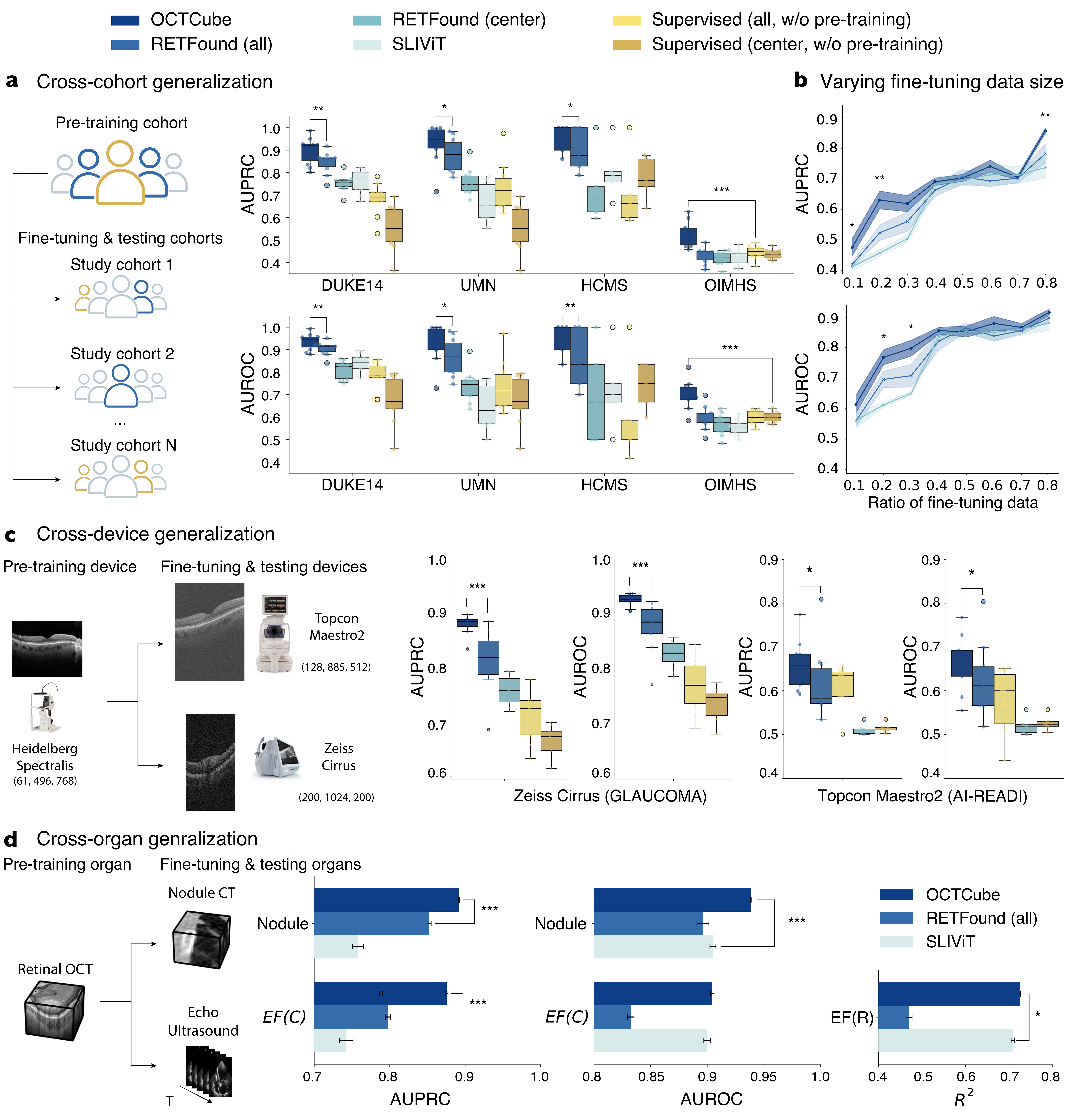

OCTCube
RETFound (all)
RETFound (center)
SLIViT
Supervised (all, w/o pre-training)
Supervised (center, w/o pre-training)
a Cross-cohort generalization
Pre-training cohort
Fine-tuning & testing cohorts
Study cohort 1
Study cohort 2
...
Study cohort N
AUPRC
AUROC
DUKE14
UMN
HCMS
OIMHS
b Varying fine-tuning data size
Ratio of fine-tuning data
c Cross-device generalization
Pre-training device
Fine-tuning & testing devices
Heidelberg Spectralis (61, 496, 768)
Topcon Maestro2 (128, 885, 512)
Zeiss Cirrus (200, 1024, 200)
Zeiss Cirrus (GLAUCOMA)
Topcon Maestro2 (AI-READI)
d Cross-organ genralization
Pre-training organ
Fine-tuning & testing organs
Retinal OCT
Nodule CT
Echo Ultrasound
T
Nodule
EF(C)
EF(R)
AUPRC
AUROC
$R^2$

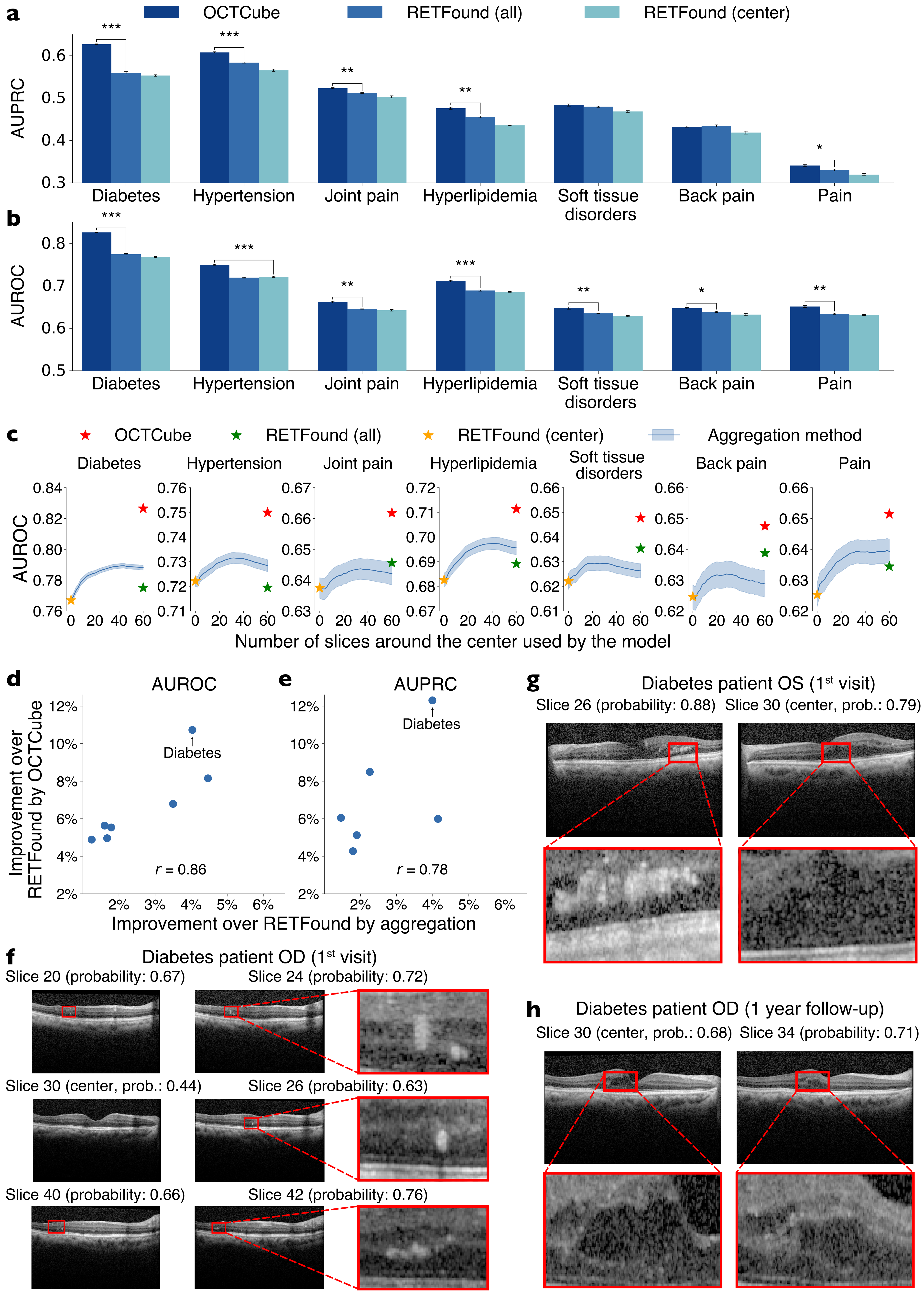

a
OCTCube
RETFound (all)
RETFound (center)
AUPRC
Diabetes
Hypertension
Joint pain
Hyperlipidemia
Soft tissue disorders
Back pain
Pain
b
AUROC
c
OCTCube
RETFound (all)
RETFound (center)
Aggregation method
Number of slices around the center used by the model
d
AUROC
Diabetes
r = 0.86
e
AUPRC
Diabetes
r = 0.78
Improvement over RETFound by OCTCube
Improvement over RETFound by aggregation
f
Diabetes patient OD (1st visit)
Slice 20 (probability: 0.67)
Slice 24 (probability: 0.72)
Slice 30 (center, prob.: 0.44)
Slice 26 (probability: 0.63)
Slice 40 (probability: 0.66)
Slice 42 (probability: 0.76)
g
Diabetes patient OS (1st visit)
Slice 26 (probability: 0.88)
Slice 30 (center, prob.: 0.79)
h
Diabetes patient OD (1 year follow-up)
Slice 30 (center, prob.: 0.68)
Slice 34 (probability: 0.71)

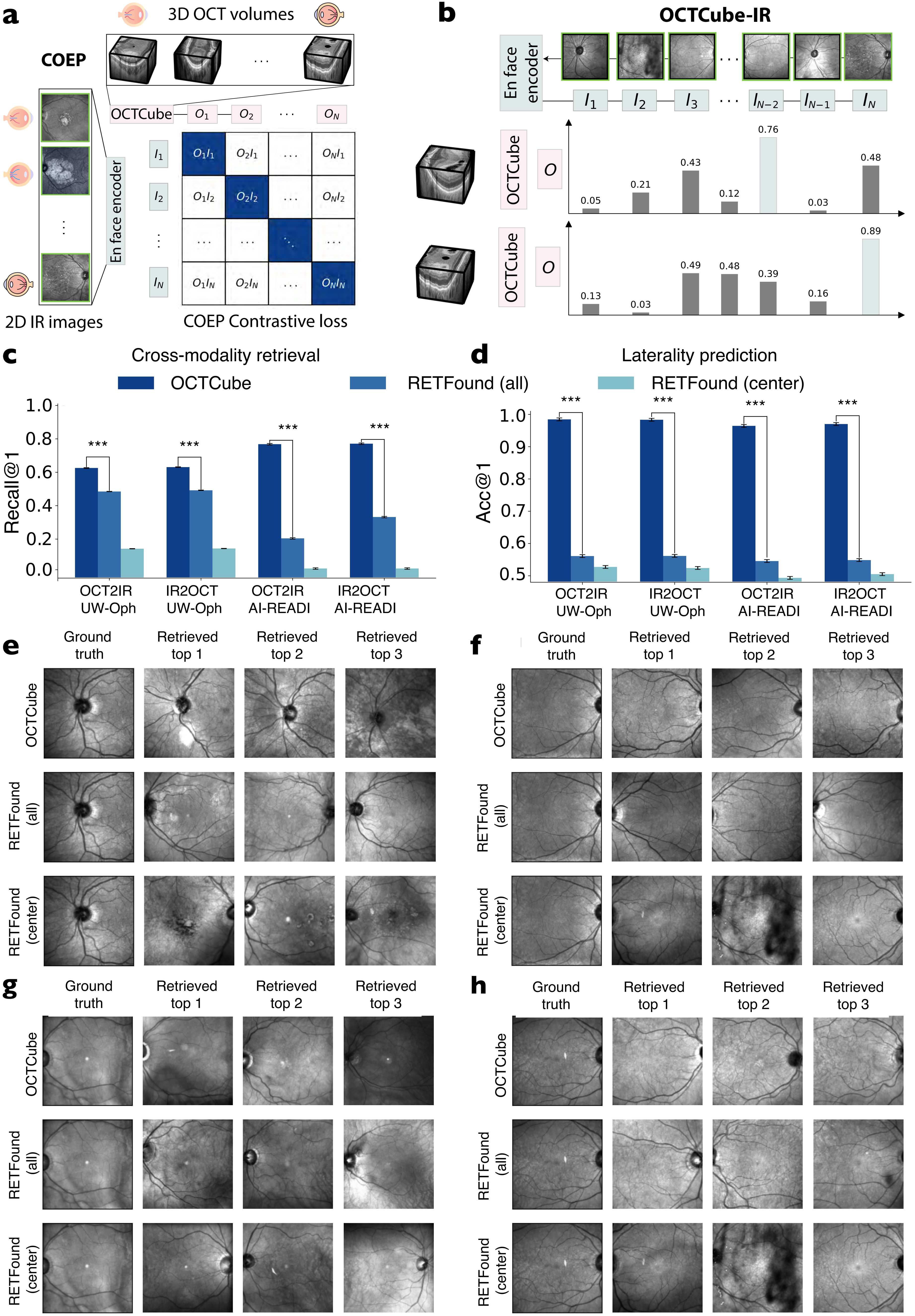

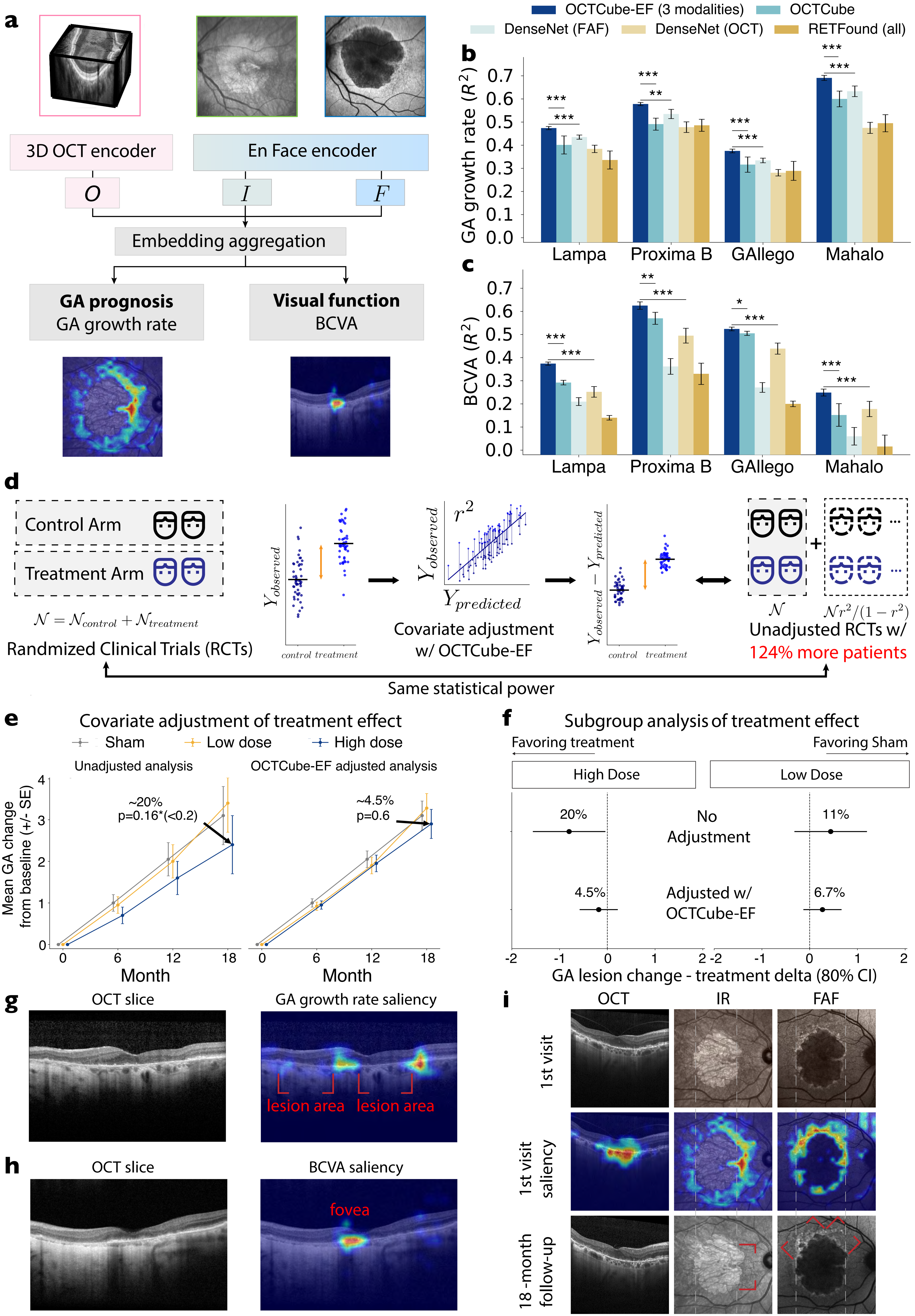
a
3D OCT encoder
En Face encoder
O
I
F
Embedding aggregation
GA prognosis
GA growth rate
Visual function
BCVA
OCTCube-EF (3 modalities)
OCTCube
DenseNet (FAF)
DenseNet (OCT)
RETFound (all)
b
GA growth rate ($R^2$)
Lampa
Proxima B
GAllego
Mahalo
c
BCVA ($R^2$)
Lampa
Proxima B
GAllego
Mahalo
d
Control Arm
Treatment Arm
$\mathcal{N} = \mathcal{N}_{control} + \mathcal{N}_{treatment}$
Randmized Clinical Trials (RCTs)
$Y_{observed}$
control treatment
$r^2$
$Y_{predicted}$
Covariate adjustment
w/ OCTCube-EF
$Y_{observed} - Y_{predicted}$
$\mathcal{N}$
$\mathcal{N}r^2/(1-r^2)$
Unadjusted RCTs w/
124% more patients
Same statistical power
e
Covariate adjustment of treatment effect
Sham
Low dose
High dose
Unadjusted analysis
OCTCube-EF adjusted analysis
Mean GA change from baseline (+/- SE)
~20%
p=0.16*(<0.2)
~4.5%
p=0.6
Month
f
Subgroup analysis of treatment effect
Favoring treatment
Favoring Sham
High Dose
Low Dose
No Adjustment
Adjusted w/ OCTCube-EF
20%
11%
4.5%
6.7%
GA lesion change - treatment delta (80% CI)
g
OCT slice
GA growth rate saliency
lesion area
lesion area
h
OCT slice
BCVA saliency
fovea
i
OCT
IR
FAF
1st visit
1st visit saliency
18-month follow-up

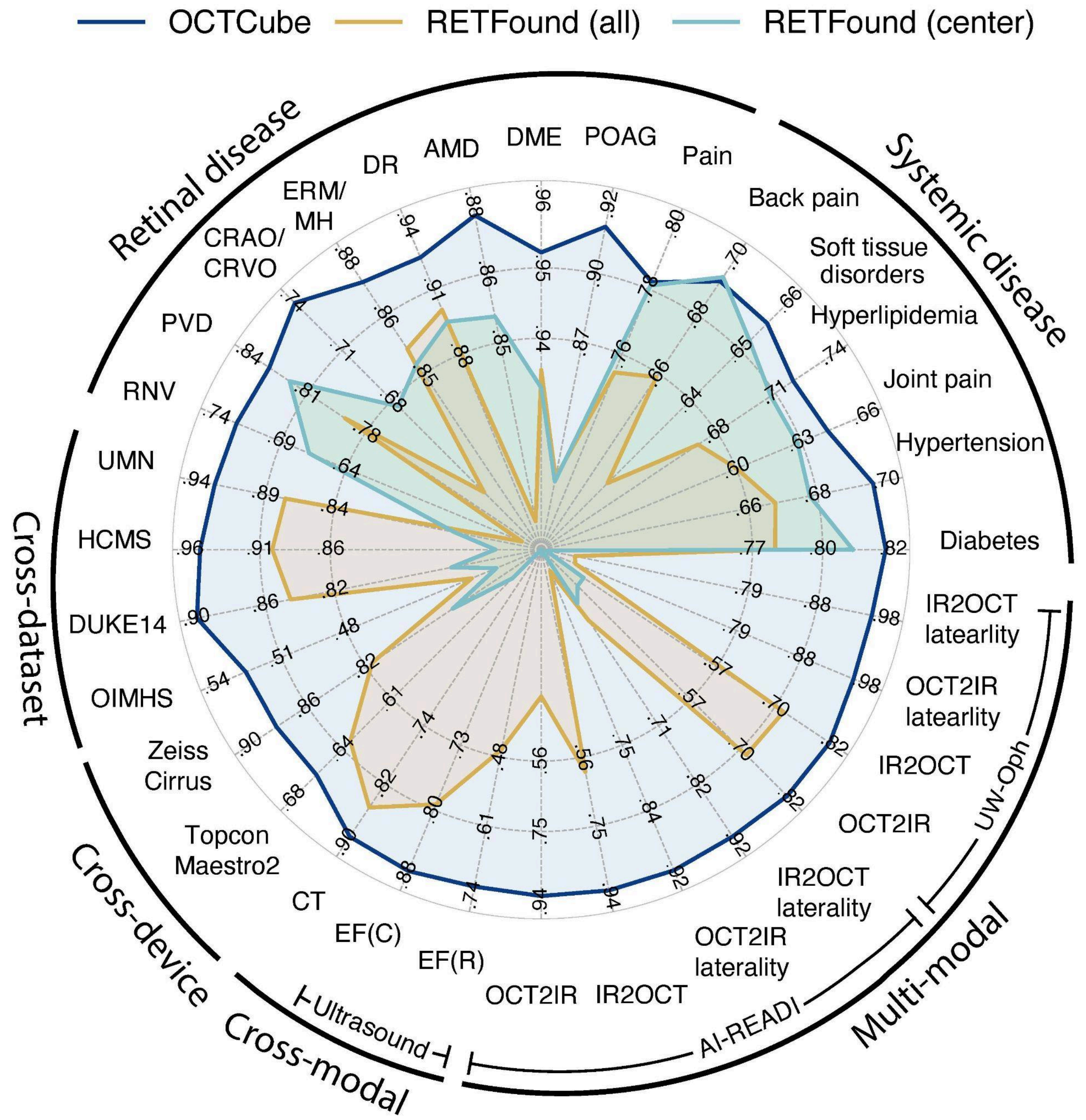

**Supplementary Fig. 1** Radar plot comparing the performance of OCTCube and competing methods on 32 tasks, including eight retinal disease prediction tasks in the inductive learning setting, four retinal disease prediction tasks in the cross-cohort learning setting, seven systemic disease prediction tasks, three cross-organ transferring prediction tasks, eight cross-modality retrieval and classification tasks and two cross-device prediction task. Recall@5 is used as the metric for the cross-modal retrieval tasks, Acc@5 is used as the metric for the cross-modal laterality prediction tasks, coefficient of determinant ($R^2$) is used as the metric for EF(R) ejection fraction prediction task, while AUPRC is used as the metric for the other tasks. UW-Oph is the abbreviation for UW Ophthalmology dataset. EF (C) is the low ejection fraction classification task, and EF (R) is the ejection fraction regression task.

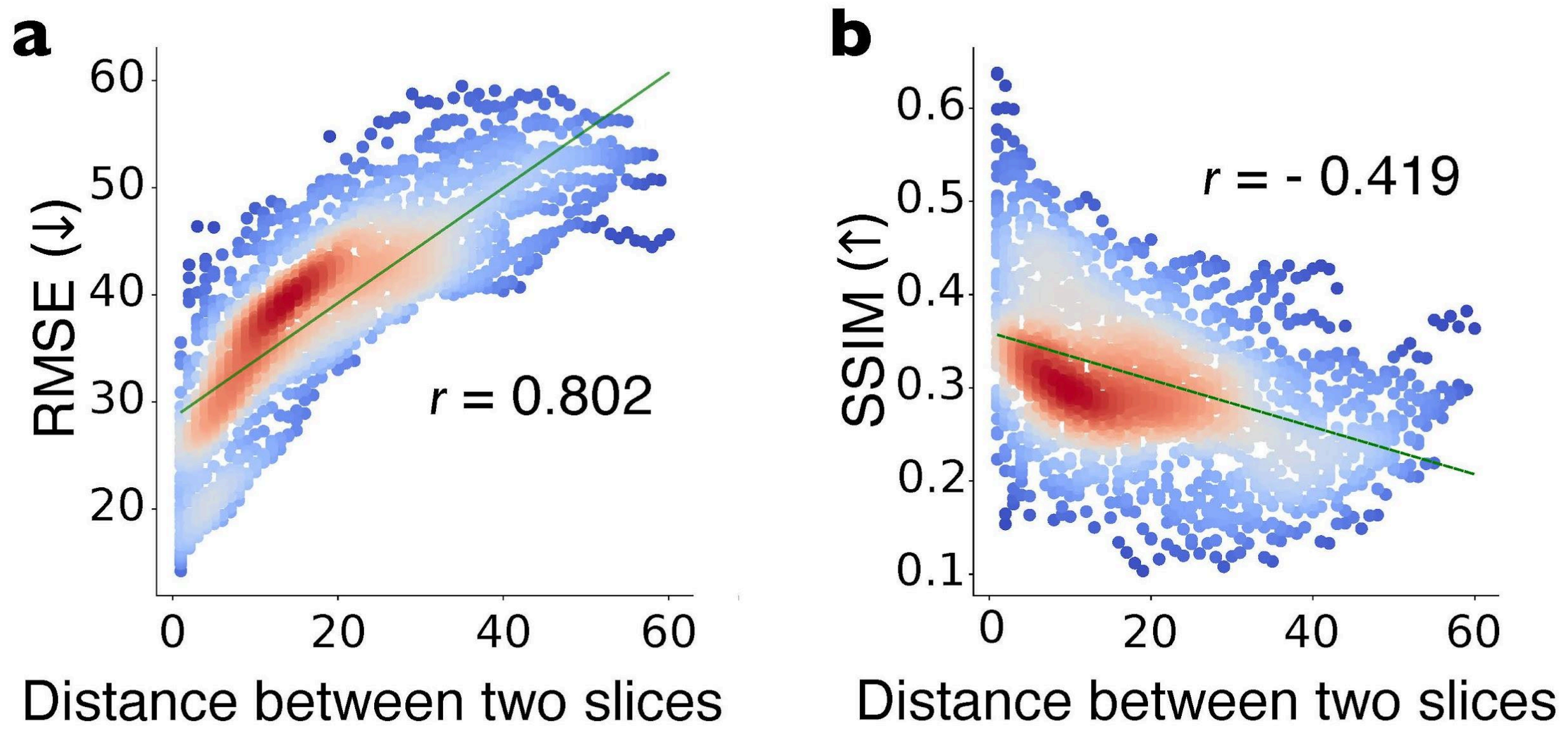

**Supplementary Fig. 2 a-b,** Scatter plots comparing the similarity of two slices in an OCT volume in terms of RMSE (**a**) and SSIM (**b**). X-axis indicates the distance between two slices in the same volume. ***r*** denotes Pearson correlation coefficient.

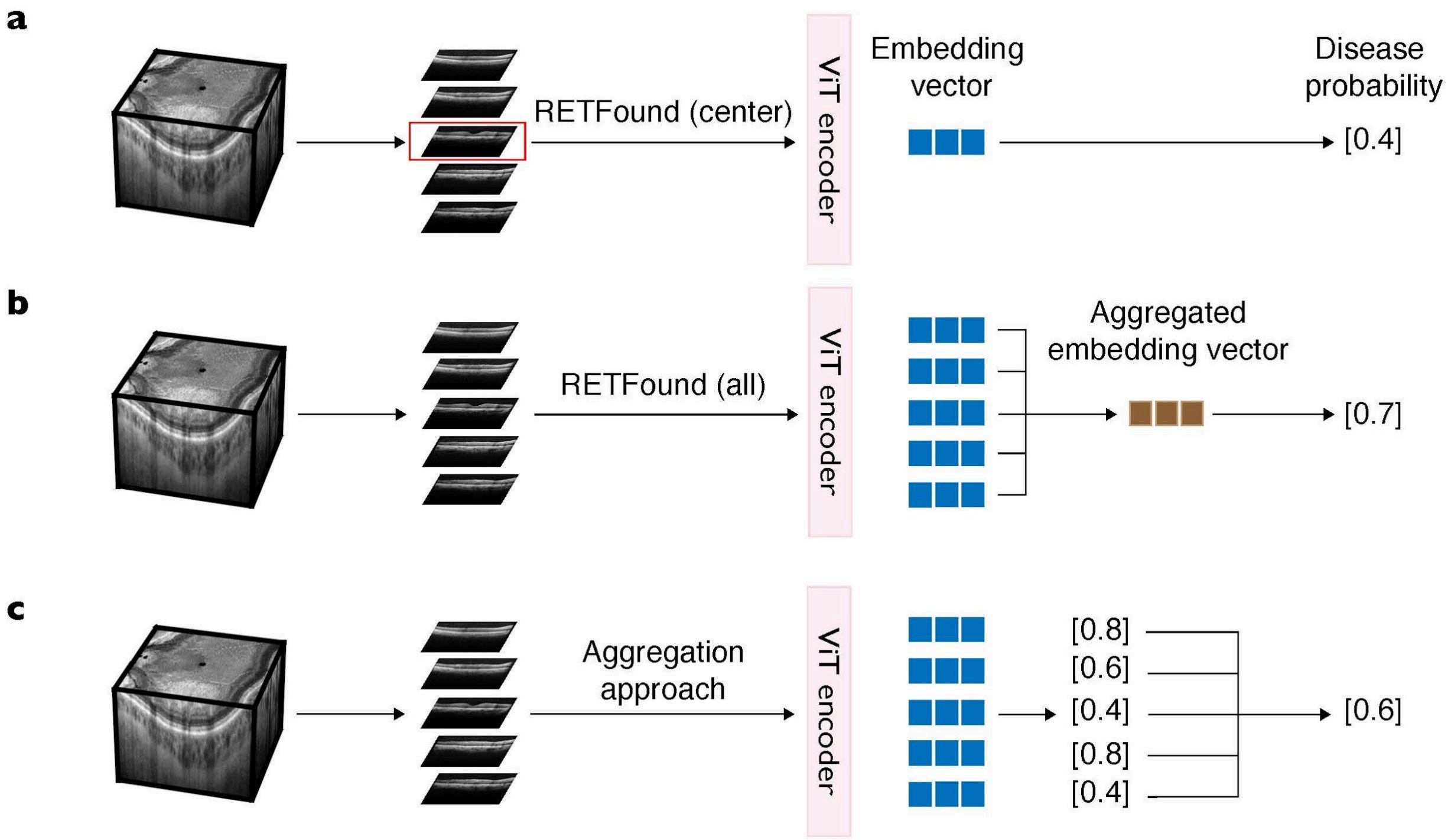

**Supplementary Fig. 3** Illustration of various competing methods. **a,** RETFound (center) extracts the center slice and generates predictions based on it. **b,** RETFound (all) takes all slices as input, extracts embeddings for each slice, averages all embeddings, and gets the final prediction score using neural networks. **c**, The aggregation approach first generates the prediction score for every slice using RETFound (center), and then averages the prediction scores to get the final prediction score.

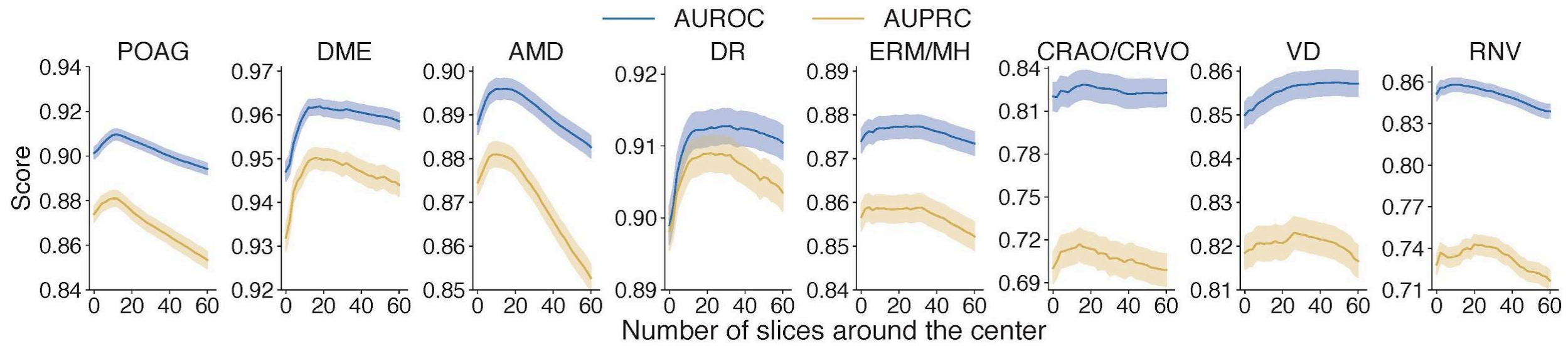

**Supplementary Fig. 4** Plots showing the AUROC and AUPRC of an aggregation approach (**Supplementary Fig. 3c**) that averages the prediction probabilities using each of the k slices around the center slices on eight retinal diseases, where k is shown in the x-axis. The prediction probability for each slice is derived using RETFound. POAG, DME, AMD, ERM/MH, DR, CRAO/CRVO, VD, RNV denote primary open-angle glaucoma, diabetic macular edema, age-related macular degeneration, epiretinal membrane or macular hole, diabetic retinography without macular edema, central retinal vein / artery occlusion, vitreous degeneration, and retinal neovascularization respectively. The metric AUROC and AUPRC are the abbreviation of Area under the Receiver Operating Characteristic Curve and the Area under the Precision-Recall Curve. RETFound, as a 2D approach, corresponds to k = 0.

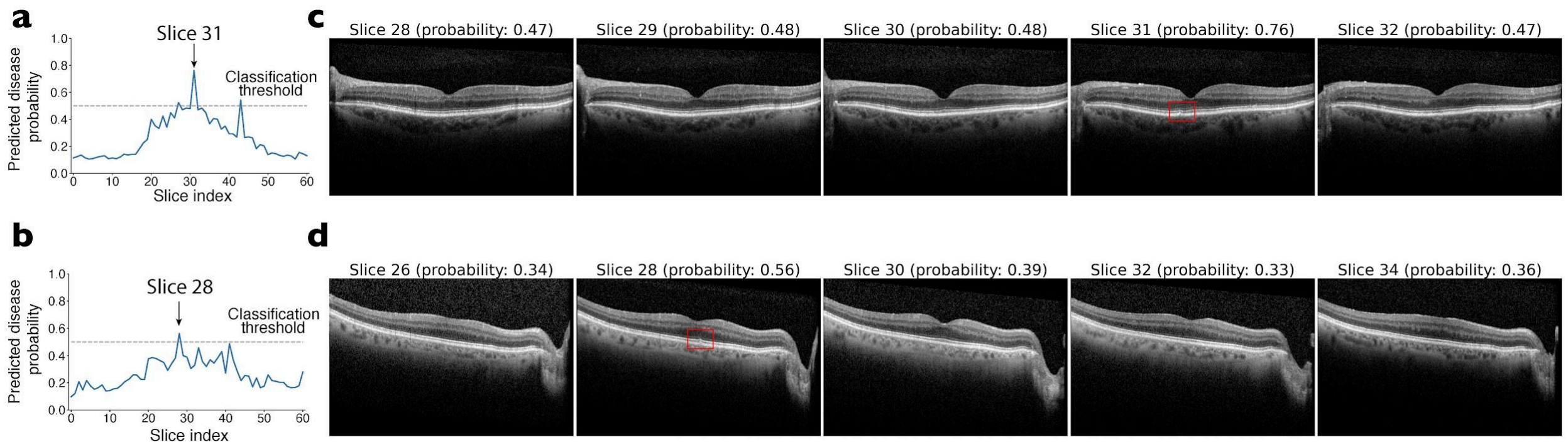

**Supplementary Fig. 5 a,b**, Plots showing the prediction probabilities of age-related macular degeneration (AMD) by using each slice as the input for RETFound on two AMD patients. While the center slice (slice 30) incorrectly predicts AMD as negative, there exist other slices near the center slice that correctly predict AMD as positive. Slice 31 for the first patient (**a**) and slice 28 for the second patient (**b**) achieve the highest prediction probability, suggesting the possibility to improve the prediction performance by considering the 3D structure. **c,d**, Visualization of the OCT slices around the center slice (slice 30) and the corresponding prediction probabilities of the AMD patient (**c** for patient 1 and **d** for patient 2). Red boxes highlight the small drusen that occurs at the slice 31 of patient 1 and the slice 28 of patient 2, indicating signals for AMD.

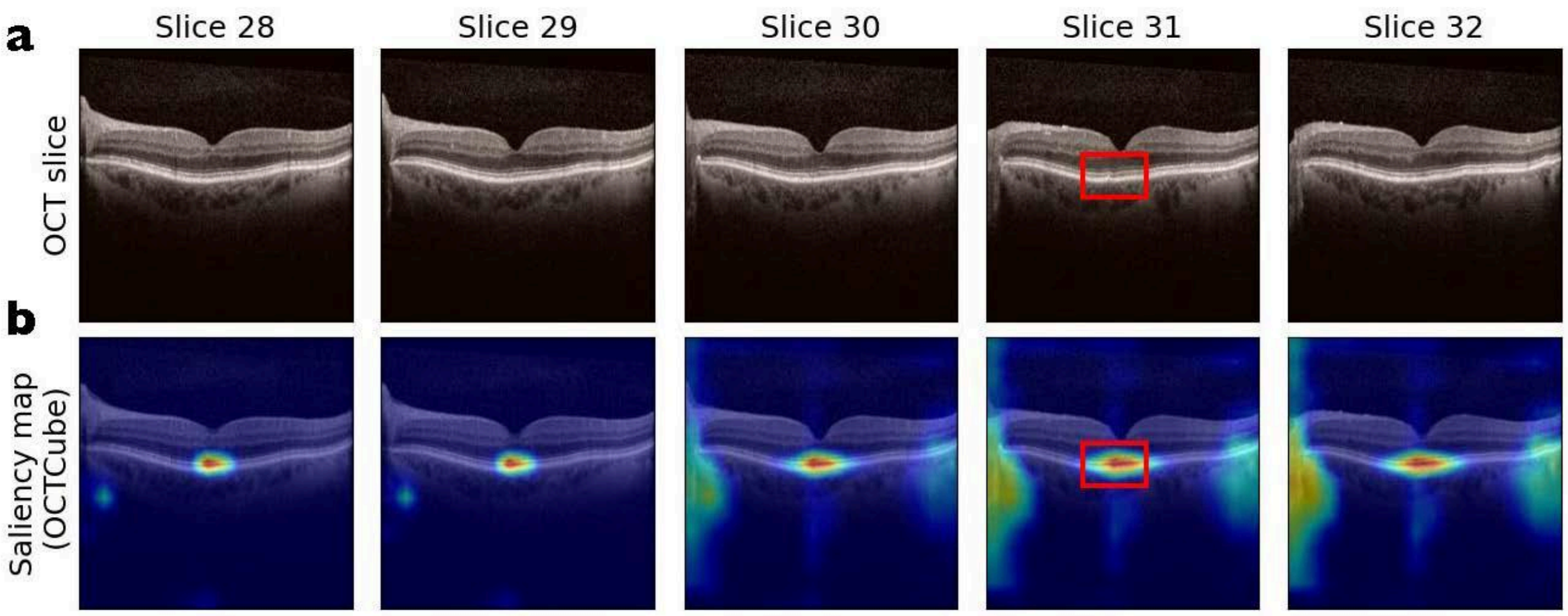

**Supplementary Fig. 6 a-b,** Visualization of multiple slices across the slow-scan dimension from the OCT volume studied in **Fig. 1e** with the OCT slices (**a**) and saliency maps based on the prediction of OCTCube (**b**). The drusen area occurs in slice 31, as also indicated by the generated saliency maps. Images are resized to (256, 256) for the purpose of visualization. Red pixels in (**b**) indicate higher importance, while blue pixels indicate lower importance. The red bounding boxes are manually drawn based on the saliency map.

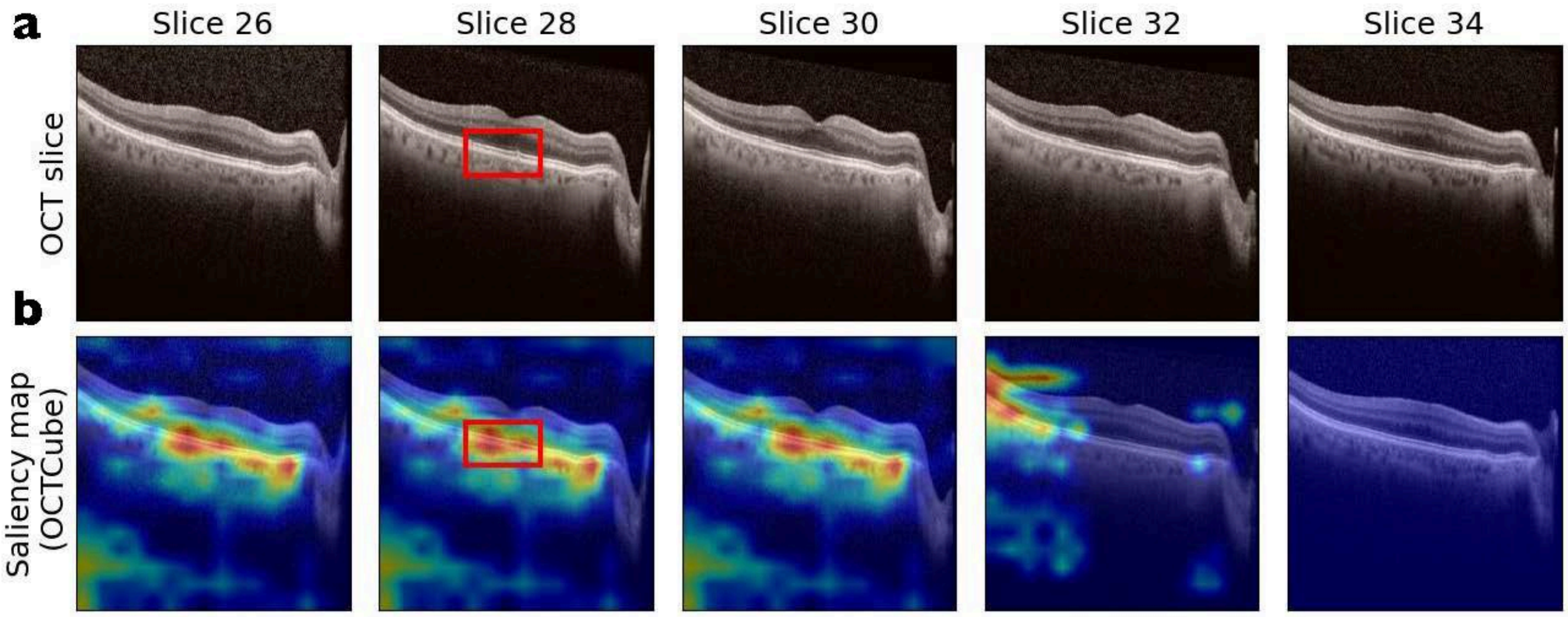

**Supplementary Fig. 7 a-b,** Visualization of multiple slices across the slow-scan dimension from the OCT volume studied in **Fig. 1f** with the OCT slices (**a**) and saliency maps based on the prediction of OCTCube (**b**). The drusen area occurs in slice 28, as also indicated by the generated saliency maps. Images are resized to (256, 256) for the purpose of visualization. Red pixels in (**b**) indicate higher importance, while blue pixels indicate lower importance. The red bounding boxes are manually drawn based on the saliency map.

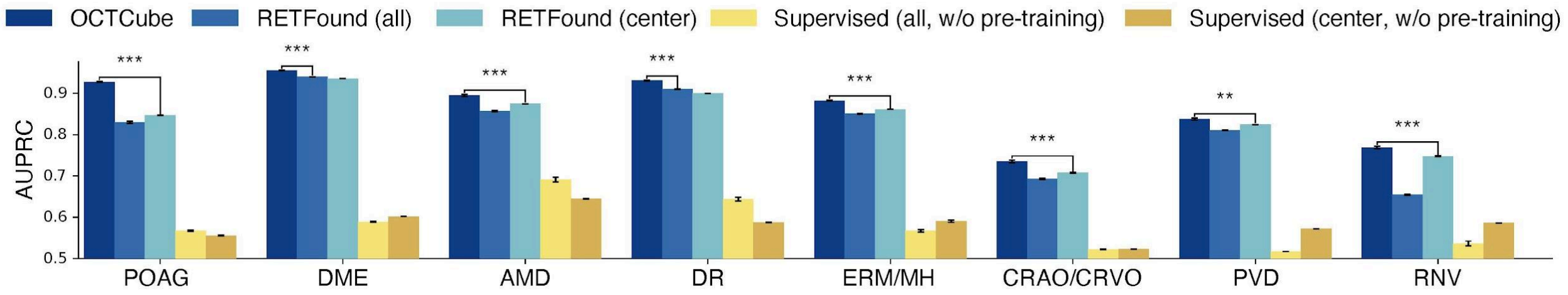

**Supplementary Fig. 8** Barplots comparing OCTCube and competing methods on disease classification of 8 retinal diseases on UW Ophthalmology dataset in terms of AUPRC. Inductive learning setting is used to ensure that test OCT volumes are not seen by OCTCube in the pre-training stage. The train:validation:test split is set to be 60%:20%:20%. POAG, DME, AMD, ERM/MH, DR, CRAO/CRVO, PVD, RNV denote primary open-angle glaucoma, diabetic macular edema, age-related macular degeneration, epiretinal membrane or macular hole, diabetic retinography without macular edema, central retinal vein / artery occlusion, posterior vitreous detachment, and retinal neovascularization respectively. Supervised approaches do not have a pre-training stage. RETFound (all) and Supervised (all) average the embeddings of all slices within a 3D volume. $*$ indicates the significance level at which OCTCube outperforms the best-competing method, with paired t-test p-value $< 5\times10^{-2}$ for *, p-value $< 1 \times 10^{-2}$ for **, p-value $< 1 \times 10^{-3}$ for ***.

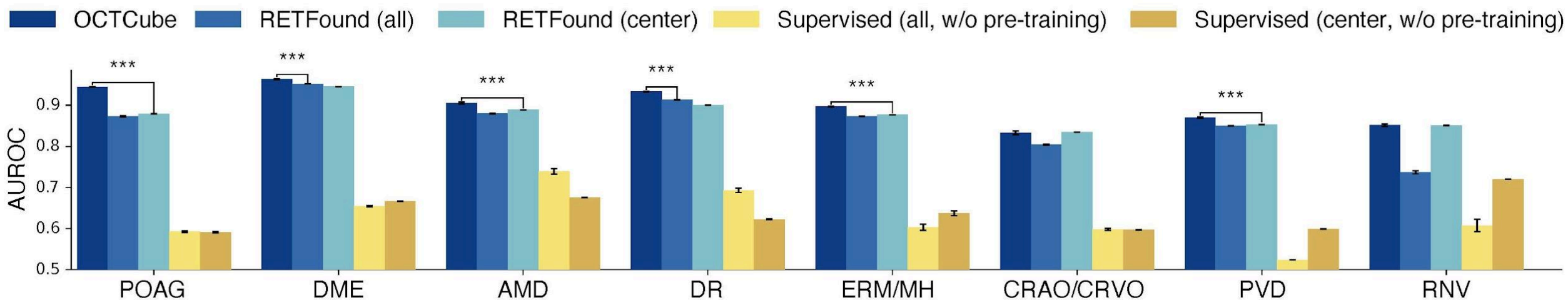

**Supplementary Fig. 9** Barplots comparing OCTCube and competing methods on disease classification of 8 retinal diseases on UW Ophthalmology dataset in terms of AUROC. Inductive learning setting is used to ensure that test OCT volumes are not seen by OCTCube in the pre-training stage. The train:validation:test split is set to be 60%:20%:20%. POAG, DME, AMD, PM, DR, CRAO/CRVO, VD, RNV denote primary open-angle glaucoma, diabetic macular edema, age-related macular degeneration, epiretinal membrane or macular hole, diabetic retinography without macular edema, central retinal vein / artery occlusion, vitreous degeneration, and retinal neovascularization respectively. Supervised approaches do not have a pre-training stage. RETFound (all) and Supervised (all) average the embeddings of all slices within a 3D volume. $*$ indicates the significance level at which OCTCube outperforms the best-competing method, with paired t-test p-value $< 5\times10^{-2}$ for *, p-value $< 1 \times 10^{-2}$ for **, p-value $< 1 \times 10^{-3}$ for ***.

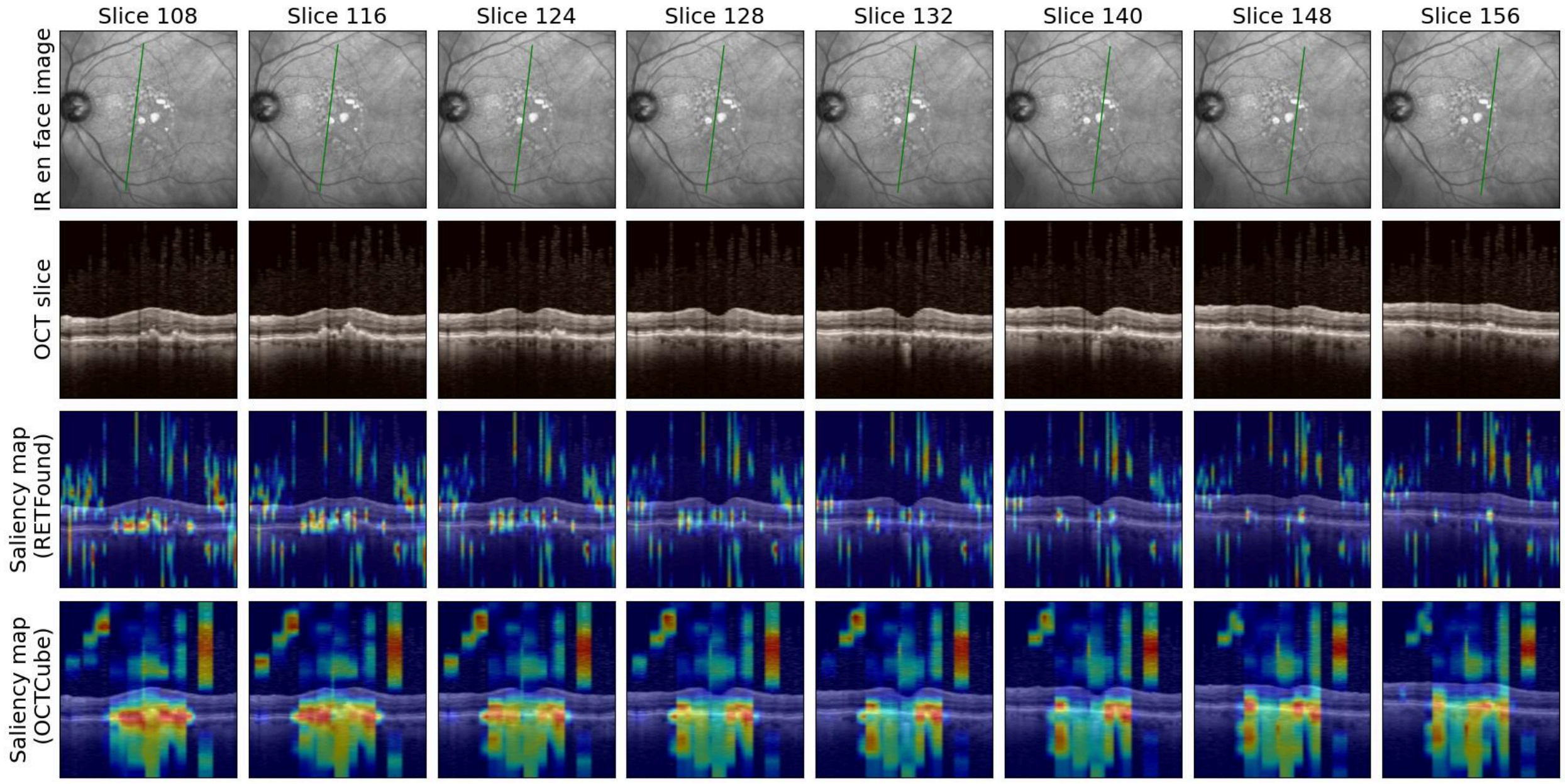

**Supplementary Fig. 10** Visualization of multiple slices across the slow-scan dimension from a single OCT volume with the sampling location (**1st row**) in the corresponding IR en face image, OCT slices (**2nd row**), saliency maps based on the prediction of RETFound (center) (**3rd row**), and saliency maps based on the prediction of OCTCube (**4th row**). OCTCube provides a more coherent saliency map across slices in the diseased area, indicating the effectiveness of the 3D modeling. The aspect ratio is adjusted from (61, 496) to (256, 256) for the purpose of visualization. Red pixels in the third and the fourth rows indicate higher saliency. Green lines in the first row are drawn with the consideration of pixel spacing of sampled OCT slices.

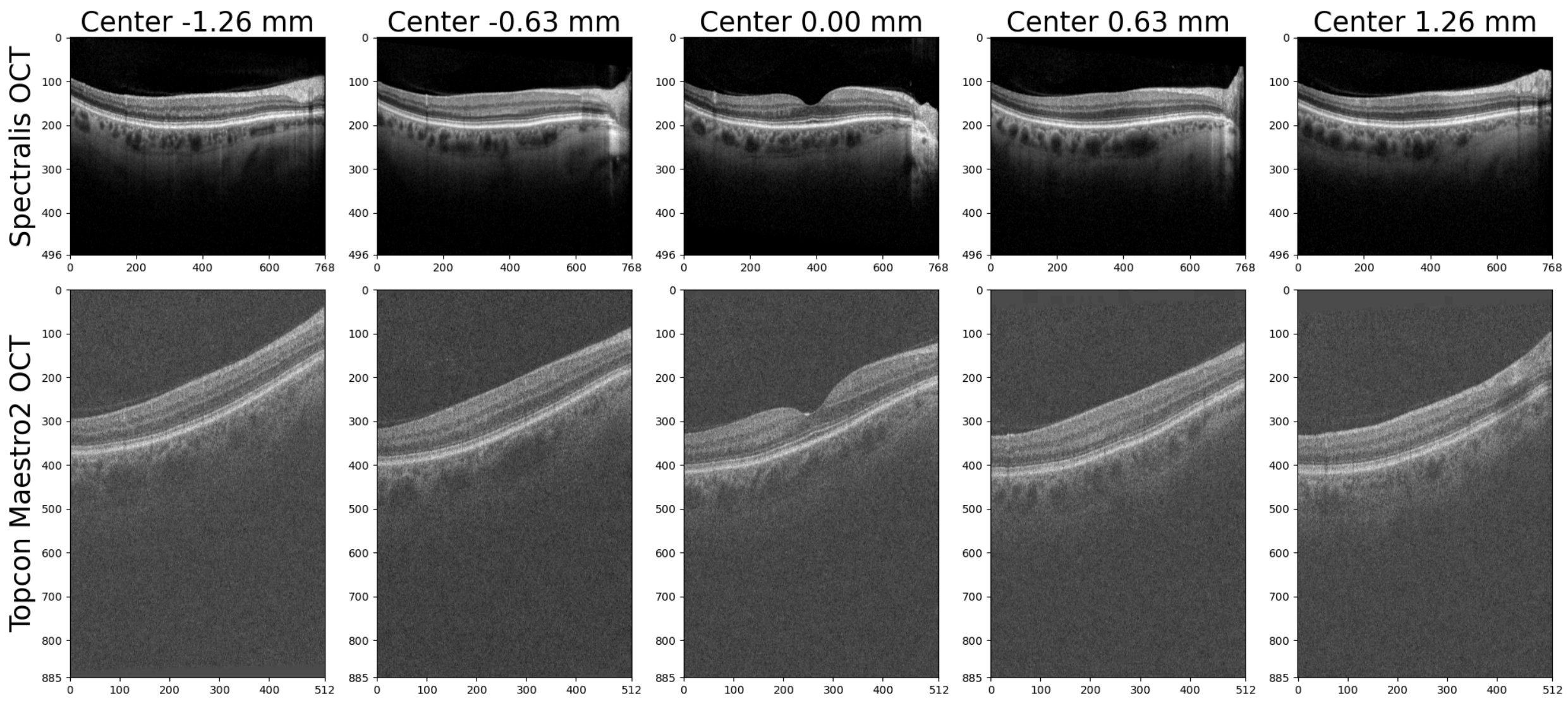

**Supplementary Fig. 11** Example OCT slices acquired by Heidelberg Spectralis device with 30 degree cube scan pattern (**1$^{st}$ row**) and Topcon Maestro2 device with 20 degree cube scan pattern (**2$^{nd}$ row**). Both OCT volumes are from the same eye of the same patient acquired at the same day in the AI-READI dataset. OCT slices are extracted based on the distance to the center slice. The Spectralis OCT slices have the resolution of (496, 768) corresponding to (1.92, 8.93) mm. The Maestro2 OCT slices have the resolution of (885, 512) corresponding to (2.56, 6) mm. Note that the width of Spectralis scan was calibrated with corneal curvature to account for geometry of different eyes, while Maestro2 by default assumes average axial length (i.e. 20 degree scan is 6 mm).

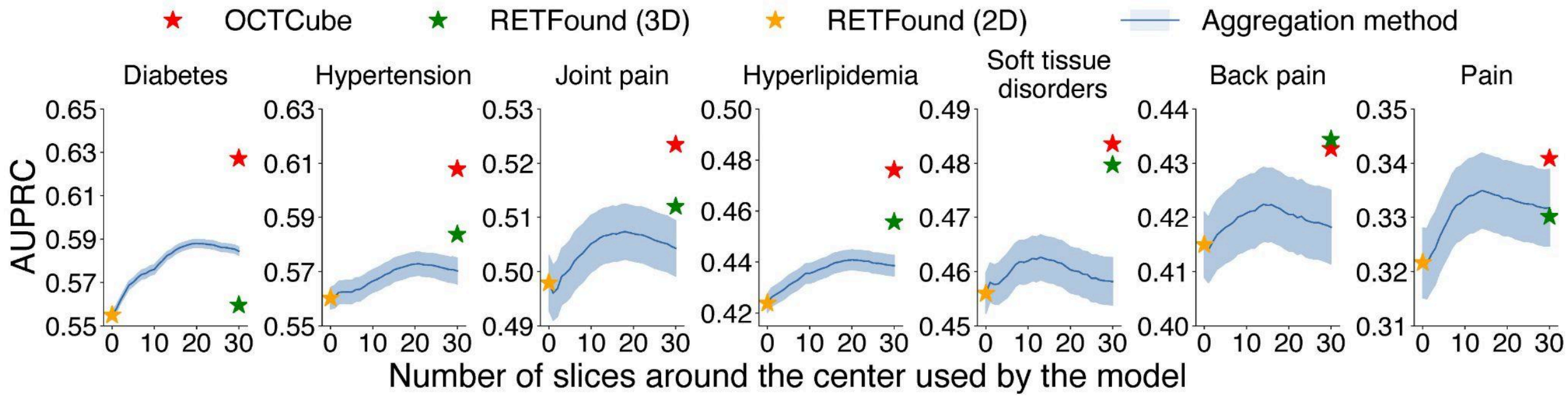

**Supplementary Fig. 12** Plots showing the AUPRC performance of an aggregation approach predicted probabilities of k slices around the center slices on seven systemic diseases, where k is shown in the x-axis. RETFound, as a 2D approach, corresponds to k = 0. The prediction probability is derived using RETFound. The improved performance by considering more slices necessitates the development of 3D models. Plots showing the AUPRC performance of an aggregation approach predicted probabilities of k slices around the center slices on seven systemic diseases, where k is shown in the x-axis. The prediction probability is derived using RETFound. The RETFound (center) model, as a 2D approach, corresponds to k = 0. Different from directly aggregating predictions, the RETFound (all) model uses a neural network to aggregate features. The improved performance by considering more slices necessitates the development of 3D models.

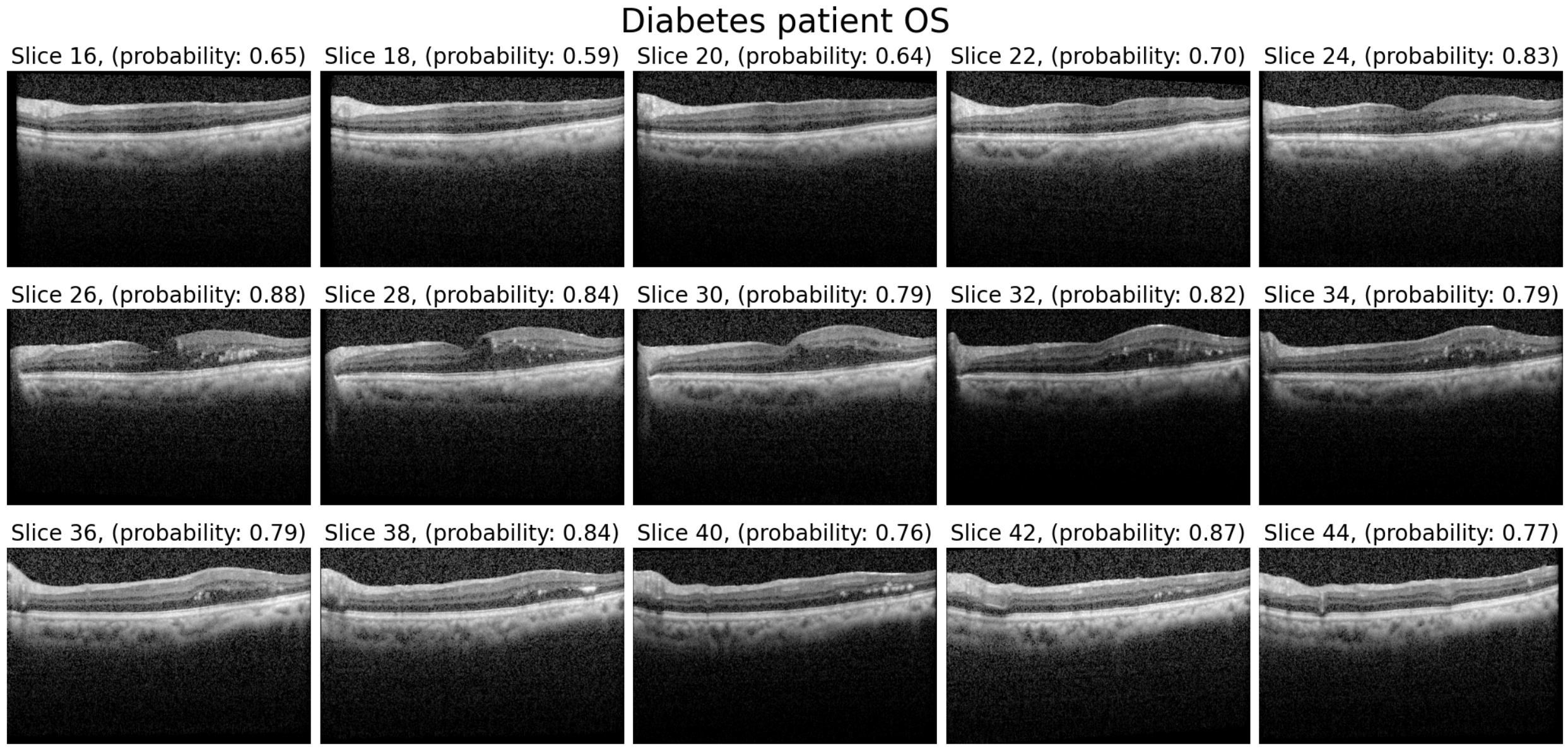

**Supplementary Fig. 13** Visualization of OCT slices from the left eye (OS) of the same patient with diabetes studied in **Fig. 4f** acquired on the same day. Slice 26 and 30 are also shown in **Fig. 4g**. Macular edema and hard exudates are more clearly observed in several slices including the center slice compared to **Fig. 4f**. Both OCTCube and RETFound (center) successfully predict diabetes using this OCT volume.

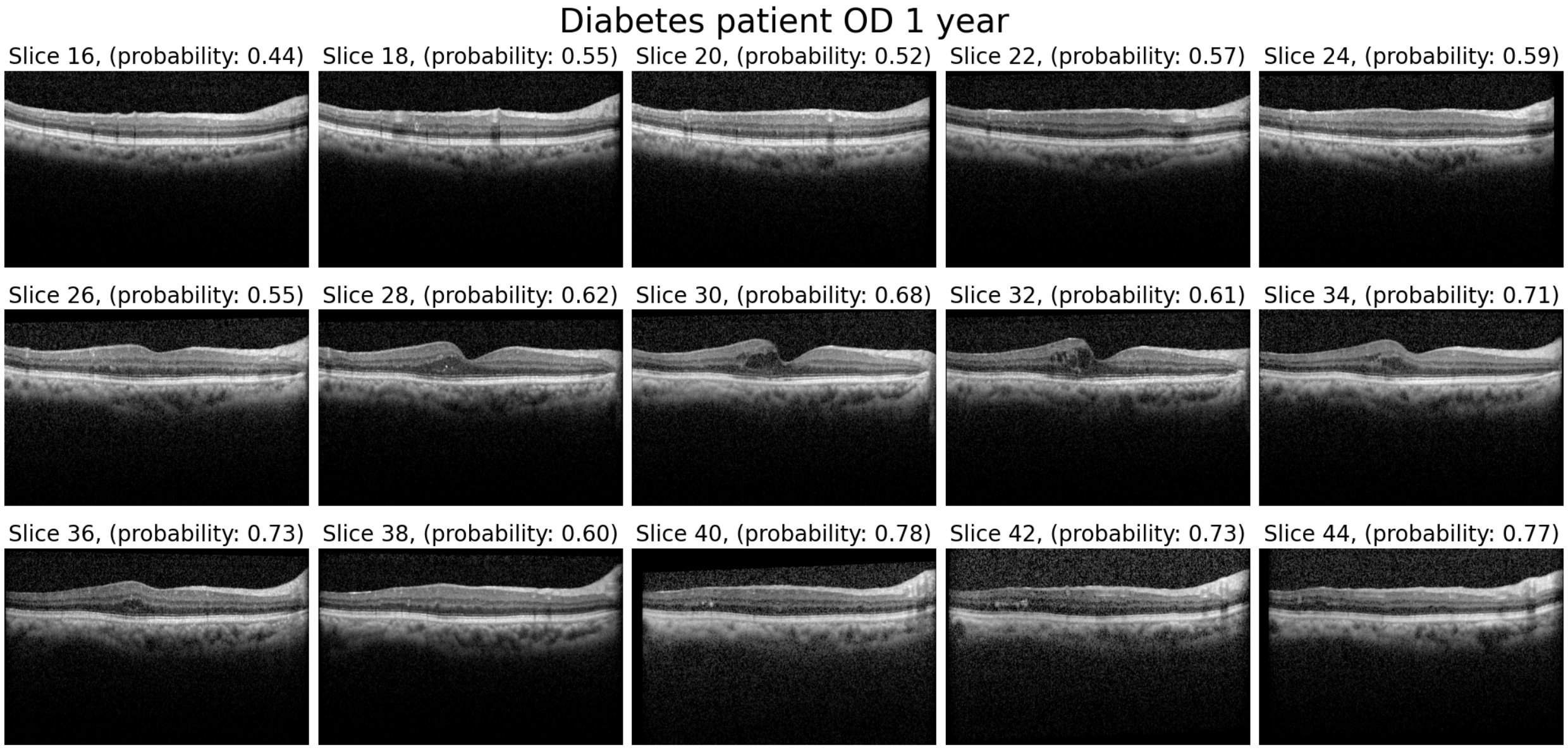

**Supplementary Fig. 14** Visualization of OCT slices from the right eye (OD) of the same patient with diabetes studied in **Fig. 4f** acquired after 1 year. Slice 30 and 34 are also shown in **Fig. 4h**. Macular edema and hard exudates are more clearly observed in several slices including the center slice compared to **Fig. 4f**. Both OCTCube and RETFound (center) successfully predict diabetes using this OCT volume.

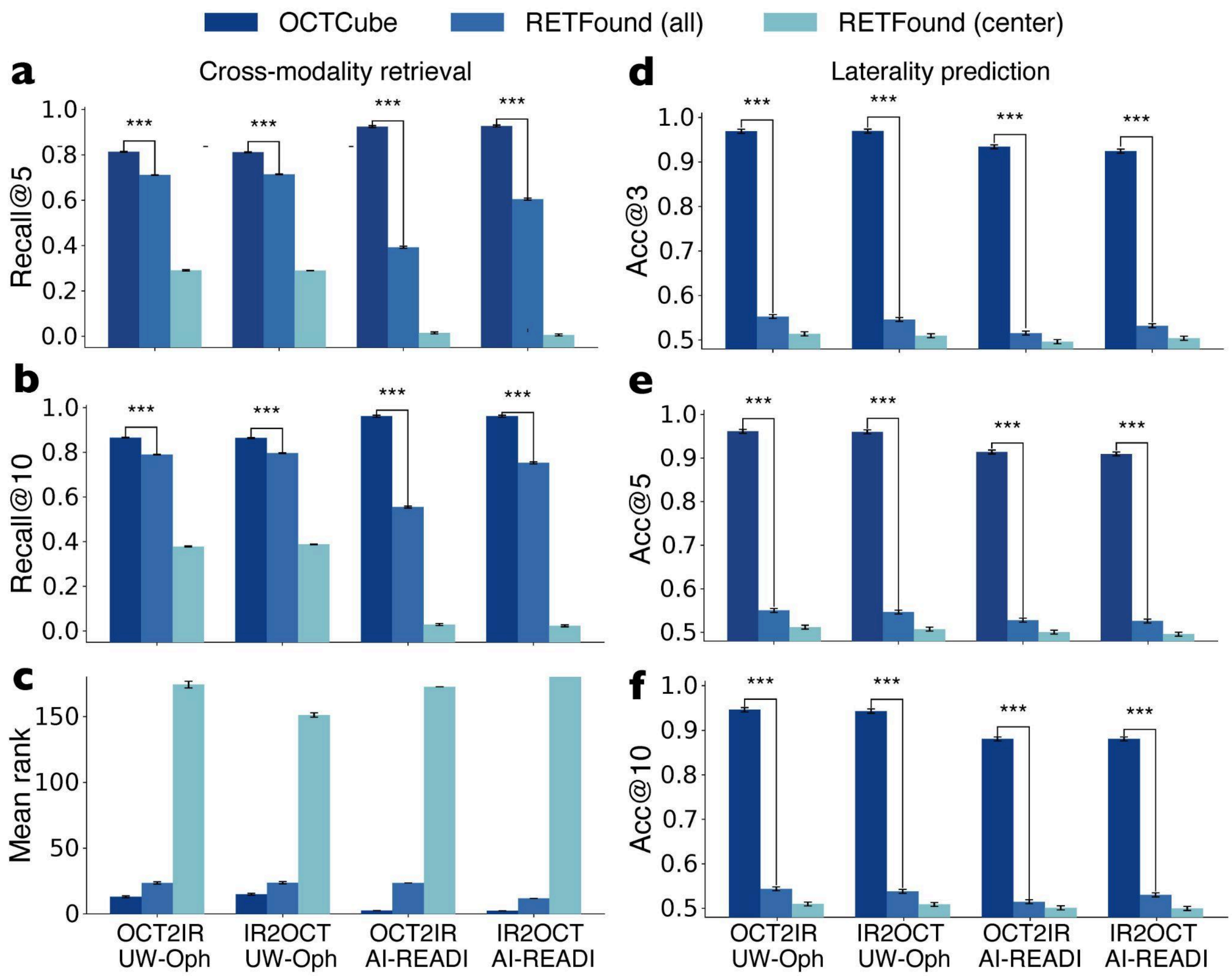

**Supplementary Fig. 15 a-c,** Barplots comparing OCTCube and competing methods on cross-modality retrieval on UW Ophthalmology and AI-READI dataset in terms of recall@5 (a), recall@10 (**b**) and mean rank (**c**) score on OCT to IR retrieval and IR to OCT retrieval. * indicates the significance level at which OCTCube outperforms the best-competing method, with paired t-test p-value < $5\times10^{-2}$ for *, p-value < $1 \times 10^{-2}$ for **, p-value < $1 \times 10^{-3}$ for ***. **d-f** Barplots comparing OCTCube and competing methods on cross-modality laterality prediction on UW Ophthalmology and cross-AI-READI dataset in terms of accuracy@3 (**d**), accuracy@5 (**e**) and accuracy@10 (**f**) on OCT to IR retrieval and IR to OCT retrieval. * indicates the significance level at which OCTCube outperforms the best-competing method, with paired t-test p-value < $5\times10^{-2}$ for *, p-value < $1 \times 10^{-2}$ for **, p-value < $1 \times 10^{-3}$ for ***.

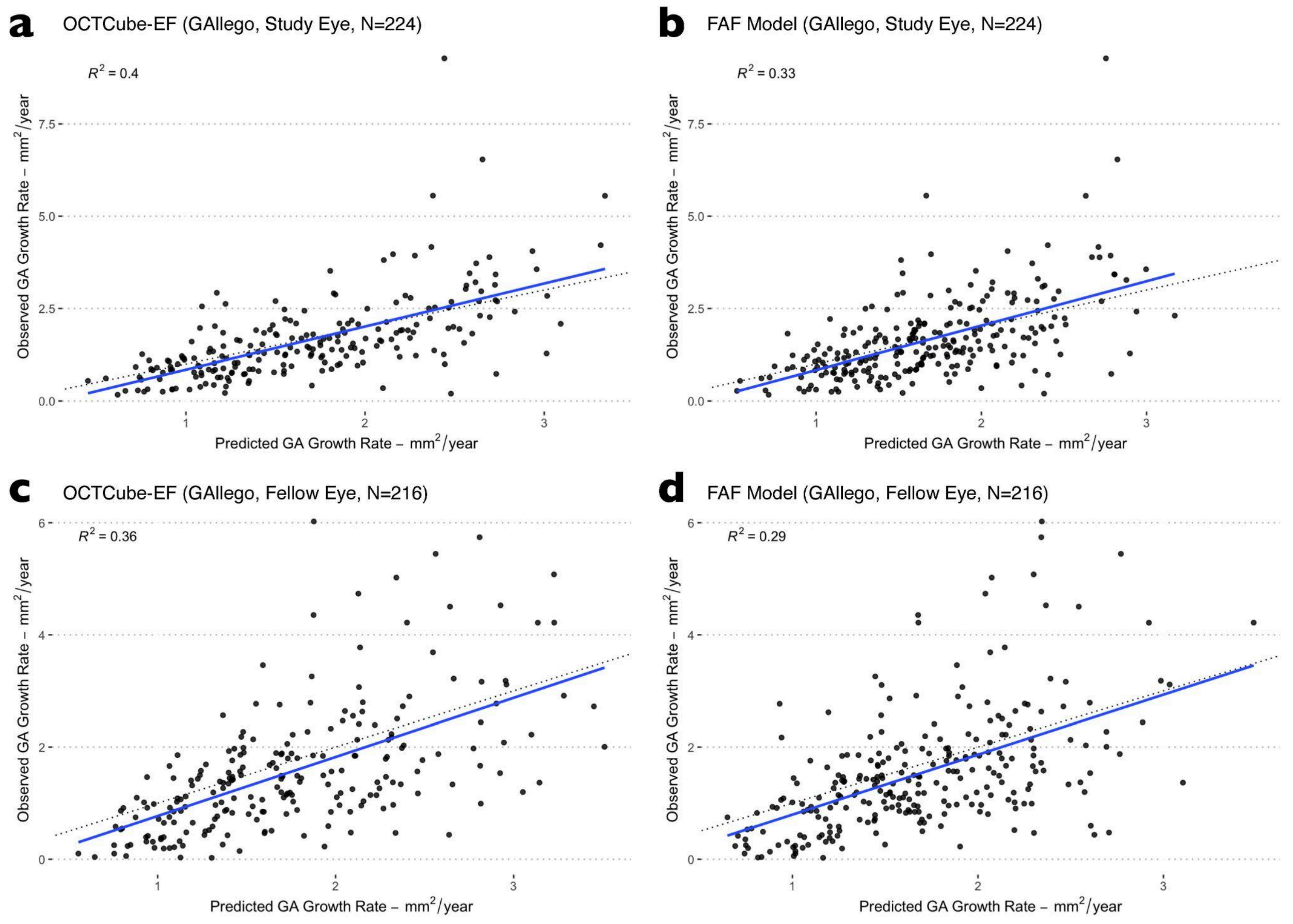

**Supplementary Fig. 16 GA lesion size growth rate prognostic model performance. a-d,** Scatter plots showing the association between ground truth GA lesion size growth rate vs. model predicted GA lesion size growth rate. OCTCube-EF GA prognostic model achieved a $R^2$ of 0.4 on the treatment arm (Study Eye, **a**) and 0.36 on the control arm (Fellow Eye, **c**), while the best baseline DenseNet FAF model only achieved a $R^2$ of 0.33 on the treatment arm (Study Eye, **b**) and 0.29 on the control arm (Fellow Eye, **d**). The blue line is the fitted line using linear regression. The gray dotted line reveals y=x. The $R^2$ is reported as the square of the Pearson Correlation Coefficient.

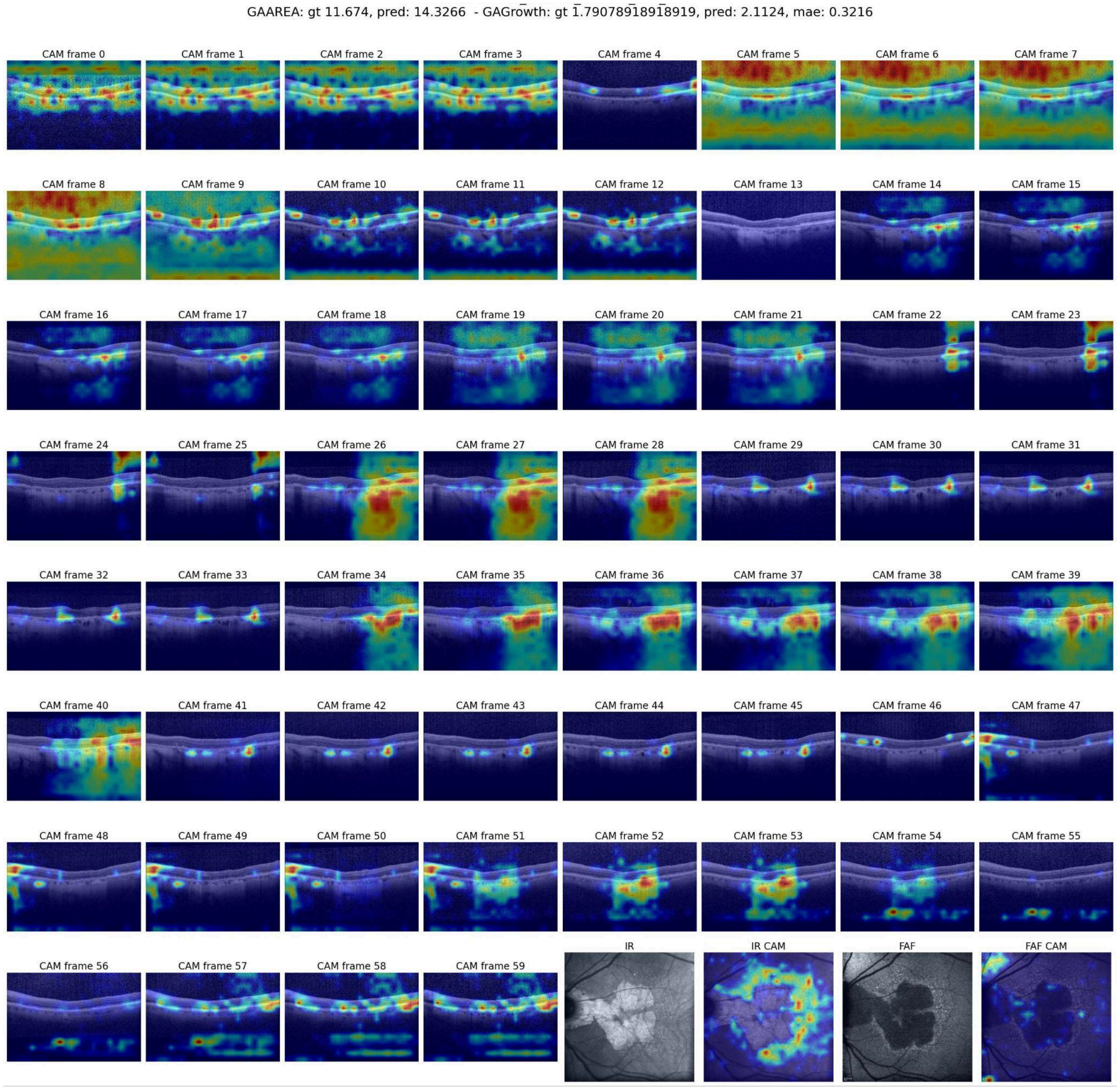

**Supplementary Fig. 17** Visualization of the saliency map generated by OCTCube-EF GA growth rate prognostic model on a GA eye. This eye has a large GA lesion (11.7 $mm^2$) surrounding but spared part of the fovea. Multiple OCT B scan slices across the slow-scan dimension from a single OCT volume were presented (the 30-th slice is shown in **Fig. 6g**), as well as IR, FAF and their saliency map (the last 4 columns of the last row). On B-scans away from the GA lesion (slide 0-8, 54-59), the saliency was dispersed, which may indicate the model assessed the overall status. On B-scans near or in the GA lesion, two classes of anatomical structures got high saliency; on slice 9 to 40, the saliency generally focused on sub-healthy photoreceptor and RPE cells near the GA lesion, while on slices 41 to 53 the saliency located more on the thinned choroidal layer. The saliency on IR focused mainly on the GA lesion

border, and on FAF highlighted a region with higher autofluorescence signals. The predicted GA growth rate is 2.1, close to the actual rate of 1.8 $mm^2$/year.

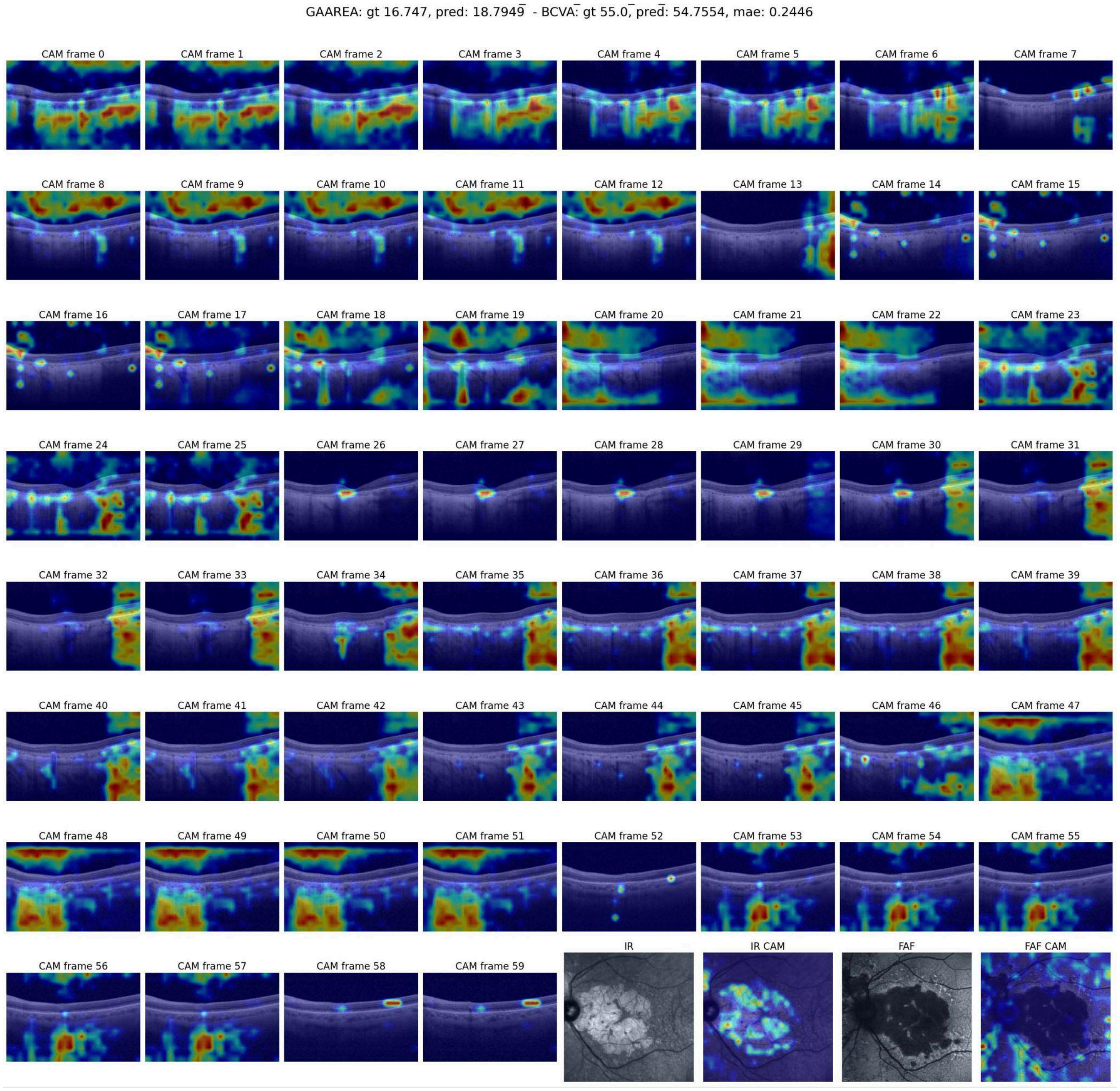

**Supplementary Fig. 18** Visualization of the saliency map generated by OCTCube-EF GA best corrected visual acuity (BCVA) prediction model on a GA eye. This eye has a large GA lesion (16.7 mm$^2$) surrounding but spared part of the fovea. Multiple OCT B scan slices across the slow-scan dimension from a single OCT volume were presented (the 28-th slice is shown in **Fig. 6h**). For the region at fovea (slice 26-30), the saliency focused on the few but remaining photoreceptors that explained the eye moderate vision (BCVA of 55.0, while model predicted accurately at 54.8) despite its large GA lesion. For regions away from fovea, the saliency was generally dispersed, which may indicate the model assessed overall structure to decide its non-fovea location.

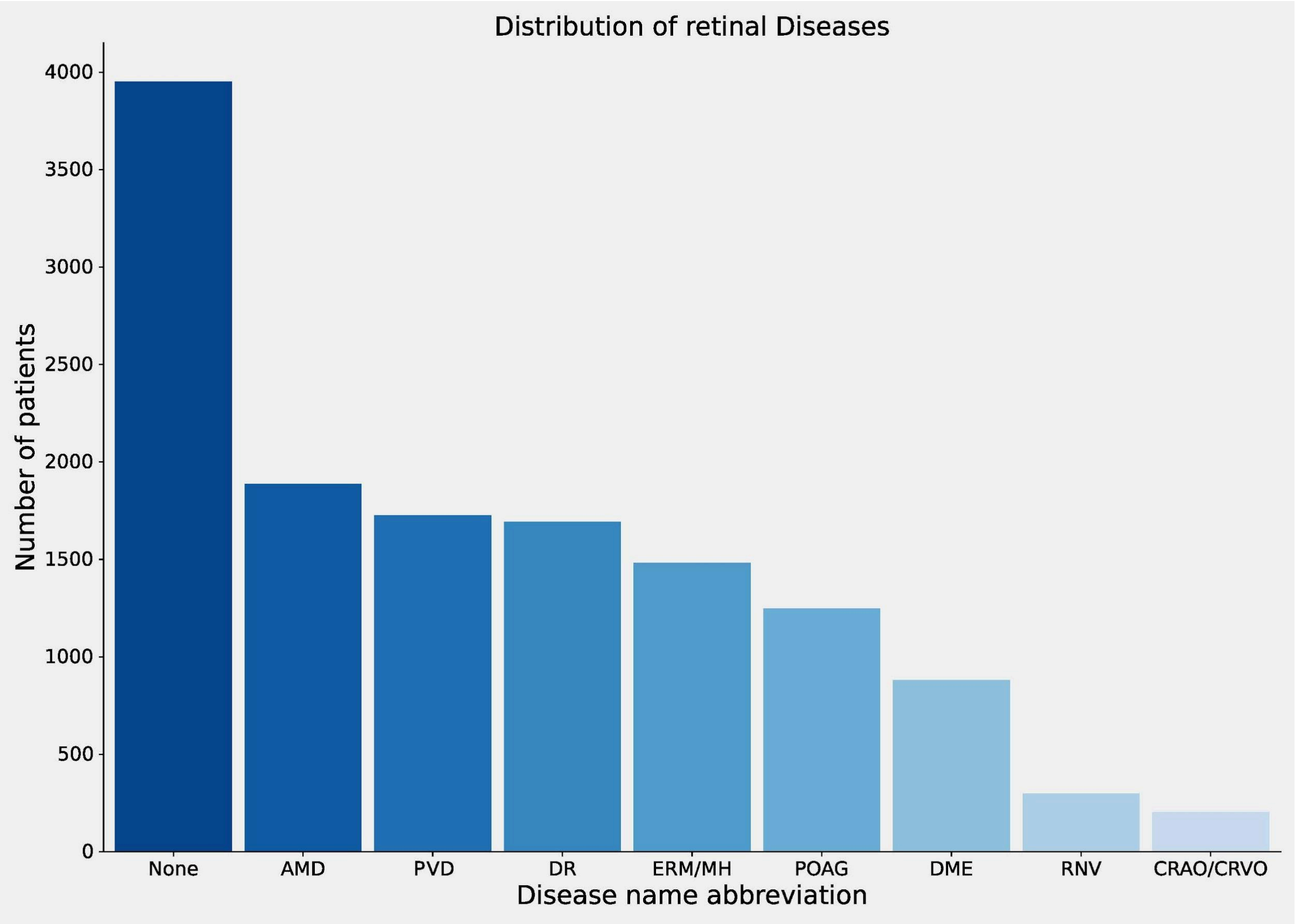

**Supplementary Fig. 19 Disease distribution of retinal disease patients.** Barplots showing number of included patients for the retinal disease prediction task. Explicit class imbalance is observed across different retinal diseases. POAG, DME, AMD, ERM/MH, DR, CRAO/CRVO, PVD, RNV denote primary open-angle glaucoma, diabetic macular edema, age-related macular degeneration, epiretinal membrane or macular hole, diabetic retinography without macular edema, central retinal vein / artery occlusion, posterior vitreous detachment, and retinal neovascularization respectively.

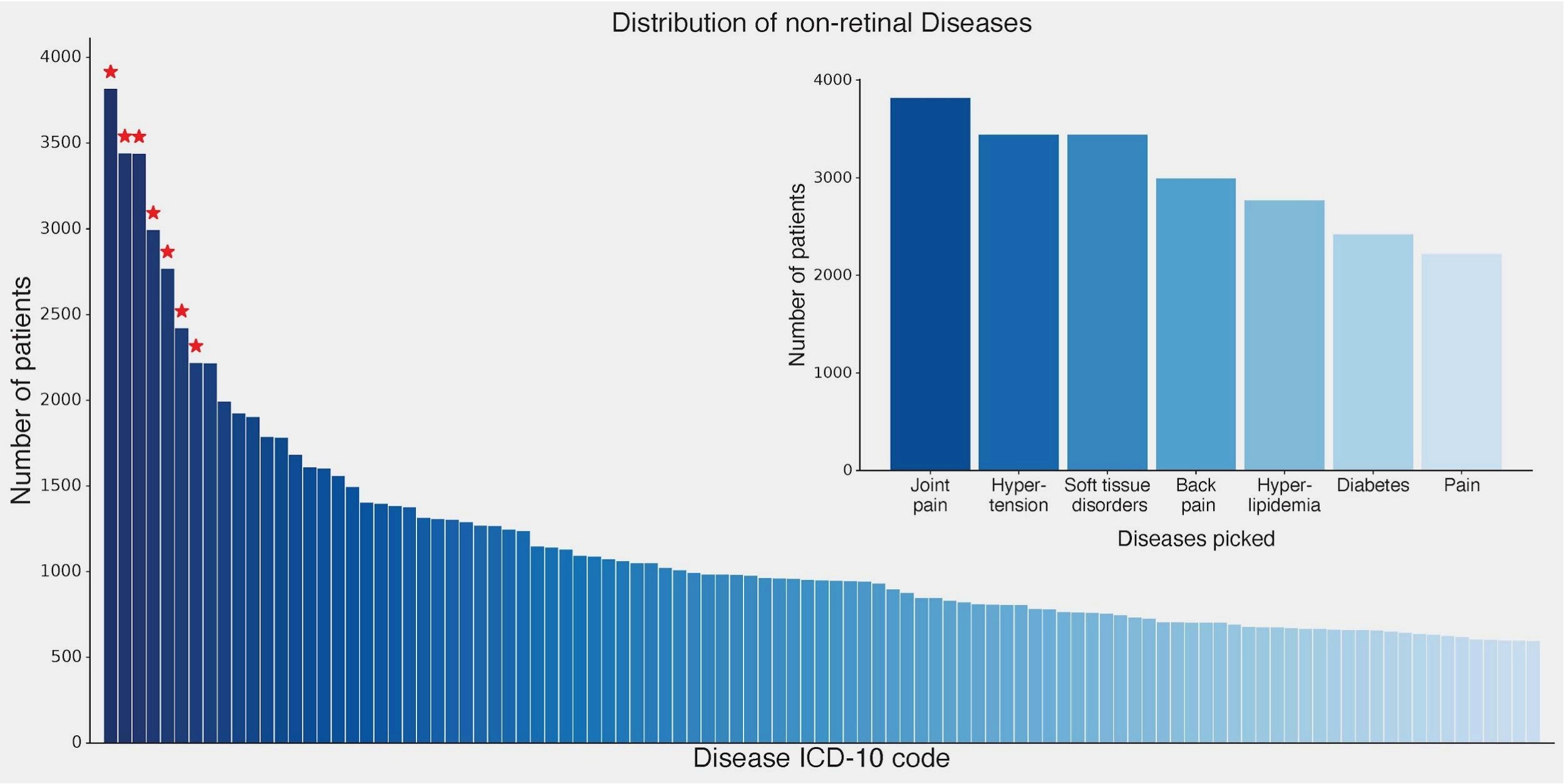

**Supplementary Fig. 20 Disease distribution of systemic disease patients. Left,** Barplots of the top 100 frequent diseases included in the non-oph disease prediction tasks. Each disease is collected and aggregated based on their level 1 ICD-10 code. **Right,** Barplots of the number of patients for the seven selected systemic diseases. M25, I10, M79, M54, E78, E11, G89 refers to joint pain, hypertension, soft tissue disorders, back pain, hyperlipidemia, diabetes, pain. The diseases are selected if their balanced accuracy on the validation set is significantly larger than random guess with p-value $< 0.001$.

**Supplementary Table. 1 Table listing ICD-9 and ICD-10 code of 8 retinal diseases considered in within-dataset retinal disease prediction task.** Level 1 code clustered in brackets (e.g., (E08-E11, E13)) indicates the same level 2 code or code series. x denotes any digit between 0-9. Codes linked with `-' indicate consecutive increased digits. POAG, DME, AMD, ERM/MH, DR, CRAO/CRVO, PVD, RNV denote primary open-angle glaucoma, diabetic macular edema, age-related macular degeneration, epiretinal membrane or macular hole, diabetic retinography without macular edema, central retinal vein / artery occlusion, posterior vitreous detachment, and retinal neovascularization respectively. Please see Supplementary Table 1 in the uploaded spreadsheet.